\documentclass[aps,prd,amssymb, amsmath,preprint,nofootinbib,nobibnotes]{revtex4}
\usepackage{graphicx}
\usepackage{dcolumn}
\usepackage{bm}
\def\lsim{\mathrel{\rlap{\lower3pt\hbox{\hskip0pt$\sim$}}
    \raise1pt\hbox{$<$}}}         
\def\gsim{\mathrel{\rlap{\lower4pt\hbox{\hskip1pt$\sim$}}
    \raise1pt\hbox{$>$}}}         
\def\simlt{\mathrel{\raise.3ex\hbox{$<$\kern-.75em\lower1ex\hbox{$\sim$}}}}
\def\simgt{\mathrel{\raise.3ex\hbox{$>$\kern-.75em\lower1ex\hbox{$\sim$}}}}

\def\slashchar#1{\setbox0=\hbox{$#1$}           
   \dimen0=\wd0                                 
   \setbox1=\hbox{/} \dimen1=\wd1               
   \ifdim\dimen0>\dimen1                        
      \rlap{\hbox to \dimen0{\hfil/\hfil}}      
      #1                                        
   \else                                        
      \rlap{\hbox to \dimen1{\hfil$#1$\hfil}}   
      /                                         
   \fi}                                         %

\begin{document}
\draft
\title{Squark Pair Production in the MSSM with Explicit CP Violation}
\author{Ahmet T. Alan$^{a}$, Kerem Canko{\c c}ak$^{b}$ and Durmu{\c s} A.
Demir$^{c}$}
\affiliation{$^{a}$Department of Physics, Abant Izzet
Baysal University, Turkey, TR14280} \affiliation{$^{b}$Department
of Physics, Mu{\~g}la University, Turkey, TR48000}
\affiliation{$^{c}$ Department of Physics, Izmir Institute of
Technology, IZTECH, Turkey, TR35430}
\preprint{IZTECH-P01/2007\; October 2007}
\begin{abstract}

We analyze effects of the CP-odd soft phases in the MSSM on the
pair-productions of colored superpartners in $p\, p$ collisions at
the LHC energies. We find that, among all pair-production
processes, those of the scalar quarks in the first and second
generations are particularly sensitive to the CP-odd phases, more
precisely, to the phases of the gluinos and neutralinos. We
compute pair-production cross sections, classify various
production modes according to their dependencies on the gluino and
neutralino phases, perform a detailed numerical analysis to
determine individual as well as total cross sections, and give
a detailed discussion of EDM bounds. We find that
pair-productions of first and second generation squarks serve as a
viable probe of the CP violation sources in the gaugino sector of
the theory even if experiments cannot determine chirality, flavor
and electric charge of the squarks produced.

\end{abstract}

\maketitle

\section{Introduction and Motivation}
Supersymmetric extension of the standard model of particle physics
(SM) is one of the most plausible scenarios for new physics to be
discovered at the LHC. Supersymmetry is to be a blatantly broken
symmetry of nature as indicated by negative searches at LEP (and
also at Tevatron). This breaking should occur softly $i.e.$ in a
way not regenerating the quadratic divergences. The soft-breaking
sector of the theory consists of a number of dimensionful
parameters (see the review volume \cite{Chung:2003fi}): the
gaugino masses $M_{\widetilde{g}, \widetilde{W}, \widetilde{B}}$,
trilinear couplings ${\bf Y_{u,d,e}^{A}}$ and sparticle masses $M_{H_u}^2 \cdots {\bf
M_{E}}^2$. These parameters, forming up the soft-breaking lagrangian
\begin{eqnarray}
\label{soft} -{\bf L}_{soft} &=& \frac{1}{2} \left[ M_{\widetilde{g}}
\lambda_{\widetilde{g}}^{a} \lambda_{\widetilde{g}}^{a} + M_{\widetilde{W}}
\lambda_{\widetilde{W}}^{i} \lambda_{\widetilde{W}}^{i} + M_{\widetilde{B}} \lambda_{\widetilde{B}}
\lambda_{\widetilde{B}} + \mbox{h.c.}\right]\nonumber\\
&+&M_{H_d}^{2} H_d^{\dagger} H_d + M_{H_u}^{2} H_u^{\dagger} H_u
-\left[\mu B {H}_d \cdot {H}_u + \mbox{h.c.}\right]\nonumber\\
&+&\left[\widetilde{Q} \cdot {H}_u {\bf Y_u^A} \widetilde{U}^c
+{H}_d \cdot \widetilde{Q} {\bf Y_d^A} \widetilde{D}^c + {H}_d \cdot
\widetilde{L} {\bf Y_e^A}
\widetilde{E}^c + \mbox{h.c.}\right]\nonumber\\
&+& \widetilde{Q}^{\dagger} {\bf M_Q}^{2} \widetilde{Q} +
\widetilde{U}^{c \dagger} {\bf M_U}^{2} \widetilde{U}^c
+\widetilde{D}^{c \dagger} {\bf M_D}^{2} \widetilde{D}^c
+\widetilde{L}^{\dagger} {\bf M_L}^{2} \widetilde{L} +
\widetilde{E}^{c \dagger} {\bf M_E}^{2} \widetilde{E}^c
\end{eqnarray}
are the main goal of measurements to be carried out at the LHC.
The $\mu$ parameter, gaugino masses, trilinear couplings  and
off-diagonal entries of the squark and slepton masses are the
sources of CP violation beyond the SM. Likewise, sfermion
mass-squareds and trilinear couplings are sources of flavor
violation beyond the SM.

The experiments at the LHC are expected to confirm massive
superpartners if nature is supersymmetric around Fermi energies.
Clearly, if measurements at the LHC will suffice to construct the
soft-breaking lagrangian (\ref{soft}) or if measurements will ever
lead to a unique supersymmetric model requires a dedicated
analysis of collider signals and model predictions \cite{nima}
(see also the dedicated review volume \cite{pape}). A further
question concerns role of the soft masses in CP and flavor
violating phenomena, and this can be achieved after a full
experimentation of various meson decays and mixings
\cite{flavor-cp}. In this work, we intend to contribute this
enormous project by a detailed analysis of squark pair production
at $p p$ collisions. These processes have been analyzed in the
past \cite{cross,cross1} (with increasing precision in
\cite{Beenakker:1996ch,Beenakker3,tilman}) by considering only the
(dominant) SUSY QCD contributions. Moreover, spin asymmetries
have been analyzed in \cite{wyler}. Our motivation for and certain
salient features of squark pair-production processes can be
summarized as follows:
\begin{enumerate}

\item Squarks and gluinos are expected to be produced copiously
(via several channels as depicted in Fig. \ref{dia} for a generic
squark flavor $\widetilde{q}$). This is not the case for leptons,
neutralinos and charginos whose direct productions are initiated
by quark--anti-quark annihilation, only.

\item Squark pair production is dominated by SUSY QCD
contributions $i.e.$ gluino exchange. However, it is important to
consider also the exchange of electroweak gauginos  since
interference of gluino-- and gaugino--mediated amplitudes can be
sizeable.

\item As illustrated in Fig. \ref{dia}, $\widetilde{q} \widetilde{q}^{\star}$
production receives significant SM contributions from photon, $Z$
and gluon exchanges in the $s$-channel. In addition, there are
purely supersymmetric contributions from $t$-channel gluino and
neutralino exchanges. The whole process is $p$-wave. The
dependence on the supersymmetric CP-odd phases involves only the
difference between neutralino phases, in particular, there is no
dependence on the phase of the gluino mass. The reason is that the
vertices connecting to $\widetilde{q}$ and $\widetilde{q}^{\star}$
interfere destructively due to complex conjugation.

The production of $\widetilde{q} \widetilde{q}$ pairs proceeds
solely with the sparticle mediation, as shown in Fig.1 part (c).
This purely supersymmetric amplitude describes an $s$-wave
scattering, and thus, near the two-squark threshold $\widetilde{q}
\widetilde{q}$ events can dominate $\widetilde{q}
\widetilde{q}^{\star}$ ones. Moreover, unlike $\widetilde{q}
\widetilde{q}^{\star}$ production amplitude, $\widetilde{q}
\widetilde{q}$ production involves both neutralino and gluino
phases due to constructive interference between the two vertices
connecting to $\widetilde{q}$ lines. Thus, number of such events
must exhibit a stronger sensitivity to CP-odd phases than those
pertaining to $\widetilde{q} \widetilde{q}^{\star}$ production.

\item
Productions of various squark pairs are governed by Feynman
diagrams in Fig. \ref{dia}. It is clear that pair-production of
third generation squarks, stop and sbottom
\cite{beenakker2,stoppair} (as well as charm squark, to a lesser
extent) receives only a tiny contribution from the third diagram
in Fig. \ref{dia} (a) and from the two diagrams in Fig. \ref{dia}
(c). In other words, stops and sbottoms are produced dominantly in
$\widetilde{t}_i \widetilde{t}_j^{\star}$ and $\widetilde{b}_i
\widetilde{b}_j^{\star}$ modes (with a tiny contamination of
$\widetilde{t}_i \widetilde{t}_j$ and $\widetilde{b}_i
\widetilde{b}_j$ final states). The reason is that heavy quarks
form an exceedingly small fraction of the proton substructure and
flavor mixings (especially between the first and other two
generations) are suppressed by the FCNC bounds \cite{flavor-cp}
(Various observables, including the ones pertaining to the Higgs
sector, can be significantly affected if sizeable flavor violation
effects are allowed in sfermion soft mass-squareds
\cite{flavor-cp2}). This then, however, implies the absence of CP
violation effects in production of third generation squarks; more
explicitly, given physical masses and mixings of stops and
sbottoms then their production rates do not depend on any
additional parameter, in particular, the CP-odd phases. This
observation holds for all stop and sbottom pair-production modes
including $\widetilde{t}_1 \widetilde{t}_2$ and $\widetilde{b}_1
\widetilde{b}_2$ since the only contributing phase, the phase of
the LR block in their mass-squared matrices, factors out.

Unlike sbottoms and stops, squarks in first and second generations
can be produced with significant rates via all the diagrams in
Fig. \ref{dia}. Therefore, one expects up, down, strange and charm
scalar quarks, especially scalar up and down quarks, to be
produced in significant amounts at the LHC such that
\begin{itemize}
\item their mass and gauge eigenstates (especially for scalar up
and down quarks) are identical due to their exceedingly small
Yukawa couplings,

\item flavor and gauge eigenstates of scalar up and down quarks
are identical whereas scalar strange quark might possesses
significant flavor mixings with scalar bottom quark,

\item they feel only gaugino contributions and hence their
production rates are viable probes of CP violation in the gaugino
sector.
\end{itemize}

\item In general, in SUSY-QCD sector, pair-production of colored particles
involves not only the squark pairs  but also gluino pairs and gluino-squark
events. The reason we focus mainly on the squark pair-production is
that gluino pair-production and gluino-squark associated production are not
sensitive to CP-odd phases in the theory. In this sense, given the
pair-productions of colored particles then one knows that it is only
the pairs of first and second  generation squarks that can have a
significant sensitivity to the CP-odd phases.

\end{enumerate}
In accord with these observations, in this work we discuss
pair-productions of the squarks belonging to first and second
generations only, and focus on their sensitivities to SUSY CP-odd
phases by considering $\widetilde{q} \widetilde{q}$ and
$\widetilde{q} \widetilde{q}^{\star}$ events in a comparative
fashion.

There is no doubt that electric dipole moments (EDMs) are the prime observables which determine
the allowed sizes of the supersymmetric phases. However, the fact that EDMs can cancel out for a
wide range of CP-odd phases (except for the $\mu$ parameter which must be nearly real)
\cite{cancel,cancel1,cancel2}, the fact that EDMs and squark pair-production processes depend on
different combinations of the phases (see for instance the slepton pair-production \cite{thomas}),
the fact that EDMs can receive sizeable contributions from Higgs exchange while squark production
cannot \cite{biz}, the fact that EDMs are sensitive to CP-violating new physics beyond the MSSM
\cite{beyondmssm}, and finally the fact that EDMs are sensitive to even the phases occurring at
two-loop level \cite{2loop} all encourage us to study impact of the CP-odd phases on squark pair-
production since, despite ${\cal{O}}(1)$ values for phases, EDMs can be sufficiently suppressed in
certain regions of the SUSY parameter space \cite{cancel,cancel1,cancel2}. Consequently, we
will first analyze squark pair-production in an mSUGRA-like scenario with all the phases varying
in their full ranges. Then we will discuss impact of EDMs in a separate section
by considering certain EDM-favored parameter domains already obtained in the literature.

It is expected that at energies probed by LHC experiments the SUSY
QCD corrections can be substantial. From the dedicated analysis of
\cite{Beenakker:1996ch} we know that $\sigma_{NLO} \approx
\sigma_{LO}$ when $m_{\widetilde{q}}/m_{\widetilde{g}} \sim 1$,
and this will be indeed the case at least for squarks of first two
generations. Moreover, when the decoupling/renormalization scale
$Q\sim m_{\widetilde{q}}$ again LO and NLO cross sections lie
closer. Therefore, lack of NLO corrections in cross sections which
will be computed in Sec. II below may not cause a substantial error
in estimates (at least for analyzing effects of supersymmetric CP
odd phases). However, in any event, for a precise prediction of
the event rates at the LHC it is necessary to take include NLO
corrections $i.e.$ associated K-factors \cite{Beenakker:1996ch}.

In the next section we will provide analytical expressions of the
cross sections for pair-production of squarks of first and second
generations. In Sec. III we will give a general discussion of the
phase sensitivities of the cross sections. In Sec. IV we will pick
up a specific post-WMAP benchmark point within minimal
supergravity (mSUGRA), and for the purpose of studying effects of
CP-odd phases, fold its universality pattern by switching on
non-universal CP-odd phases for gaugino masses at the unification
scale. In here we will also provide a detailed
numerical/analytical discussion of events with squark pairs at the
LHC. In Sec. V we will discuss impact of the EDMs on the cross sections by
considering EDM-favored parameter domains already present in the literature.
In Sec. VI we conclude.

\section{Squark pair production in $p p$ collisions}
In this section we discuss pair production of squarks of varying
chirality and flavor. We will analyze $\widetilde{q}_{a}\,
\widetilde{\hat q}^{\star}_{a}$ and $\widetilde{q}_{a}\,
\widetilde{\hat q}_{a'}$ type final states ($a, a' = L, R$) where
$\hat q$ and $q$ may or may not be identical. Here $\widetilde{q}$
stands for any of the squarks $\widetilde{u}$, $\widetilde{d}$,
$\widetilde{s}$, $\widetilde{c}$. Therefore, the $2 \rightarrow 2$
scatterings
\begin{eqnarray}
{ p\, p  \rightarrow \widetilde{q}_a \widetilde{\hat
q}^{\star}_{a'}\ + X}\label{sqsqbar}
\end{eqnarray}
and
\begin{eqnarray}
{ p\, p  \rightarrow \widetilde{q}_a\; \widetilde{\hat q}_{a'} +
X}\label{sqsq}
\end{eqnarray}
are the main processes to be investigated. Below we discuss these
scatterings one by one by including both SUSY QCD and SUSY
electroweak contributions.

\subsection{$\widetilde{q}_a\, \widetilde{q}^{\star}_{a'}\ $ production}
\begin{figure}
\includegraphics[width=10cm]{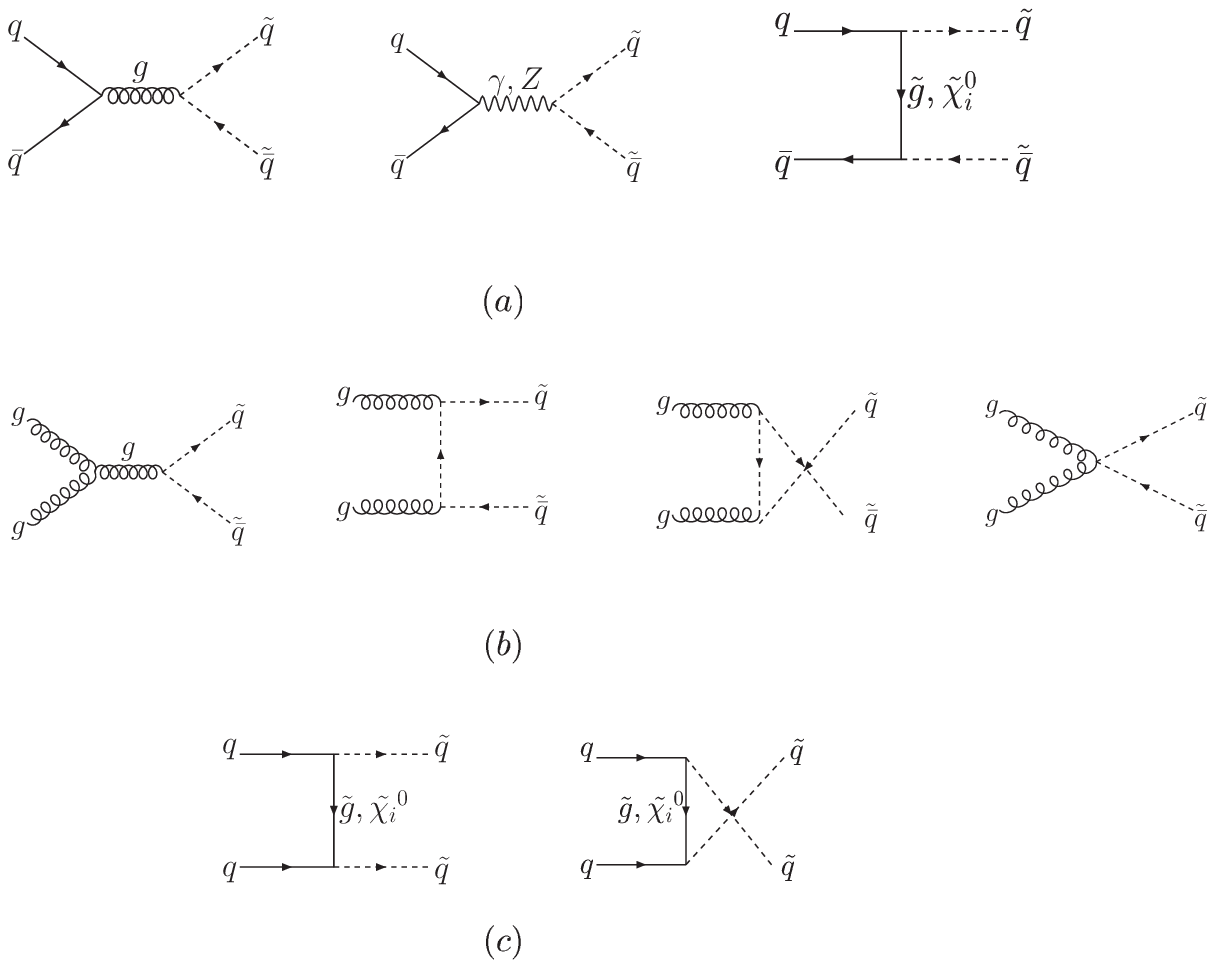}
\caption{Feynman diagrams for squark pair production. Part (a)
stands for processes started by quark--anti-quark annihilation,
(b) for gluon fusion and (c) for quark-quark scattering into
squark pairs.}\label{dia}
\end{figure}

The squark pair production via $p\, p \rightarrow \widetilde
q\widetilde {q}^{\star} +X$ is initiated either by $q\,
\overline{q}$ annihilation or by gluon fusion. The relevant
Feynman diagrams are depicted in Fig.~\ref{dia}.

The production of squark pairs, as initiated by $q\, \overline{
q}$ annihilation, involves gluon, photon and $Z$ boson exchanges
in the $s$-channel as well as  gluino and neutralino exchanges in
the $t$-channel (see Fig.1 part (a)). After color and spin
averaging the differential cross section for $\widetilde{q}_L
\widetilde{q}^{\star}_{L}$ production takes the form
\begin{eqnarray}\label{eq1}
\frac{d\hat\sigma(q^{\prime}\,
\overline{q}^{\prime}\rightarrow\widetilde q_L\,
\widetilde{q}^{\star}_L)}{d\hat t}&=&\frac{2\pi}{9\hat s^2}
\left(\hat t\hat u-m_{\widetilde{q}_L}^4\right) \left[ \delta_{q'
q} {\cal T}_{\mbox{FC}} + \left(1 - \delta_{q' q}\right) {\cal
T}_{\mbox{FV}}\right]
\end{eqnarray}
where FC and FV stand, respectively, for flavor-conserving and
flavor-violating, and associated quantities are given by
\begin{eqnarray} \label{fc}
{\cal T}_{\mbox{FC}} &=&\alpha_s^2\Big[\frac{2}{\hat
s^2}+\frac{1}{(\hat t- M_{\widetilde{g}}^2)^2}-\frac{2}{3\hat
s(\hat t-M_{\widetilde{g}}^2)}\Big]\nonumber\\
&+&\alpha^2\Big[\frac{9 e_q^4}{\hat s^2}+\frac{9|A_{q_L i}|^2|A_{q_L
j}|^2}{2s_W^4(\hat t- M_{\widetilde{\chi}_i^0}^2)(\hat t-
M_{\widetilde{\chi}_j^0}^2)}\Big]+\frac{2\alpha_s\alpha e_q^2}{\hat
s}\Big(\frac{2}{(\hat t-M_{\widetilde{g}}^2)}\Big)\nonumber\\
&-&\frac{\alpha}{s_W^2}\Big(\frac{4\alpha_s}{\hat s(\hat t-
M_{\widetilde{\chi}_i^0}^2)}+\frac{3\alpha e_q^2}{\hat s(\hat
t-M_{\widetilde{\chi}_i^0}^2)}\Big)|A_{q_L i}|^2\nonumber\\
&+&\frac{\alpha C_{qLL}(\hat s-M_Z^2)}{s_W^2[(\hat
s-M_Z^2)^2+\Gamma_Z^2 M_Z^2]}\Big[\frac{9\alpha
C_{qLL}\left(C_{qLL}^2+C_{qRR}^2\right)}{2 s_W^2(\hat
s-M_z^2)}\nonumber\\
&+&\frac{4\alpha_s C_{qLL}}{(\hat t-
M_{\widetilde{g}}^2)}+\frac{9\alpha e_q^2
\left(C_{qLL}+C_{qRR}\right)}{\hat s} -\frac{3\alpha
C_{qLL}}{s_W^2(\hat t- M_{\widetilde{\chi}_i^0}^2)}|A_{q_L
i}|^2\Big]
\end{eqnarray}
and
\begin{eqnarray}
\label{fv} {\cal T}_{\mbox{FV}} &=& \frac{1}{\hat
s^2}\Big[2\alpha_s^2 +9\alpha^2 e_{q'}^2 e_q^2\Big] +\frac{(\hat
s-M_Z^2)C_{qLL}}{s_W^2[(\hat s-M_Z^2)^2+\Gamma_Z^2
M_Z^2]}\Big[\frac{9\alpha^2 \left(C_{q'LL}^2+C_{q'RR}^2\right)
C_{qLL}}{2 s_W^2(\hat s-M_Z^2)}\nonumber\\
 &+&9\alpha^2 e_{q} e_{q'}\frac{(C_{q'LL}+C_{q'RR})}{\hat
s}\Big]
\end{eqnarray}
where summations over $i, j$ ( which label the neutralino
eigenstates) are implied. The couplings $C_{q LL}$ are given by
\begin{eqnarray}
C_{u LL}&=&\frac{1}{\cos\theta_W}(-\frac{1}{2}+ \frac{2}{3} \sin^2
\theta_{W})\;,\;\;C_{d LL}=\frac{1}{\cos\theta_W}(\frac{1}{2}-
\frac{1}{3} \sin^2 \theta_{W})
\end{eqnarray}
for up-- and down--type quarks, respectively. The
neutralino-quark-squark couplings are collected in $A_{q_L i}$,
and their explicit expressions are given below.

The differential cross section for $\widetilde{q}_R\,
\widetilde{q}^{\star}_{R}$ production is obtained from (\ref{eq1})
by replacing $A_{q_L i}$ by $A_{q_R i}$, $C_{u LL}$ and $C_{d LL}$
by
\begin{eqnarray}
C_{u RR}= \frac{1}{\cos\theta_W}(\frac{2}{3} \sin^2 \theta_W)
\;,\;\; C_{d RR} = \frac{1}{\cos\theta_W}(-\frac{1}{3} \sin^2
\theta_W)\,,
\end{eqnarray}
and finally $m_{\widetilde{q}_L}$ by $m_{\widetilde{q}_R}$.

A short glance at (\ref{fc}) and (\ref{fv}) reveals that flavors
of annihilating quarks differ from those of the produced squarks
thanks to gauge boson mediation, only. The reason is that such
$s$-channel diagrams do not communicate flavor information from
$\mid \mbox{in} \rangle$ to $\mid \mbox{out} \rangle$ states. One
also observes that FC-scatterings proceed solely with the gauge
boson mediation because of the fact that quark-squark-gaugino
vertices are taken strictly flavor-diagonal. This is an excellent
approximation given the bounds on such vertices from FCNC
processes \cite{flavor-cp}. However, one keeps in mind that, in
principle, $q'\, \overline{q}$ annihilation can produce third
generation squarks first, and they might subsequently get
converted into second generation squarks to the extent that $B$
and $D$ physics permit. This possibility is neglected in our
analysis.

Since gauge bosons cannot couple to (s)quarks of distinct
chirality, $\widetilde{q}_L\, \widetilde{q}^{\star}_{R} +
\widetilde{q}_R\, \widetilde{q}^{\star}_{L}$ production proceeds
solely with sparticle exchange
\begin{eqnarray}\label{eq3}
\frac{d\hat\sigma(q \overline{q}\rightarrow\widetilde q_L\widetilde{
q}^{\star}_R+\widetilde {q}_R\widetilde{q}^{\star}_L)}{d\hat
t}&=&\frac{2\pi}{9\hat s^2}\Biggm\{\frac{2\alpha_s^2
M_{\widetilde{g}}^2 \hat s}{(\hat t- M_{\widetilde{g}}^2
)^2}\nonumber\\&+&\frac{9\alpha^2 \hat s M_{\widetilde{\chi}_i^0}
M_{\widetilde{\chi}_j^0}}{s_W^4(\hat t-
M_{\widetilde{\chi}_i^0}^2)(\hat t- M_{\widetilde{\chi}_j^0}^2
)}(A_{q_R i}A_{q_R j}^{\star}A_{q_L j}A_{q_L i}^{\star})\Biggm\}
\end{eqnarray}
where FV transitions are forbidden by the reasons mentioned above.

The neutralino-quark-squark couplings $A_{Li, Ri}$ appearing in
(\ref{fc}), (\ref{fv}) and (\ref{eq3}) are given by
\begin{eqnarray}
A_{q_R i}=\tan\theta_W e_q N_{i1}^{\star}\;,\;\; A_{q_L i}=[T_{3q}
N_{i2}-\tan\theta_W (T_{3q}-e_q) N_{i1}]
\end{eqnarray}
where $N_{i j}$ are obtained by diagonalizing the neutralino mass
matrix
\begin{eqnarray}
M_{\widetilde\chi^0}=\left(%
\begin{array}{cccc}
  M_{\widetilde{B}} e^{i \varphi_{\widetilde{B}}}& 0 & -M_Z \cos\beta \sin\theta_W & M_Z \sin\beta \sin\theta_W \\
  0 & M_{\widetilde{W}} e^{i \varphi_{\widetilde{W}}} & M_Z \cos\beta \cos\theta_W & -M_Z \sin\beta \cos\theta_W \\
  -M_Z \cos\beta \sin\theta_W & M_Z \cos\beta \cos\theta_W & 0 & -\mu \\
   M_Z \sin\beta \sin\theta_W&  -M_Z \sin\beta \cos\theta_W &-\mu& 0
\end{array}%
\right)
\end{eqnarray}
via $N^* M_{\widetilde\chi^0} N^{-1}=\mbox{diag.}
(M_{\widetilde\chi_1^0},...,M_{\widetilde\chi_4^0})$. Here
$M_{\widetilde{B}}$ and $M_{\widetilde{W}}$ designate absolute
magnitudes of the U(1)$_Y$ and SU(2)$_L$ gaugino masses, and
$\varphi_{\widetilde{B}}$ and $\varphi_{\widetilde{W}}$ their
phases. We find it useful to separate modulus and phase of the
gaugino masses for ease of analysis. It is clear that Higgsinos
contribute to squark pair production via only Higgsino--gaugino
mixings $i.e.$ the off-diagonal entries $N_{31, 41}$ and $N_{32,
42}$ in $A_{q_L i, q_R i}$.

Having completed quark--anti-quark annihilation, we now analyze
$\widetilde{q}_{L,R} \widetilde{q}^{\star}_{L,R}$ productions
initiated by  gluon fusion. From Fig.\ref{dia} (b) the
differential cross section is found to read
\begin{eqnarray}\label{eq4}
\frac{d\hat\sigma(g g \rightarrow
\widetilde{q}\widetilde{q}^{\star})}{d\hat
t}&=&\frac{\pi\alpha_s^2}{\hat s^2}\bigg\{\frac{3(\hat t-\hat
u)^2}{16\hat s^2}+\frac{(\hat t+m^2)^2}{6(\hat
t-m^2)^2}+\frac{(\hat u+m^2)^2}{6(\hat u-m^2)^2}+\frac{3(\hat
t-\hat u)(4m^2-\hat s+4\hat t)}{64\hat s(\hat
t-m^2)}\nonumber\\
&-&\frac{(4m^2-\hat s)^2}{96(\hat t-m^2)(\hat u-m^2)}+\frac{3(\hat
u-\hat t)(4m^2-\hat s+4\hat u)}{64\hat s(\hat
u-m^2)}-\frac{7(4m^2-\hat s+4\hat t)}{192(\hat
t-m^2)}\nonumber\\
&-&\frac{7(4m^2-\hat s+4\hat u)}{192(\hat
u-m^2)}+\frac{7}{24}\bigg\}
\end{eqnarray}
after color and spin averaging. Here $m$ stands for
$m_{\widetilde{q}_L}$ or $m_{\widetilde{q}_R}$, whichever is
produced.

\subsection{$\widetilde{q}_a\, \widetilde{q}_{a'}$ production}
In obvious contrast to $\widetilde{q}^{\star}\widetilde q$
production, the partonic process that leads to $\widetilde
q\widetilde q$ production proceeds with sole sparticle mediation.
Indeed, at tree level $pp \rightarrow \widetilde q \widetilde q
+X$ scattering is initiated by quarks, and proceeds with
$t$-channel gaugino exchanges as shown in Fig.\ref{dia} (c).
Fermion number violating $qq \rightarrow \widetilde q \widetilde
q$ reaction occurs because of the Majorana nature of gauginos. The
color and spin averaged parton level differential cross section
for $\widetilde{q}_L\, \widetilde{q}_{L}$ production is given by
\begin{eqnarray}\label{eq5}
\frac{d\hat\sigma(qq \rightarrow \widetilde q_L \widetilde
q_L)}{d\hat t}&=&\frac{2\pi}{9\hat
s^2}\Biggm\{\alpha_s^2\Bigg[\frac{ M_{\widetilde{g}}^2\hat
s}{(\hat t- M_{\widetilde{g}}^2)^2}+\frac{M_{\widetilde{g}}^2 \hat
s}{(\hat u- M_{\widetilde{g}}^2 )^2}-\frac{2}{3}\frac{
M_{\widetilde{g}}^2
\hat s}{(\hat t-M_{\widetilde{g}}^2)(\hat u- M_{\widetilde{g}}^2)}\Bigg]\nonumber\\
&+&\frac{9\alpha^2}{2 s_W^4}\Bigg[\Big(\frac{1}{(\hat
t-M_{\widetilde{\chi}_i^0}^2)(\hat t-
M_{\widetilde{\chi}_j^0}^2)}+\frac{1}{(\hat u-
M_{\widetilde{\chi}_i^0}^2)(\hat
u- M_{\widetilde{\chi}_j^0}^2)}\Big)(A_{q_L i}^{\star})^2(A_{q_L j})^2\nonumber\\
&+&\frac{1}{3(\hat t- M_{\widetilde{\chi}_i^0}^2)(\hat u-
M_{\widetilde{\chi}_j^0}^2 )}[(A_{q_L i})^2(A_{q_L
j}^{\star})^2+(A_{q_L i}^{\star})^2(A_{q_L j})^2] \Bigg](\hat s
M_{\widetilde{\chi}_i^0} M_{\widetilde{\chi}_j^0})\nonumber\\
&+&\frac{\alpha_s \alpha}{2 s_W^2} \Biggm[\frac{4}{(\hat t- M_{\widetilde{g}}^2)(\hat u- M_{\widetilde{\chi}_i^0}^2)}\nonumber\\
&+&\frac{4}{(\hat u- M_{\widetilde{g}}^2)(\hat t-
M_{\widetilde{\chi}_i^0}^2)}\Biggm](\hat s M_{\widetilde{g}}
M_{\widetilde{\chi}_i^0})[(A_{q_L
i})^2e^{-i\varphi_{\widetilde{g}}}+(A_{q_L
i}^{\star})^2e^{i\varphi_{\widetilde{g}}}]\Biggm\}
\end{eqnarray}
which bears a manifest sensitivity to the gluino phase
$\varphi_{\widetilde{g}}$, as indicated by the last term. This is
one of the most important differences between $\widetilde{q}_a\,
\widetilde{q}_{a'}$ and $\widetilde{q}_a\,
\widetilde{q}_{a'}^{\star}$ productions: while the former involves
phases of each neutralino and gluino exchanged the latter does
only the relative phases among neutralino states.

The cross section for $\widetilde{q}_R\, \widetilde{q}_{R}$
production follows directly from (\ref{eq5}) after replacing
$A_{q_L i}$ by $A_{q_R i}$.

The spin and color averaged $\widetilde{q}_L\, \widetilde{q}_{R} +
\widetilde{q}_R\, \widetilde{q}_{L} $ production cross section is
given by
\begin{eqnarray}\label{eq6}
\frac{d\hat\sigma(q\, q \rightarrow \widetilde q_L \widetilde
q_R+\widetilde q_R \widetilde q_L)}{d\hat t}&=&\frac{2\pi}{9\hat
s^2}(\hat t \hat u-m_{\widetilde q_L}^2m_{\widetilde
q_R}^2)\Biggm\{2\alpha_s^2\Bigg[\frac{1}{(\hat
t-M_{\widetilde{g}}^2)^2}+\frac{1}{(\hat
u- M_{\widetilde{g}}^2)^2}\Bigg]\nonumber\\
&+&\frac{9\alpha^2}{s_W^4}\Bigg[\Big(\frac{1}{(\hat t-
M_{\widetilde{\chi}_i^0}^2)(t-
M_{\widetilde{\chi}_j^0}^2)}\nonumber\\&+&\frac{1}{(\hat u-
M_{\widetilde{\chi}_i^0}^2)(\hat u-
M_{\widetilde{\chi}_j^0}^2)}\Big)(A_{q_L i}^{\star}A_{q_R
i}^{\star}A_{q_L j}A_{q_R j})\Bigg] \Biggm\}.
\end{eqnarray}
whose dependence on the $CP$-odd phases is similar to
Eq.~(\ref{eq3}).
\subsection{$\widetilde{q}^{\star}_a\, \widetilde{q}'_{a'}\ $ production}
Having completed analyses of $\widetilde{q}^{\star}\widetilde q$
and $\widetilde q\widetilde q$ productions, we now focus on
$\widetilde{q}^{\star}\widetilde q'$ type final states with $\widetilde
q \in \left\{ \widetilde u, \widetilde c\right\}$, $\widetilde
q'\in \left\{ \widetilde d, \widetilde s\right\}$ and vice versa.
Such final states, carrying $\pm 1$ electric charge, receive
contributions from $s$-channel $W^{\pm}$ plus $t$-channel gaugino
exchanges. The differential cross section for $\widetilde
q_L\widetilde {q'}^{\star}_L$ production is given by
\begin{eqnarray}
\label{qlqlp} \frac{d\hat\sigma}{d\hat t}(q\bar {q'}\rightarrow
\widetilde q_L\widetilde {q'}^{\star}_L)&=&\frac{2\pi}{9\hat
s^2}\Bigg\{\frac{1}{[(\hat
s-M_W^2)^2+\Gamma_W^2M_W^2]}\Big[\frac{9\alpha^2\mid C_{q' q }\mid^2
}{8s_W^4}+\frac{2\alpha_s\alpha(\hat
s-M_W^2)C_{q' q }}{s_W^2(\hat t-M_{\widetilde{g}}^2)}\\
&-&\frac{9\alpha^2C_{q' q }[(\hat
s-M_W^2)Re(A_{q'Li}^{\star}A_{qLi})+\Gamma_WM_WIm(A_{q'_L
i}^{\star}A_{q_L i})]}{2s_W^4(\hat
t-M_{\widetilde{\chi}_i^0}^2)}\Big]+\frac{\alpha_s^2}{(\hat t-M_{\widetilde{g}}^2)^2}\nonumber\\
&+&\frac{9\alpha^2A_{q_L i}^{\star}A_{q_L j}A_{q'_L
j}^{\star}A_{q'_L i}}{2s_W^4(\hat t-M_{\widetilde{\chi}_i^0}^2)(\hat
t-M_{\widetilde{\chi}_j^0}^2)}\Bigg\}(\hat t\hat
u-m_{\widetilde{q}_L}^2 m_{\widetilde{q}'_L}^2)\nonumber
\end{eqnarray}
where $C_{q' q}$ are the elements of the CKM matrix having the
experimental mid-point values  $\mid C_{u d} \mid = 0.9745, \mid
C_{u s} \mid = 0.2240, \mid C_{u b} \mid = 0.037, \mid C_{c d}
\mid = 0.2240, \mid C_{c s} \mid = 0.9737, \mid C_{c b} \mid =
0.0415, \mid C_{t d} \mid =0.094, \mid C_{t s} \mid = 0.040, \mid
C_{t b}\mid  = 0.999$. As before, sum over $i,j=1,2,3,4$ is
implied.

Similarly, the differential cross section for $\widetilde
q_R\widetilde {q'}^{\star}_R$ production is given by
\begin{eqnarray}
\frac{d\hat\sigma}{d\hat t}(q\bar {q'}\rightarrow \widetilde
q_R\widetilde {q'}^{\star}_R)&=&\frac{2\pi}{9\hat
s^2}\Bigg\{\frac{\alpha_s^2}{(\hat
t-M_{\widetilde{g}}^2)^2}+\frac{9\alpha^2[A_{q_R i}^{\star}A_{q_R
j}A_{q'_R j}^{\star}A_{q'_R i}]}{2s_W^4(\hat
t-M_{\widetilde{\chi}_i^0}^2)(\hat
t-M_{\widetilde{\chi}_j^0}^2)}\Bigg\}(\hat t\hat u-\widetilde
m_{\widetilde{q}_R}^2 m_{\widetilde{q}'_R}^2)
\end{eqnarray}
which differs from (\ref{qlqlp}) by the absence of $W^{\pm}$
contribution. Indeed, $\widetilde q_R\widetilde {q'}^{\star}_R$
production proceeds via only the gluino and neutralino exchanges.

Finally, squarks with distinct electric charges and
chiralities possess the following differential cross section:
\begin{eqnarray}
\frac{d\hat\sigma}{d\hat t}(q\bar {q'}\rightarrow \widetilde
q_L\widetilde {q'}^{\star}_R+\widetilde q_R\widetilde
{q'}^{\star}_L)&=&\frac{2\pi}{9\hat
s^2}\Bigg\{\frac{2\alpha_s^2{M_{\widetilde{g}}^2}\hat s}{(\hat
t-M_{\widetilde{g}}^2)^2}\nonumber\\&+&\frac{9\alpha^2\hat
sM_{\widetilde{\chi}_i^0}M_{\widetilde{\chi}_j^0}[A_{q_L
i}^{\star}A_{q'_R i}A_{q'_R j}^{\star}A_{q_L j}+A_{q_R
i}^{\star}A_{q'_L i}A_{q'_L j}^{\star}A_{q_R j}]}{2s_W^4(\hat
t-M_{\widetilde{\chi}_i^0}^2)(\hat t-M_{\widetilde{\chi}_j^0}^2)}
\Bigg\}
\end{eqnarray}
which is generated by gluino and neutralino exchanges, only.

\subsection{$\widetilde{q}_a\, \widetilde{q}'_{a'}\ $ production}
In this subsection we discuss production of squarks having
distinct electric charges and chiralities with no involvement of
anti-squarks. We start with $\widetilde {q_L} \widetilde {q'_L}$
production
\begin{eqnarray}
\frac{d\hat \sigma}{d\hat t}(q q'\rightarrow \widetilde {q_L}
\widetilde {q'_L})&=&\frac{\pi}{36\hat
s^2}\Bigg\{\frac{9\alpha^2U_{k1}^{\star}V_{k1}^{\star}U_{l1}V_{l1}}{s_W^4(\hat
u-M_{\widetilde{\chi}_k^{\pm}}^2)(\hat
u-M_{\widetilde{\chi}_l^{\pm}}^2)}\hat
sM_{\widetilde{\chi}_k^{\pm}}M_{\widetilde{\chi}_l^{\pm}}
\nonumber\\&+&\frac{8\alpha_s^2}{(\hat t-M_{\widetilde{g}}^2)^2}\hat
sM_{\widetilde{g}}^2+\frac{36\alpha^2[A_{q_L i}^{\star}A_{q_L
j}A_{q'_L i}^{\star}A_{q'_L j}]}{s_W^4(\hat
t-M_{\widetilde{\chi}_i^0}^2)(\hat
t-M_{\widetilde{\chi}_j^0}^2)}\hat
sM_{\widetilde{\chi}_i^0}M_{\widetilde{\chi}_j^0}\nonumber\\&+&\frac{8\alpha_s\alpha(U_{k1}^{\star}V_{k1}^{\star}e^{i\varphi_{\widetilde{g}}}+U_{k1}V_{k1}e^{-i\varphi_{\widetilde{g}}})}{s_W^2(\hat
u-M_{\widetilde{\chi}_k^{\pm}}^2)(\hat t-M_{\widetilde{g}}^2)}\hat
sM_{\widetilde{\chi}_k^{\pm}}M_{\widetilde{g}}\nonumber\\&+&\frac{12\alpha^2Re[U_{k1}^{\star}V_{k1}^{\star}A_{q_L
i}A_{q'_L i}]}{s_W^4(\hat u-M_{\widetilde{\chi}_k^{\pm}}^2)(\hat
t-M_{\widetilde{\chi}_i^0}^2)}\hat
sM_{\widetilde{\chi}_k^{\pm}}M_{\widetilde{\chi}_i^0} \Bigg\}
\end{eqnarray}
which receives contributions from gluino, neutralino as well as
chargino exchanges. It is the left-chirality nature of squarks
that involves $t$-channel chargino contribution.

In contrast to $\widetilde {q_L} \widetilde {q'_L}$ production,
$\widetilde {q_R} \widetilde {q'_R}$ production does not receive
contributions from chargino exchange since first and second
generation squarks do not have significant couplings to Higgsinos.
Indeed, one finds
\begin{eqnarray}
\frac{d\hat \sigma}{d\hat t}(q q'\rightarrow \widetilde {q_R}
\widetilde {q'_R})&=&\frac{\pi}{36\hat
s^2}\Bigg\{\frac{8\alpha_s^2}{(\hat t-M_{\widetilde{g}}^2)^2}\hat
sM_{\widetilde{g}}^2+\frac{36\alpha^2[A_{q_R i}^{\star}A_{q_R
j}A_{q'_R i}^{\star}A_{q'_R j}]}{s_W^4(\hat
t-M_{\widetilde{\chi}_i^0}^2)(\hat
t-M_{\widetilde{\chi}_j^0}^2)}\hat
sM_{\widetilde{\chi}_i^0}M_{\widetilde{\chi}_j^0} \Bigg\}
\end{eqnarray}
which is a pure $t$-channel effect.

Finally, squarks with unequal charges and chiralities are
produced with the cross section
\begin{eqnarray}
\frac{d\hat \sigma}{d\hat t}(qq'\rightarrow \widetilde q_L
\widetilde q'_R+\widetilde q_R\widetilde q'_L)&=&\frac{2\pi}{9\hat
s^2}(\hat t\hat
u-m_{\widetilde q_L}^2 m_{\widetilde q_R}^2)\Bigg\{\frac{2\alpha_s^2}{(\hat t-M_{\widetilde{g}}^2)^2}\nonumber\\
&+&\frac{9\alpha^2[A_{q_L i}^{\star}A_{q'_R i}^{\star}A_{q_L
j}A_{q'_R j}+A_{q_R i}^{\star}A_{q'_L i}^{\star}A_{q_R j}A_{q'_L
j}]}{2s_W^4(\hat t-M_{\widetilde{\chi}_i^0}^2)(\hat
t-M_{\widetilde{\chi}_j^0}^2)}\Bigg\}
\end{eqnarray}
which involves only gluino and neutralino exchanges. Therefore,
$\widetilde {q_L} \widetilde {q'_L}$ production is unique in that
it is the only pair-production process which involves chargino
$i.e.$ wino  mediation.

In the expressions above, $i,j=1,2,3,4$ are neutralino indices
with implied summations. The charginos are designated by $k,l=1,2$
indices with again implied summations. The chargino mixing
matrices $U$ and $V$ are obtained via
\begin{eqnarray}
U^{\star} M_{\chi^{\pm}} V^{-1} = \mbox{diag.} \left(
M_{\widetilde{\chi}_1^{\pm}},  M_{\widetilde{\chi}_2^{\pm}}
\right)
\end{eqnarray}
where
\begin{eqnarray}
M_{\chi^{\pm}} = \left(\begin{array}{cc} M_{\widetilde{W}} e^{i
\varphi_{\widetilde{W}}} &
\sqrt{2} M_W \cos \beta \\
\sqrt{2} M_W \sin \beta & \mu \end{array} \right)
\end{eqnarray}
is the mass matrix of charged gauginos and Higgsinos.

\section{Phase Sensitivities of Individual Cross Sections}
In this section we perform a comparative analysis of various cross
sections in terms of their dependencies on the CP-odd phases. Our
discussions will be mainly schematic as we leave exact numerical
analysis to the next section.

\begin{table}[tbp]

\begin{center}

\begin{tabular}{|c|c|c|c|}
\hline Squark Pair & Insensitive to & Directly sensitive to &
Indirectly sensitive to\\\hline\hline $\widetilde{q}_L\,
\widetilde{q}^{\star}_{L}$ & $\varphi_{\widetilde{g}}$ & -- &
$\varphi_{\widetilde{W}}$, $\varphi_{\widetilde{B}}$,
$\varphi_{\mu}$\\\hline $\widetilde{q}_R\,
\widetilde{q}^{\star}_{R}$ & $\varphi_{\widetilde{g}}$ & -- &
$\varphi_{\widetilde{B}}$, $\varphi_{\widetilde{W}}$,
$\varphi_{\mu}$\\\hline $\widetilde{q}_L\,
\widetilde{q}^{\star}_{R}$ & $\varphi_{\widetilde{g}}$ & -- &
$\varphi_{\widetilde{B}}$,  $\varphi_{\widetilde{W}}$,
$\varphi_{\mu}$\\\hline $\widetilde{q}_L\, \widetilde{q}_{L}$ & -- &
$\varphi_{\widetilde{g}}$, $\varphi_{\widetilde{W}}$,
$\varphi_{\widetilde{B}}$ & $\varphi_{\mu}$\\\hline
$\widetilde{q}_R\, \widetilde{q}_{R}$ & -- &
$\varphi_{\widetilde{g}}$, $\varphi_{\widetilde{B}}$ &
$\varphi_{\widetilde{W}}$, $\varphi_{\mu}$\\\hline
$\widetilde{q}_L\, \widetilde{q}_{R}$ & $\varphi_{\widetilde{g}}$ &
$\varphi_{\widetilde{B}}$ & $\varphi_{\widetilde{W}}$,
$\varphi_{\mu}$\\\hline $\widetilde{q}_L\,
\widetilde{q'}^{\star}_{L}$ & $\varphi_{\widetilde{g}}$ & -- &
$\varphi_{\widetilde{W}}$, $\varphi_{\widetilde{B}}$,
$\varphi_{\mu}$\\\hline $\widetilde{q}_R\,
\widetilde{q'}^{\star}_{R}$ & $\varphi_{\widetilde{g}}$ & --
&$\varphi_{\widetilde{B}}$, $\varphi_{\widetilde{W}}$,
$\varphi_{\mu}$\\\hline $\widetilde{q}_L\,
\widetilde{q'}^{\star}_{R}$ & $\varphi_{\widetilde{g}}$ & -- &
$\varphi_{\widetilde{B}}$, $\varphi_{\widetilde{W}}$,
$\varphi_{\mu}$\\\hline $\widetilde{q}_L\, \widetilde{q'}_{L}$ & --&
$\varphi_{\widetilde{g}}$, $\varphi_{\widetilde{W}}$,
$\varphi_{\widetilde{B}}$ & $\varphi_{\mu}$\\\hline
$\widetilde{q}_R\, \widetilde{q'}_{R}$ & $\varphi_{\widetilde{g}}$ &
$\varphi_{\widetilde{B}}$ & $\varphi_{\widetilde{W}}$,
$\varphi_{\mu}$\\\hline $\widetilde{q}_L\, \widetilde{q'}_{R}$ &
$\varphi_{\widetilde{g}}$ & $\varphi_{\widetilde{B}}$ &
$\varphi_{\widetilde{W}}$, $\varphi_{\mu}$\\\hline
\end{tabular}
\end{center}
\caption{\label{table1}{ Productions of various squark pairs of
varying chirality and flavor. Shown in the second column are the
phases to which production cross section is insensitive. The third
column shows phases coming from the gauginos exchanged. The last
column, the fourth column, shows those phases which enter the
cross section via only the mixings among neutral and charged
gauginos and Higginos $i.e.$ mixings in neutralino and chargino
mass matrices. }}
\end{table}

Table \ref{table1} shows phase dependencies of pair-production cross
sections for various chirality and flavor combinations. It is
clear from the table that each cross section possesses a specific
dependence on gaugino phases, and in future collider
studies (like LHC or NLC) this may be used to establish
existence/absence of CP-violating sources in the gaugino sector in
a way independent of the phases of the trilinear couplings as well
as Higgs mediation effects.

For a comparative analysis, consider first $\widetilde{q}_L
\widetilde{q}^{\star}_{L}$ production. This process is initiated
by quark--anti-quark annihilation or by gluon fusion whose cross
sections are given in (\ref{eq1}) and (\ref{eq4}). The latter is
completely blind to CP--odd phases. The former, on the other hand,
is independent of $\varphi_{\widetilde{g}}$, and feels
$\varphi_{\widetilde{W}}$, $\varphi_{\widetilde{B}}$,
$\varphi_{\mu}$ via mixings in the neutralino mass matrix, only.
The reason is that the two quark-squark-gaugino vertices, which
arise in $t$-channel gaugino exchange diagrams, are complex
conjugate of each other. Similar observations also hold for
$\widetilde{q}_R\, \widetilde{q}^{\star}_{R}$ production.

Notably, this phase-dependence of $\widetilde{q}_L\,
\widetilde{q}^{\star}_{L}$ (and of $\widetilde{q}_R\,
\widetilde{q}^{\star}_{R}$ ) production radically differs from that
of $\widetilde{q}_L\, \widetilde{q}_{L}$ (and of $\widetilde{q}_R\,
\widetilde{q}_{R}$) production. First of all, there is no
$s$-channel vector boson exchange contributions to
$\widetilde{q}_L\, \widetilde{q}_{L}$ production; it is a pure
$t$-channel process mediated solely by the gauginos. Next, and more
importantly, the phases of the two quark-squark-gaugino vertices
interfere constructively giving thus a pronounced phase sensitivity
to $\widetilde{q}_L\, \widetilde{q}_{L}$ (and $\widetilde{q}_R\,
\widetilde{q}_{R}$) production. These observations hold also for
charged final states $i.e.$ $\widetilde{u}_L \widetilde{d}_L$ type
squark pairs.

The main point is that for $\widetilde{q} \widetilde{\hat
q}^{\star}$ type final states the sensitivity to CP-odd phases is
restricted to those in the neutralino/chargino sector, and depends
crucially on how strong the gauginos/higgsinos mix with each
other. In particular, when $M_{\widetilde{W},\widetilde{B}} \gg
M_{W, Z}$ the cross section for $\widetilde{q} \widetilde{\hat
q}^{\star}$ production becomes independent of the CP-odd phases.
On the other hand, for for $\widetilde{q} \widetilde{ \hat q}$
production sensitivity to CP-odd phases is maximal, and is
independent of the strength of mixing in neutralino/chargino
system. For instance, $\widetilde{q}_L\,
\widetilde{q}_{L}/\widetilde{q}_R\, \widetilde{q}_{R}$ production
is a sensitive probe of $\varphi_{\widetilde{g}, \widetilde{W},
\widetilde{B}}/\varphi_{\widetilde{g}, \widetilde{B}}$. Clearly,
the difference between $\widetilde{q}_L\, \widetilde{q}_{L}$ and
$\widetilde{q}_R\, \widetilde{q}_{R}$ production cross sections,
with known masses of squarks, is a viable measure of
$\varphi_{\widetilde{W}}$.

In general, depending on chirality, charge and flavor structures
of the squark pairs produced, the squark pair-production cross
sections exhibit different types of sensitivities to CP-odd phases
(and various soft masses as well). The main advantage of the
pair-productions of squarks belonging to first and second
generations is their potential of isolating the gaugino masses in
a way independent of the Higgs sector parameters and triliear
couplings.

In the next section we will study squark pair production cross
sections at the LHC for a specific yet phenomenologically viable
supersymmetric parameter space.

\section{Squark Pair Production at LM1}

In this section we perform a detailed numerical study of squark
pair production at the LHC with special emphasis on the effects of
the gaugino phases.

The subprocess cross sections which were calculated in the previous
section will be used to estimate squark pair production events at a
$p\, p$ collider with $\sqrt {\cal{S}}=14\, {\rm TeV}$ appropriate
for LHC experiments. For instance, the total hadronic cross section
for $\widetilde{q}_a\,\widetilde{q}^{\star}_{a}$ production takes
the form
\begin{eqnarray}
\label{cross1} \sigma\left({ p\, p \rightarrow
\widetilde{q}_a\,\widetilde{q}^{\star}_{a} + X}\right) =
\int_{\frac{4 m_{\widetilde{q}_a}^2}{{\cal{S}}}}^{1} d \tau
\int_{\tau}^{1} \frac{d x}{x} \Bigg\{ f_{g}\left(x, Q^2\right)
f_{g}\left(\frac{\tau}{x}, Q^2\right)
\sigma\left(g\, g \rightarrow \widetilde{q}_a\,\widetilde{q}^{\star}_{a}\right)+\nonumber\\
\left(f_{q'}\left(x, Q^2\right)
f_{\overline{q}'}\left(\frac{\tau}{x}, Q^2\right) +
f_{\overline{q}'}\left(x, Q^2\right) f_{q'}\left(\frac{\tau}{x},
Q^2\right) \right) \sigma\left(q'\, \overline{q}' \rightarrow
\widetilde{q}_a\,\widetilde{q}^{\star}_{a} \right) \Bigg\}
\end{eqnarray}
where structure functions $f_i(x,Q^2)$ represent the number
density of the parton $i$ which carries the fraction $x$ of the
longitudinal proton momentum. The initial state partons scatter
with a center--of--mass energy $\hat{s}=\tau\, {\cal{S}}$. All couplings
and masses in the partonic reactions are defined at the scale $Q$,
the renormalization and factorization scale, which has to lie
around $m_{\widetilde{q}_a}$. The QCD corrections give rise to
scale dependence of the structure functions, and $f_i(x,Q^2)$ can
be evaluated at any scale $Q$ using the Altarelli-Parisi
equations. In our calculations we use CTEQ5 parton distributions
\cite{CTQ5}. All pair-production processes are calculated in a way
similar to (\ref{cross1}).

The explicit expressions for cross sections in Sec. III show that, for
analyzing the pair-productions of squarks in first and second generations
one needs only a subset of the model parameters be fixed. The relevant
parameter set includes
\begin{eqnarray}
\mid \mu \mid\,,\; M_{\widetilde{g}}\,,\; M_{\widetilde{W}} \,,\;
M_ {\widetilde{B}} \,,\; m_{\widetilde{q}_{L,R}}^2\,,\; \tan\beta\,,\;
\varphi_{\mu}\,,\;\varphi_{\widetilde{g}}\,,\;
\varphi_{\widetilde{W}}\,,\; \varphi_{\widetilde{B}}
\end{eqnarray}
where it is understood that each parameter is evaluated at energy scale
where experiments are carried out $i.e.$ around a ${\rm TeV}$
for the LHC experiments.

None of these parameters is known a priori. All one can do is to determine
their allowed ranges via various laboratory ($e.g.$ the
lower bound on chargino mass, $b\rightarrow s \gamma$ decay rate
etc.) and astrophysical ($e.g.$ the WMAP results for cold dark
matter density) observations. In this sense, mSUGRA (the
constrained MSSM) serves as a prototype model where several
phenomenological bounds can be analyzed with minimal number
of parameters. The mSUGRA scheme is achieved by postulating certain
unification relations among the soft mass parameters in (\ref{soft}). In explicit
terms,
\begin{eqnarray}
\label{gut}
&&M_{\widetilde{g}}^0 = M_{\widetilde{W}}^0 = M_{\widetilde{B}}^0 =
M_{1/2}\nonumber\\
&&\left(M_{H_u}^{2}\right)^0 =\left(M_{H_d}^{2}\right)^0 = m_0^2\nonumber\\
&&\left({\bf Y _{u,d,e}^A}\right)^{0} = A_0 \left({\bf Y_{u,d,e}}\right)^0\nonumber\\
&&\left({\bf M_{Q}}^{2}\right)^0 = \left({\bf M_{U}}^{2}\right)^0 = \left({\bf M_{D}}^{2}\right)^0 =
\left({\bf M_{L}}^{2}\right)^0 = \left({\bf M_{E}}^{
2}\right)^0= m_0^2 {\bf 1}
\end{eqnarray}
so that not only the gauge couplings but also the scalar masses (into a common value $m_0$),
the trilinear couplings (into a common value $A_0$ times the corresponding Yukawa matrix), and
the gaugino masses (into a common value $M_{1/2}$) are unified. The bilinear Higgs coupling $B$
is traded for the pseudoscalar Higgs boson mass, and $\mu$ parameter is determined by the
requirement of correct electroweak breaking. The superscript $^0$ on a parameter in (\ref{gut})
implies the GUT-scale value of that parameter $e.g.$ $\left({\bf M_{Q}}^{2}\right)^0 \equiv {\bf
M_{Q}}^2(Q=M_{GUT})$ where $Q$ is the energy scale.

Under the assumptions made in (\ref{gut}), a general MSSM (parameterized by (\ref{soft}) given
in Introduction) reduces to mSUGRA (the constrained
MSSM) which involves only a few unknown parameters. Within the framework of mSUGRA, after LEP
\cite{post-lep} as well as WMAP \cite{post-wmap} a set of
benchmark points (at which all the existing bounds are satisfied)
has been constructed. In the language of experimentalists \cite{CMS},
there exist a number of benchmark points LM1, LM2, $\cdots$ LM9 at
which detector simulations are carried out. For instance, the
benchmark point LM1 (similar to point B in \cite{post-lep} and
identical to point B' in \cite{post-wmap}) corresponds to
\begin{eqnarray}
\label{lm1} m_{0}= 60\ {\rm GeV}\,,\; M_{1/2}=250\ {\rm GeV}\,,\;
A_0 = 0\,,\; \tan\beta= 10
\end{eqnarray}
with $\mu>0$. The parameter space we consider is wider than this
mSUGRA pattern in that universality pattern is respected in moduli
but not in phases. In other words, the gaugino masses (possibly
also the trilinear couplings in a setting with $A_0 \neq 0$) are
universal in size but not in phase (see also \cite{tarek}). To
this end we fold LM1 to a new point LM1' by switching on
CP-violating phases of gaugino masses at the GUT scale. More
explicitly,
\begin{eqnarray}
\label{lm1p} M_{\widetilde{g}}^0 e^{i \varphi_{\widetilde{g}}^0}=
M_{1/2}\ e^{i \varphi_3}\,,\; M_{\widetilde{W}}^0 e^{i
\varphi_{\widetilde{W}}^0} = M_{1/2}\ e^{i \varphi_2}\,,\;
M_{\widetilde{B}}^0 e^{i \varphi_{\widetilde{B}}^0}= M_{1/2}\ e^{i
\varphi_1}
\end{eqnarray}
at the GUT scale. The $\mu$ parameter is necessarily complex: $\mu
= \mid \mu \mid e^{i \varphi_{\mu}}$. The rest of the parameters,
including $M_{1/2}$ in (\ref{lm1p}), are fixed to their values in
(\ref{lm1}). For determining the impact of these GUT-scale phases
on SUSY parameter space at the electroweak scale it is necessary
to solve their RGEs with the boundary conditions (\ref{lm1p}).
With two-loop accuracy, one finds for gaugino masses
\begin{eqnarray}
\label{gaugino-soln} M_{\widetilde{B}}e^{i
\varphi_{\widetilde{B}}} &=&105.088\, e^{i\varphi_1} -0.229\,
e^{i\varphi_2}-2.811\, e^{i\varphi_3}\nonumber\\
M_{\widetilde{W}} e^{i \varphi_{\widetilde{W}}}&=&-0.074\,
e^{i\varphi_1} +198.763\,
e^{i\varphi_2} -7.410\, e^{i\varphi_3}\nonumber\\
M_{\widetilde{g}} e^{i \varphi_{\widetilde{g}}} &=&-0.332\,
e^{i\varphi_1} -2.684\, e^{i\varphi_2} + 605.705\, e^{i\varphi_3}
\end{eqnarray}
and for squark soft mass-squareds
\begin{eqnarray}
\label{squark-soln} {m}_{\widetilde{Q}}^2&=& (560.725)^2
\left(1 - 4.66\, 10^{-5} \cos\varphi_{12}-8.86\, 10^{-4}
\cos\varphi_{13}- 1.19\, 10^{-2}
\cos\varphi_{23}\right)\nonumber\\
{m}_{\widetilde{u}_R}^2&=& (542.317)^2 \left(1 - 6.49\,
10^{-6} \cos\varphi_{12}-1.64\, 10^{-3} \cos\varphi_{13}- 6.65\,
10^{-3}
\cos\varphi_{23}\right)\nonumber\\
{m}_{\widetilde{d}_R}^2&=& (540.119)^2 \left(1 - 2.31\,
10^{-7} \cos\varphi_{12}-1.13\, 10^{-3} \cos\varphi_{13}- 6.44\,
10^{-3} \cos\varphi_{23}\right)
\end{eqnarray}
all of which being given in ${\rm GeV}$. The angle parameters
appearing in soft mass-squareds are defined as $\varphi_{ij} =
\varphi_i - \varphi_j$  with $i,j=1,2,3$. These semi-analytic
solutions prove quite useful while interpreting weak-scale
parameters in terms of the GUT-scale ones. For instance, as
follows from (\ref{squark-soln}), squark soft mass-squareds are
found to feel GUT-scale phases very weakly. In fact, largest
contribution comes from $\varphi_{23}$ and it remains at $1 \%$
level. The squark soft mass-squareds entering the cross sections
are obtained by including the D-term contributions:
\begin{eqnarray}
\label{squark-phys} \left({m}_{\widetilde{u}_L}^2\right)^{D-term} &=&
{{m}_{\widetilde{Q}}}^2 +
\frac{1}{6} (4 M_{W}^2 - M_{Z}^2) \cos 2\beta \nonumber\\
\left({m}_{\widetilde{d}_L}^2\right)^{D-term} &=& {m}_{\widetilde{Q}}^2 -
\frac{1}{6} (2 M_{W}^2 + M_{Z}^2) \cos 2\beta \nonumber\\
\left({m}_{\widetilde{u}_R}^2\right)^{D-term} &=& {m}_{\widetilde{u}_R}^2 +
\frac{2}{3} (M_{Z}^2 - M_{W}^2) \cos 2\beta \nonumber\\
\left({m}_{\widetilde{d}_R}^2\right)^{D-term} &=& {m}_{\widetilde{d}_R}^2 -
\frac{1}{3} (M_{Z}^2 - M_{W}^2) \cos 2\beta
\end{eqnarray}
which are the physical squared-masses of the squarks belonging to
first and second generations. (In what follows we will drop the superscript
$^{D-term}$ for simplicity of notation.).

Returning to gaugino masses (\ref{gaugino-soln}), the two-loop
contributions are found to modify both moduli and phases of the
gaugino masses at the weak scale. However, these effects do not
exceed a few percent; the largest departure from one-loop scheme
occurs in $M_{\widetilde{W}}$, due to gluino mass, and it is at
${\cal{O}}(4 \%)$ level. Similarly, $M_{\widetilde{B}}$ undergoes
an ${\cal{O}}(3 \%)$ modification due to gluino contribution. On
the other hand, running of the gluino mass is influenced in a less
significant way by the masses of electoweak gauginos.
Consequently, depending on precision with which certain
observables are measured, in some cases one can employ the
following approximate relations
\begin{eqnarray}
\label{gaugino-approx} \varphi_{\widetilde{g}} \approx
\varphi_3\;,\;\;\varphi_{\widetilde{W}} \approx \varphi_2\;,\;\;
\varphi_{\widetilde{B}} \approx \varphi_1
\end{eqnarray}
along with $M_{\widetilde{g}} \approx M_3$, $M_{\widetilde{W}}
\approx M_2$ and $M_{\widetilde{B}} \approx M_1$. Hence, though we
deal with constrained MSSM with complex gaugino masses at the GUT
scale, as given in (\ref{lm1p}), one might regard whole analysis
as being carried out in unconstrained MSSM with squark masses in
(\ref{squark-phys}) and gaugino masses in (\ref{gaugino-approx}).
In this sense, the numerical results that follow can be
interpreted within unconstrained MSSM with explicit CP violation
and fixed moduli for sparticle masses. However, one keeps in mind
that precision with which certain observables are probed can be
sensitive to two-loop effects encoded in (\ref{gaugino-soln}). In
such cases, (\ref{gaugino-approx}) represents a too bad
approximation to utilize.

In the numerical calculations below we take $\varphi_{\mu} =0$ so
that $\mu$ parameter is real positive (as required by $b
\rightarrow s \gamma$ for instance \cite{bsgam}). This choice can
be useful also for not violating the EDM bounds \cite{cancel} in
spite of ${\cal{O}}(1)$ values for rest of the phases in
(\ref{lm1p}). We test our numerical results against PYTHIA
predictions \cite{pythia} in those regions of the parameter space
where all phases vanish.

In what follows, for clarity of discussions, we divide numerical
analysis into subparts according to what squarks are observed with
what property. This kind of fine-graining of squark production
could be useful for interpretation of the signal in simulations as
well as experimentation.

\subsection{Squark Pair-Production: Definite Flavor and Definite Chirality}
In this subsection we analyze productions of squarks with a
definite flavor and chirality in regard to their dependencies on
the CP-odd soft phases of the gluinos and neutralinos.

\begin{figure}
\includegraphics[width=5cm]{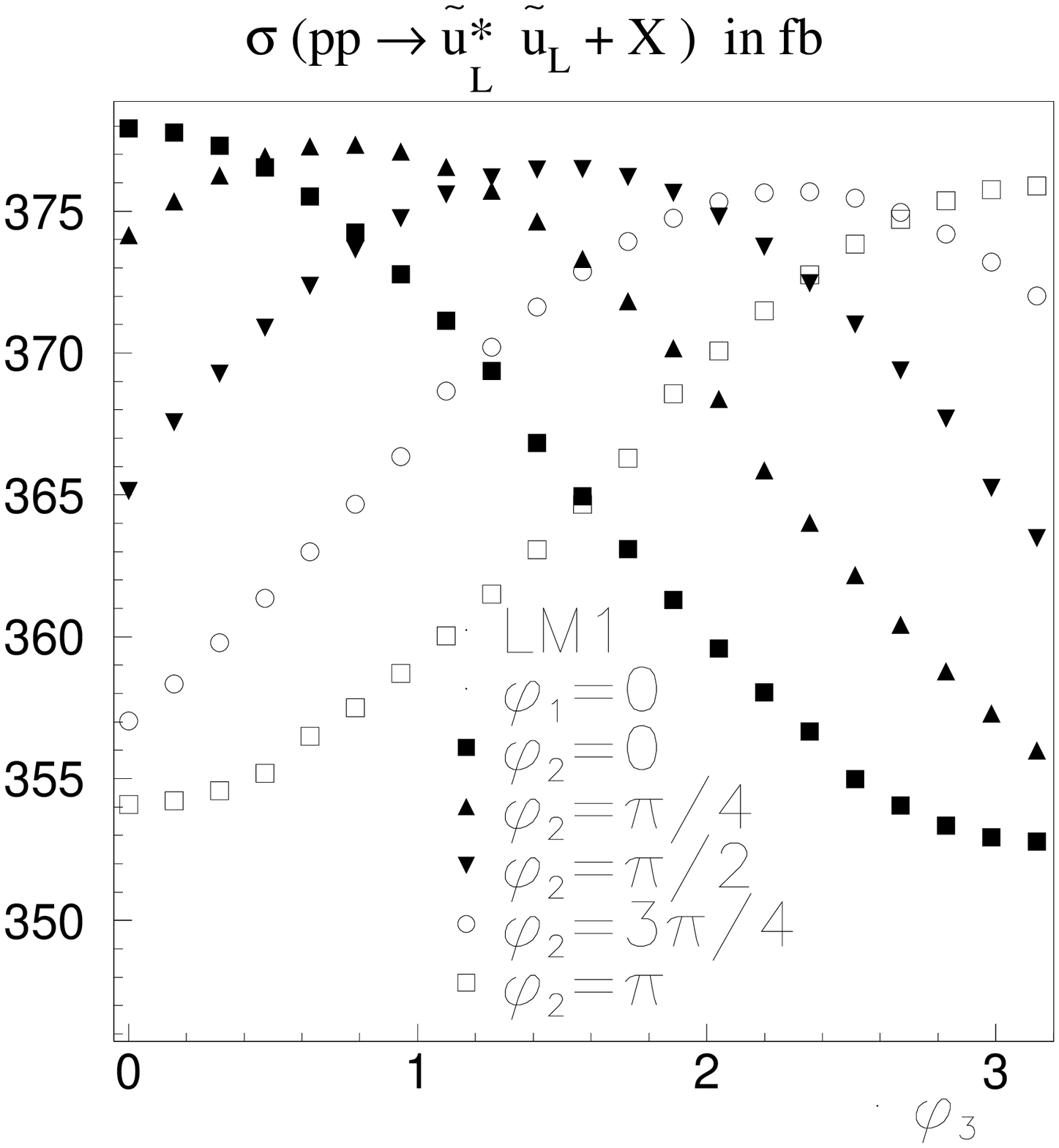}
\includegraphics[width=5cm]{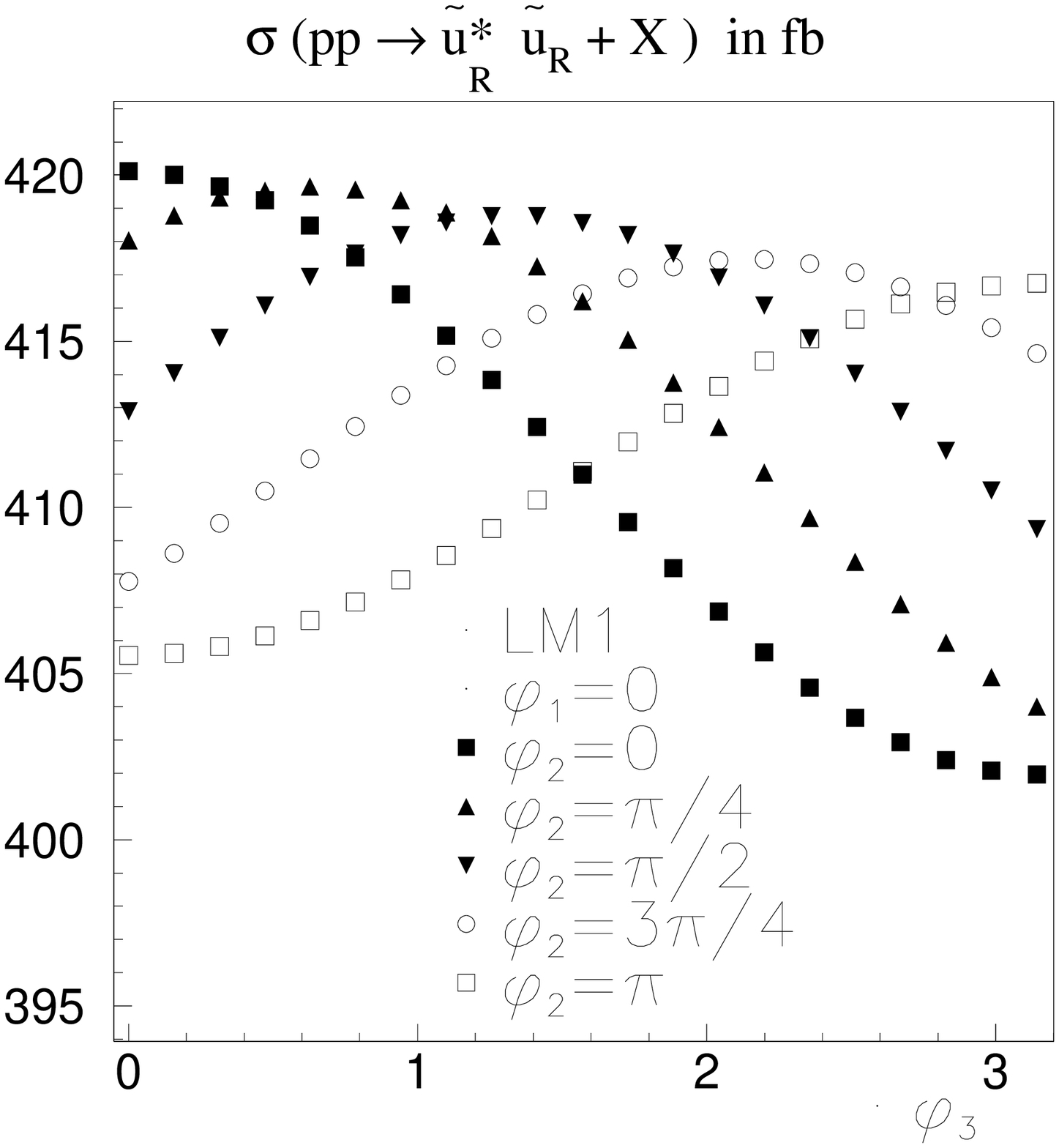}
\includegraphics[width=5cm]{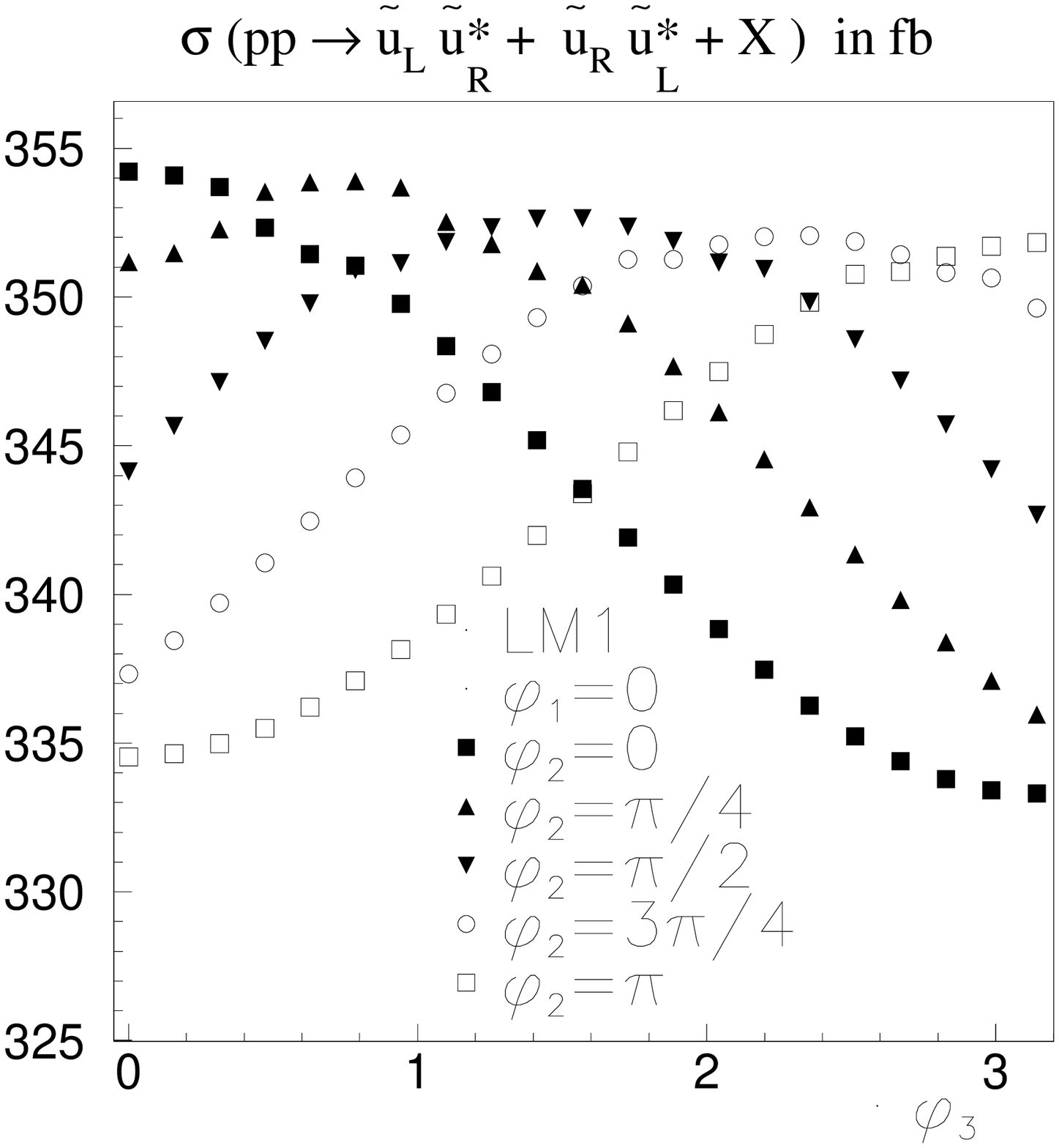}
\includegraphics[width=5cm]{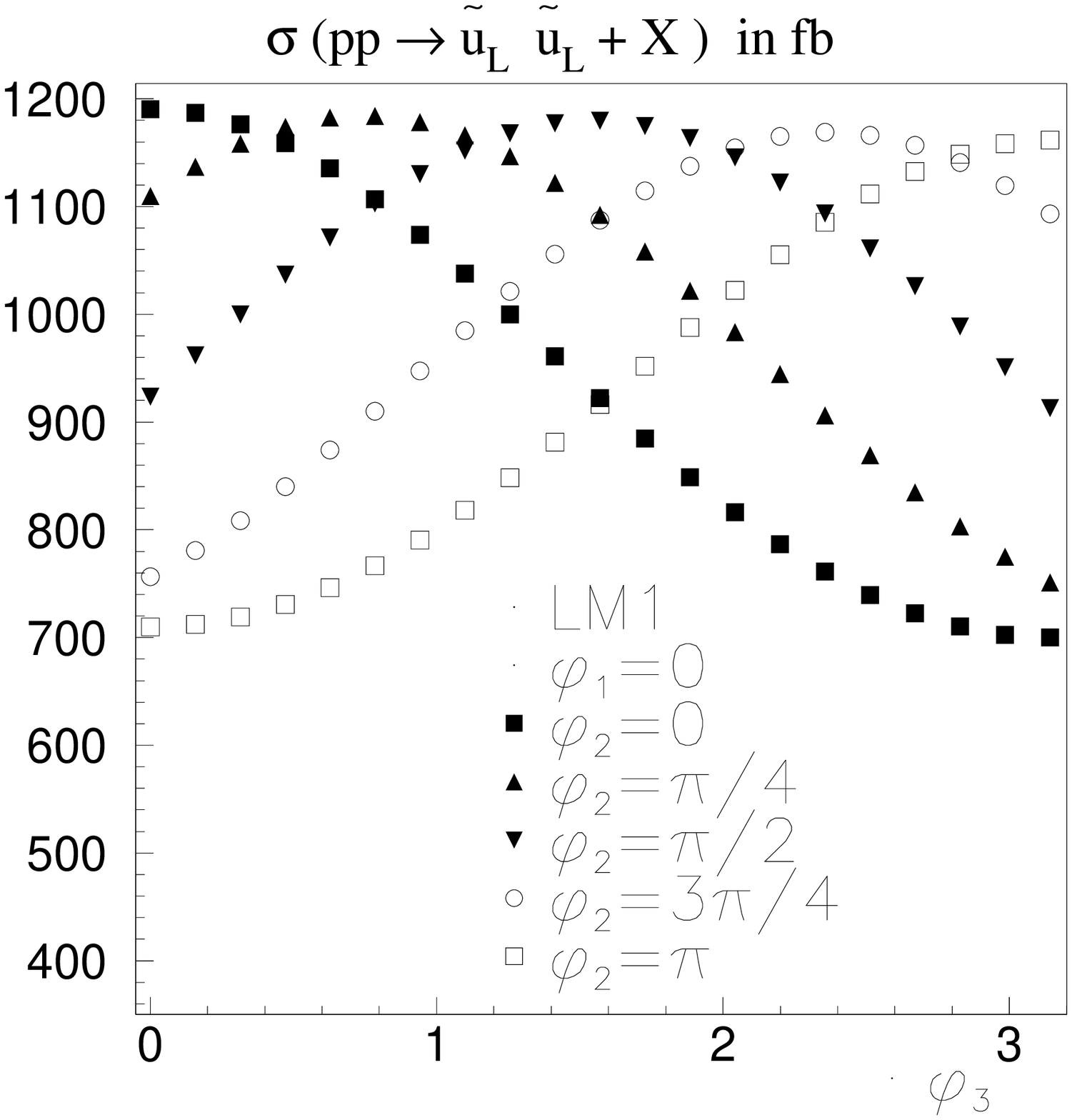}
\includegraphics[width=5cm]{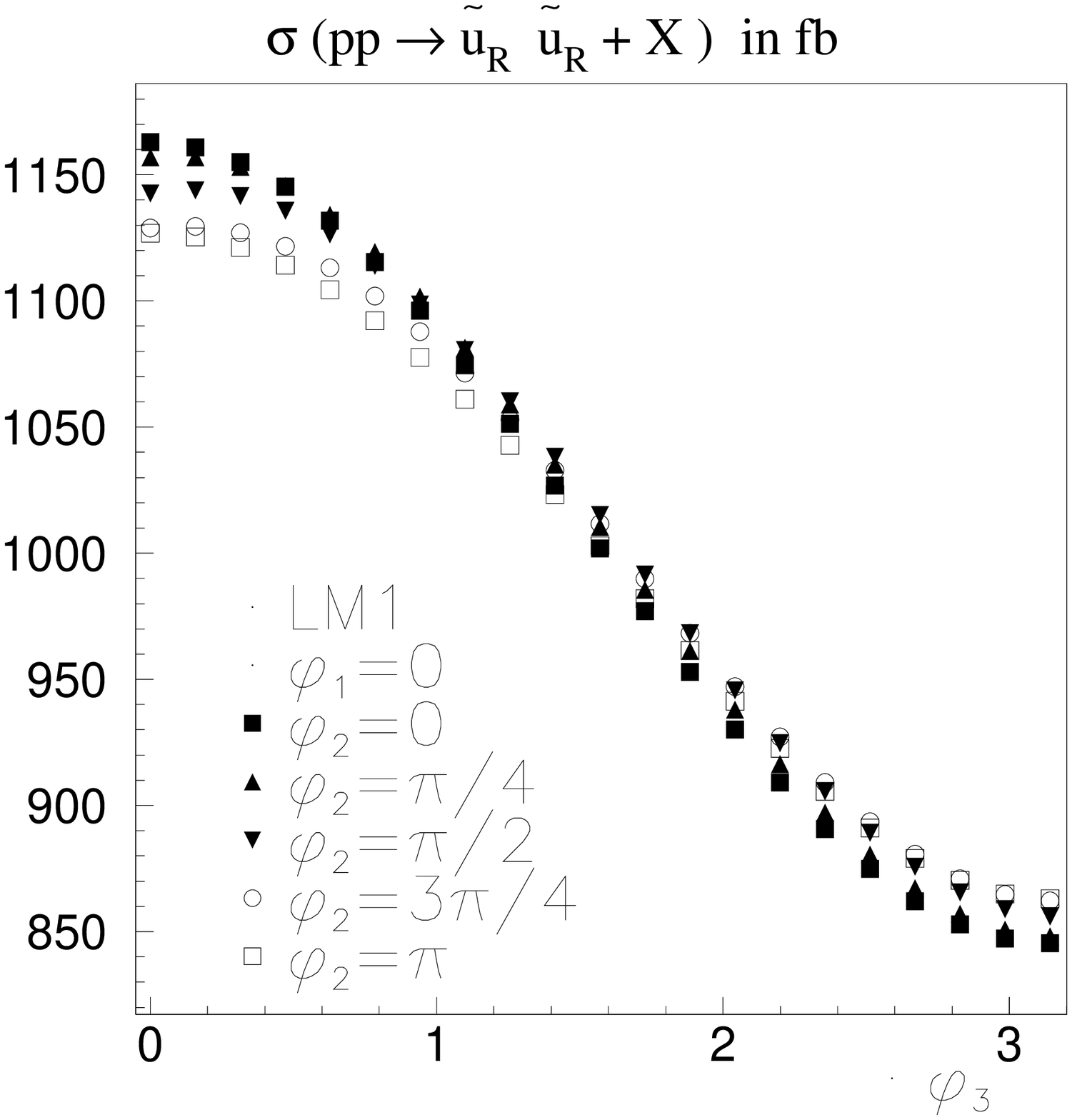}
\includegraphics[width=5cm]{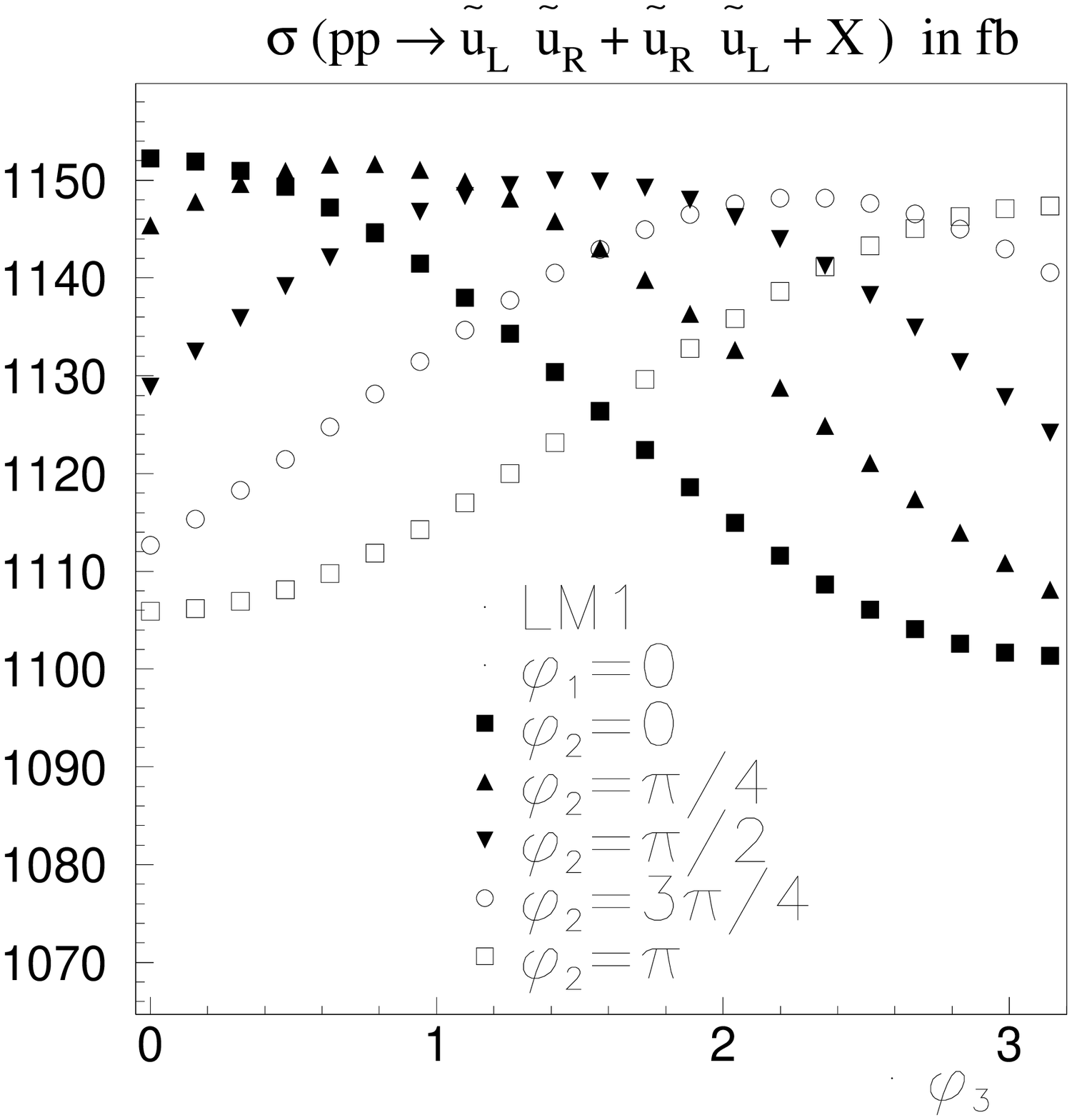}
\caption{Up squark pair-production cross sections (in fb) at the
LHC as functions of $\varphi_3$ for $\varphi_1=0$ and several
values of $\varphi_2$. Left: $\sigma(p\, p \rightarrow
\widetilde{u}_L \widetilde{u}_{L}^{\star})$ (top panel) and
$\sigma(p\, p \rightarrow \widetilde{u}_L \widetilde{u}_{L})$
(bottom panel). Middle: $\sigma(p\, p \rightarrow \widetilde{u}_R
\widetilde{u}_{R}^{\star})$ (top panel) and $\sigma(p\, p
\rightarrow \widetilde{u}_R \widetilde{u}_{R})$ (bottom panel).
Right: $\sigma(p\, p \rightarrow \widetilde{u}_L
\widetilde{u}_{R}^{\star})$ (top panel) and $\sigma(p\, p
\rightarrow \widetilde{u}_L \widetilde{u}_{R})$ (bottom
panel).}\label{fig-sig-up_0}
\end{figure}

\begin{figure}
\includegraphics[width=5cm]{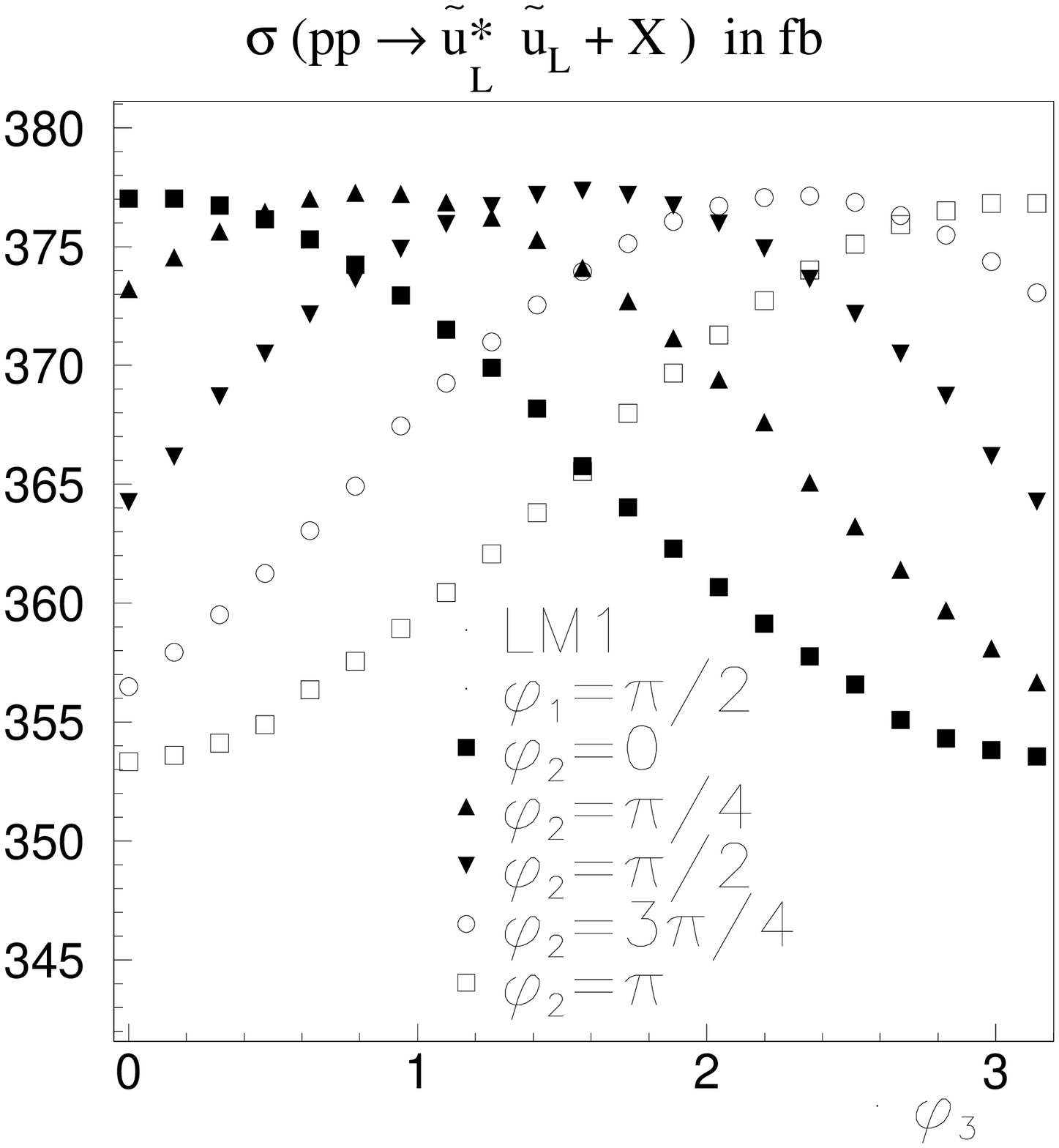}
\includegraphics[width=5cm]{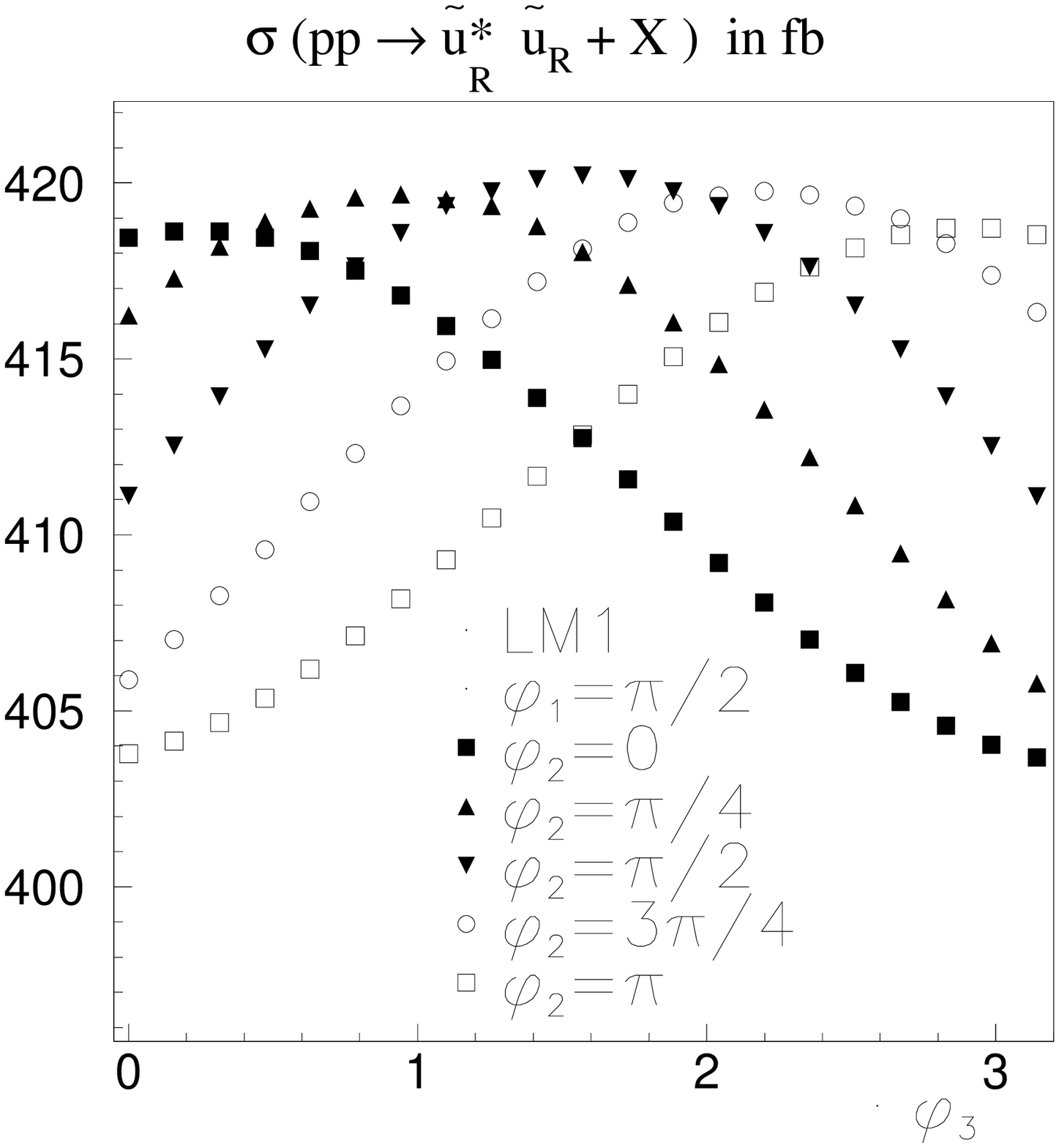}
\includegraphics[width=5cm]{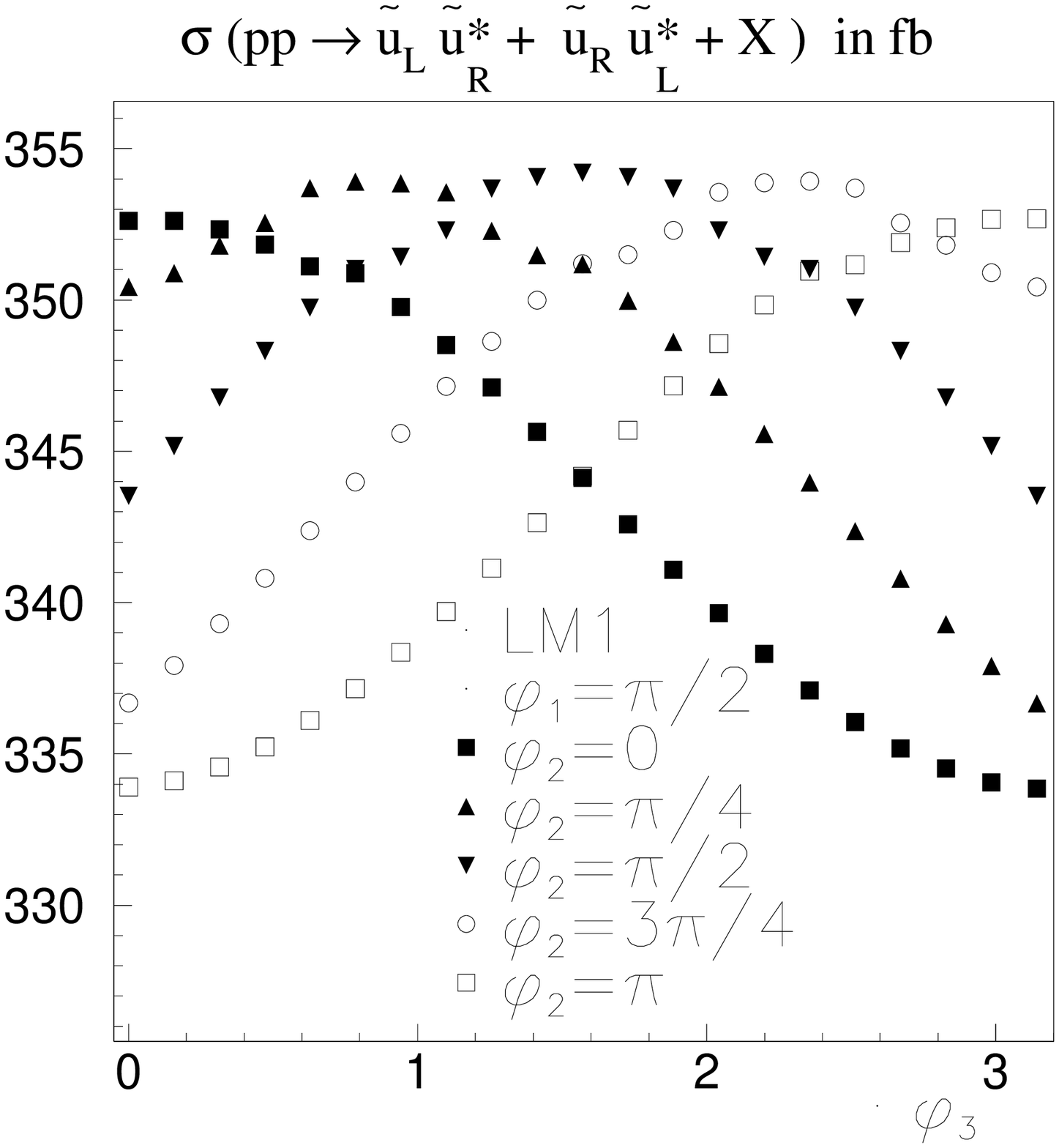}
\includegraphics[width=5cm]{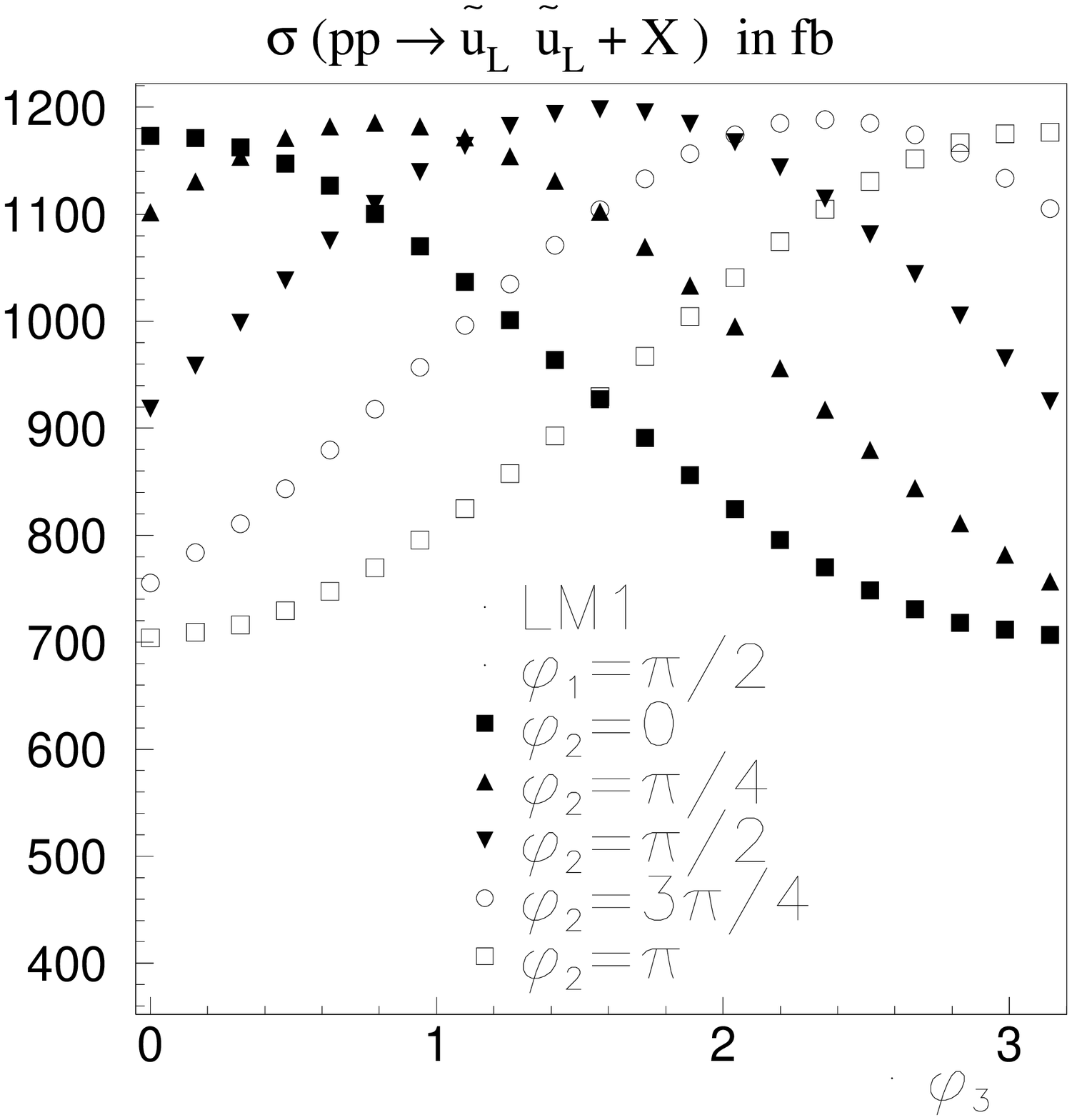}
\includegraphics[width=5cm]{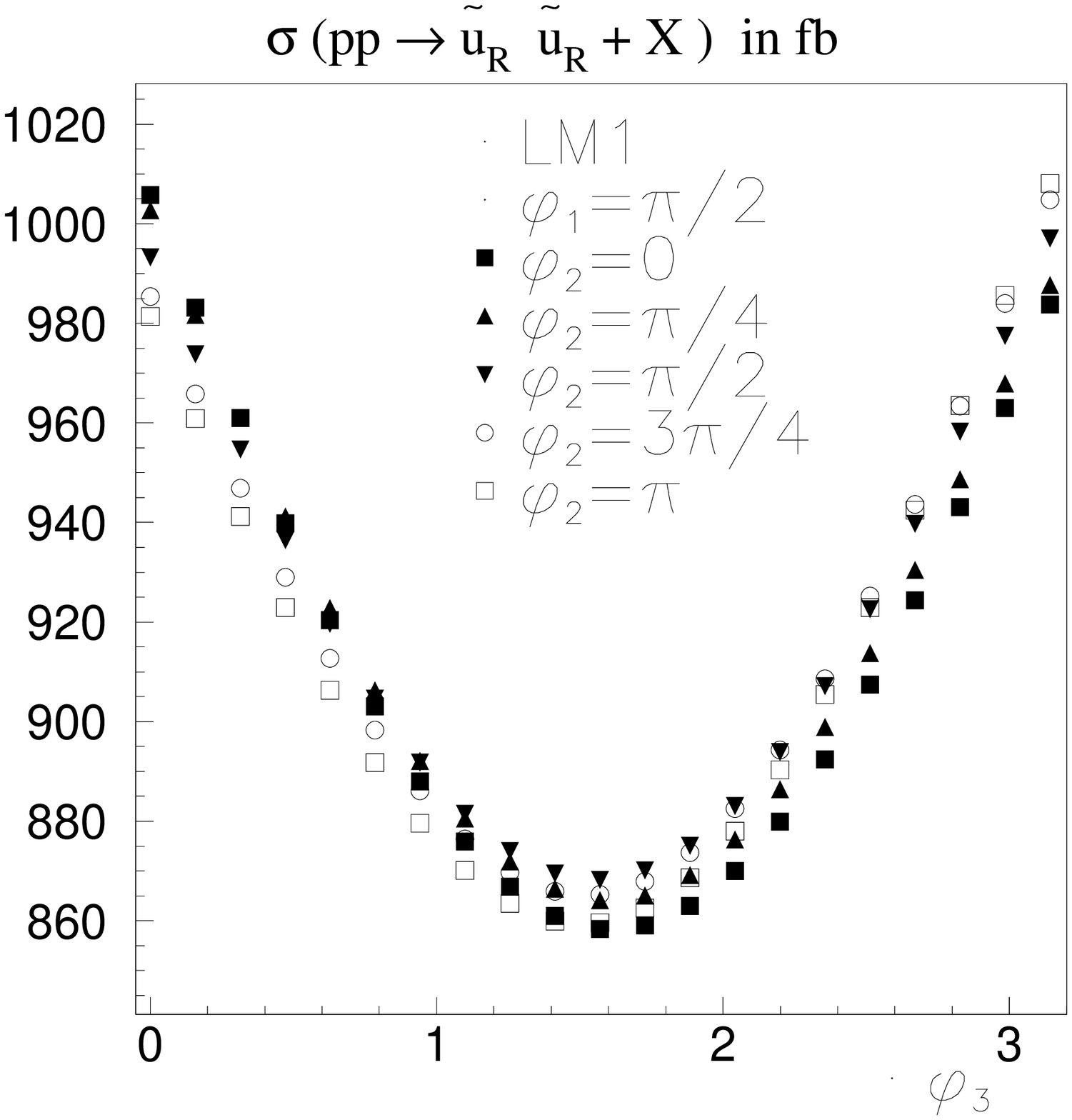}
\includegraphics[width=5cm]{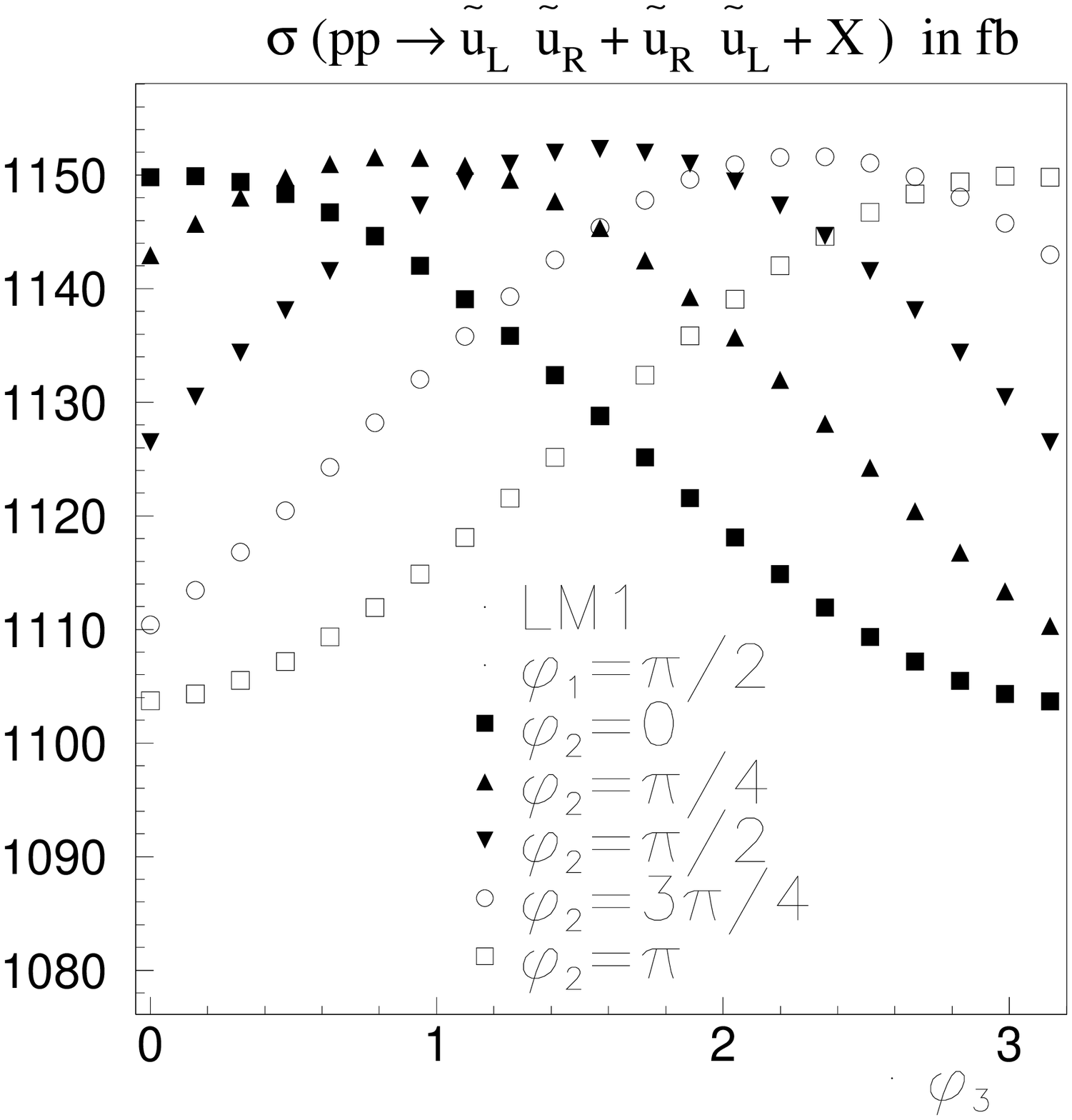}
\caption{ The same as in Fig. \ref{fig-sig-up_0} but for
$\varphi_1 = \pi/2$. }\label{fig-sig-up_pi2}
\end{figure}

We start analysis by illustrating the dependencies of $\sigma(p\,
p \rightarrow \widetilde{q} \widetilde{q}^{\star})$ and
$\sigma(p\, p \rightarrow \widetilde{q} \widetilde{q})$ upon the
CP-odd phases $\varphi_{1, 2, 3}$ for $\widetilde{q} =
\widetilde{u}$ $i.e.$ the pair-productions of scalar up (or,
effectively, scalar charm quarks). Depicted in Fig.
\ref{fig-sig-up_0} are $\sigma(p\, p \rightarrow \widetilde{q}
\widetilde{q}^{\star})$ and $\sigma(p\, p \rightarrow
\widetilde{q} \widetilde{q})$ for $\varphi_1 =0$, and the ones in
Fig. \ref{fig-sig-up_pi2} are the same cross sections for
$\varphi_1 = \pi/2$.

As observed in the top panel of the left column, $\sigma(p\, p
\rightarrow \widetilde{u}_L \widetilde{u}_{L}^{\star}) \approx
378\ {\rm fb}$ in agreement with the PYTHIA prediction
\cite{pythia}. It falls down to $\approx 352\ {\rm fb}$ as
$\varphi_{3}$ varies from 0 to $\pi$. However, as $\varphi_2$
varies from $0$ to $\pi$, in steps of $\pi/4$, $\sigma(p\, p
\rightarrow \widetilde{u}_L \widetilde{u}_{L}^{\star})$ is seen to
reverse its behavior at $\varphi_2 = 0$; it equals $\approx 354\
{\rm fb}$ at $\varphi_3 =0$ and $\approx 376\ {\rm fb}$ at
$\varphi_3 = \pi$. Clearly, explicit $\varphi_3$ dependence of
this cross section stems solely from the two-loop contributions of
the gluino mass to isospin and hypercharge gaugino masses  in
(\ref{gaugino-soln}). More explicitly, dependence on the phases
(apart from squark masses (\ref{squark-soln}) and gaugino masses
$M_{\widetilde{W}, \widetilde{B}}$ in (\ref{gaugino-soln})) comes
from $\left|A_{q_L i}\right|^2 \approx (1/4) \left|N_{i 2} - 0.18
N_{i 1}\right|^2$ which is obviously dominated by $N_{2 2}$
contribution $i.e.$ the isospin gaugino. Therefore, $\varphi_{3}$
dependencies of $M_{\widetilde{W}}$ and $\varphi_{\widetilde{W}}$
dominantly determine the sensitivity of $\sigma(p\, p \rightarrow
\widetilde{u}_L \widetilde{u}_{L}^{\star})$ on GUT-scale phases
$\varphi_{1,2,3}$. The total swing of the cross section $i.e.$ the
difference between its extrema is $\approx 26\ {\rm fb}$.

In a strict unconstrained MSSM framework (expressed by approximate
relations in (\ref{gaugino-approx})), this top panel at the left
column of Fig. \ref{fig-sig-up_0} would be plotted against
$\varphi_{\widetilde{g}}$ and it would be a series of horizontal
lines. In this sense, manifest $\varphi_3$ dependence of
$\sigma(p\, p \rightarrow \widetilde{u}_L
\widetilde{u}_{L}^{\star})$ signals departure from strict
unconstrained MSSM limit in which gluino phase is expected to give
no contribution to $\widetilde{q}\, \widetilde{q}^{\star}$
production, in general.

The observations made above hold also for $\sigma(p\, p
\rightarrow \widetilde{u}_R \widetilde{u}_{R}^{\star})$ cross
section given in the top panel in the middle column of Fig.
\ref{fig-sig-up_0}. The main difference lies in the fact that
$\sigma(p\, p \rightarrow \widetilde{u}_R
\widetilde{u}_{R}^{\star})$ involves $\left| A_{q_R i}\right|^2 =
0.14 \left|N_{i 1}\right|^2$ which is dominated by $N_{1 1}$
$i.e.$ the hypercharge gaugino. $N_{1 1}$ is less sensitive to
$\varphi_{2,3}$ than $N_{2 2}$, and hence comparatively milder
phase-dependence of $\sigma(p\, p \rightarrow \widetilde{u}_R
\widetilde{u}_{R}^{\star})$ than $\sigma(p\, p \rightarrow
\widetilde{u}_L \widetilde{u}_{L}^{\star})$. The cross section
exhibits a total swing of $18\ {\rm fb}$ which can roughly be
estimated from $\sigma(p\, p \rightarrow \widetilde{u}_L
\widetilde{u}_{L}^{\star})$ swing given the dependencies of
$M_{\widetilde{B}}$ and $M_{\widetilde{W}}$ in
(\ref{gaugino-soln}) on $\varphi_3$.

Depicted in top panel of the right column of Fig.
\ref{fig-sig-up_0} is $\sigma(p\, p \rightarrow \widetilde{u}_L
\widetilde{u}_{R}^{\star})$. Unlike $\widetilde{u}_R
\widetilde{u}_{R}^{\star}$ production, and like $\widetilde{u}_L
\widetilde{u}_{L}^{\star}$ production, $ \widetilde{u}_L
\widetilde{u}_{R}^{\star}+ \widetilde{u}_R
\widetilde{u}_{L}^{\star}$ production cross section exhibits a
strong sensitivity to phases; its total swing is $\sim 22\ {\rm
fb}$. The reason for this is clear: $\sigma(p\, p \rightarrow
\widetilde{u}_L \widetilde{u}_{R}^{\star})$ involves $A_{q_L i}
A_{q_R i}^{\star}=0.183 N_{i 1} N_{i 2} - 0.0335 N_{i 1}^2$ which
combines constructively the phases of isospin and hypercharge
gauginos. Therefore, this pair production mode, like
$\widetilde{u}_R \widetilde{u}_{R}^{\star}$ production,  exhibits
a strong sensitivity to $\varphi_3$ due to the fact that it
involves phases of both $M_{\widetilde{W}}$ and $M_{
\widetilde{B}}$ in an additive fashion.

Given in the bottom panels of Fig. \ref{fig-sig-up_0} are
$\sigma(p\, p \rightarrow \widetilde{u}_L \widetilde{u}_{L})$ (the
left panel), $\sigma(p\, p \rightarrow \widetilde{u}_R
\widetilde{u}_{R})$ (the middle panel) and $\sigma(p\, p
\rightarrow \widetilde{u}_L \widetilde{u}_{R})$ (the right panel).
The figure shows it explicitly that $\sigma(p\, p \rightarrow
\widetilde{u}_L \widetilde{u}_{L})$, compared to $\sigma(p\, p
\rightarrow \widetilde{u}_L \widetilde{u}_{L}^{\star})$ atop,
exhibits a much stronger variation with $\varphi_3$ and
$\varphi_2$. In accord with discussions in Sec. III above, this
pronounced dependence follows from $t$-channel exchange of gluino
and isospin gaugino with direct dependence on their phases (See
Table \ref{table1}). Therefore, $\widetilde{u}_L
\widetilde{u}_{L})$ production is a sensitive probe of the gluino
and wino phases (and of the hypercharge phase depending on mixing
in the neutralino mass matrix). One notes that, in unconstrained
MSSM limit one would have a similar pattern for $\sigma(p\, p
\rightarrow \widetilde{u}_L \widetilde{u}_{L})$.

A striking illustration of phase sensitivities of cross sections
is provided by $\sigma(p\, p \rightarrow \widetilde{u}_R
\widetilde{u}_{R})$ shown in the middle panel. This process
proceeds with gluino and bino exchanges in $t$ channel, and thus
it is a highly sensitive probe of $\varphi_3$. Its $\varphi_2$
dependence is rather weak as expected from (\ref{gaugino-soln}).

As shown in the right panel, $\sigma(p\, p \rightarrow
\widetilde{u}_L \widetilde{u}_{R})$ possesses the smallest swing:
its extrema differ by $49\ {\rm fb}$ which is much smaller than
$500\ {\rm fb}$ swing of $\sigma(p\, p \rightarrow \widetilde{u}_L
\widetilde{u}_{L})$ and $304\ {\rm fb}$ swing of $\sigma(p\, p
\rightarrow \widetilde{u}_R \widetilde{u}_{R})$. The reason for
this milder dependence on $\varphi_3$ results from destructive
combination of phases contained in $A_{q_L i} A_{q_R i}$. This
aspect has been detailed while analyzing $ \widetilde{u}_L
\widetilde{u}_{R}^{\star}+ \widetilde{u}_R
\widetilde{u}_{L}^{\star}$ production. One notes, in particular,
that variation of the cross section with $\varphi_3$ is extremely
suppressed for $\varphi_2= \pi$ due to the aforementioned
cancellation effects.

Fig. \ref{fig-sig-up_pi2} shows the same cross sections plotted in
Fig. \ref{fig-sig-up_0} for $\varphi_1 = \pi/2$. This repetition
is intended for determining how cross sections vary with the phase
of the hypercharge gaugino. A comparative look at these two
figures reveals some interesting aspects of $\varphi_1$
dependence. First of all, $\sigma(p\, p \rightarrow
\widetilde{u}_L \widetilde{u}_{L}^{\star})$, $\sigma(p\, p
\rightarrow \widetilde{u}_R \widetilde{u}_{R}^{\star})$,
$\sigma(p\, p \rightarrow \widetilde{u}_L
\widetilde{u}_{L}^{\star})$, $\sigma(p\, p \rightarrow
\widetilde{u}_L \widetilde{u}_{L})$ and $\sigma(p\, p \rightarrow
\widetilde{u}_L \widetilde{u}_{R})$ exhibit rather small changes
as $\varphi_1$ changes from 0 to $\pi/2$. This is actually
expected since these production modes are dominated by GUT-scale
isospin and gluino phases. On the other hand, $\sigma(p\, p
\rightarrow \widetilde{u}_R \widetilde{u}_{R})$ undergoes
observable modifications as $\varphi_1$ changes from 0 to $\pi/2$.
This process  proceeds exclusively with $t$-channel gluino and
bino exchanges. Indeed, while $\sigma(p\, p \rightarrow
\widetilde{u}_R \widetilde{u}_{R}) \sim 1000 + 150 \cos \varphi_3\
{\rm fb}$ at $\varphi_1 =0$ it changes to $\sigma(p\, p
\rightarrow \widetilde{u}_R \widetilde{u}_{R}) \sim 1000 - 140
\sin \varphi_3\ {\rm fb}$ at $\varphi_1 =\pi/2$. This strong
variation with $\varphi_{1,3}$ makes $\widetilde{u}_R
\widetilde{u}_{R}$ a viable probe for hunting CP-odd phases. One
notes that similar $\varphi_{\widetilde{g}, \widetilde{B}}$
dependencies are also expected in the unconstrained MSSM.

\begin{figure}
\includegraphics[width=5cm]{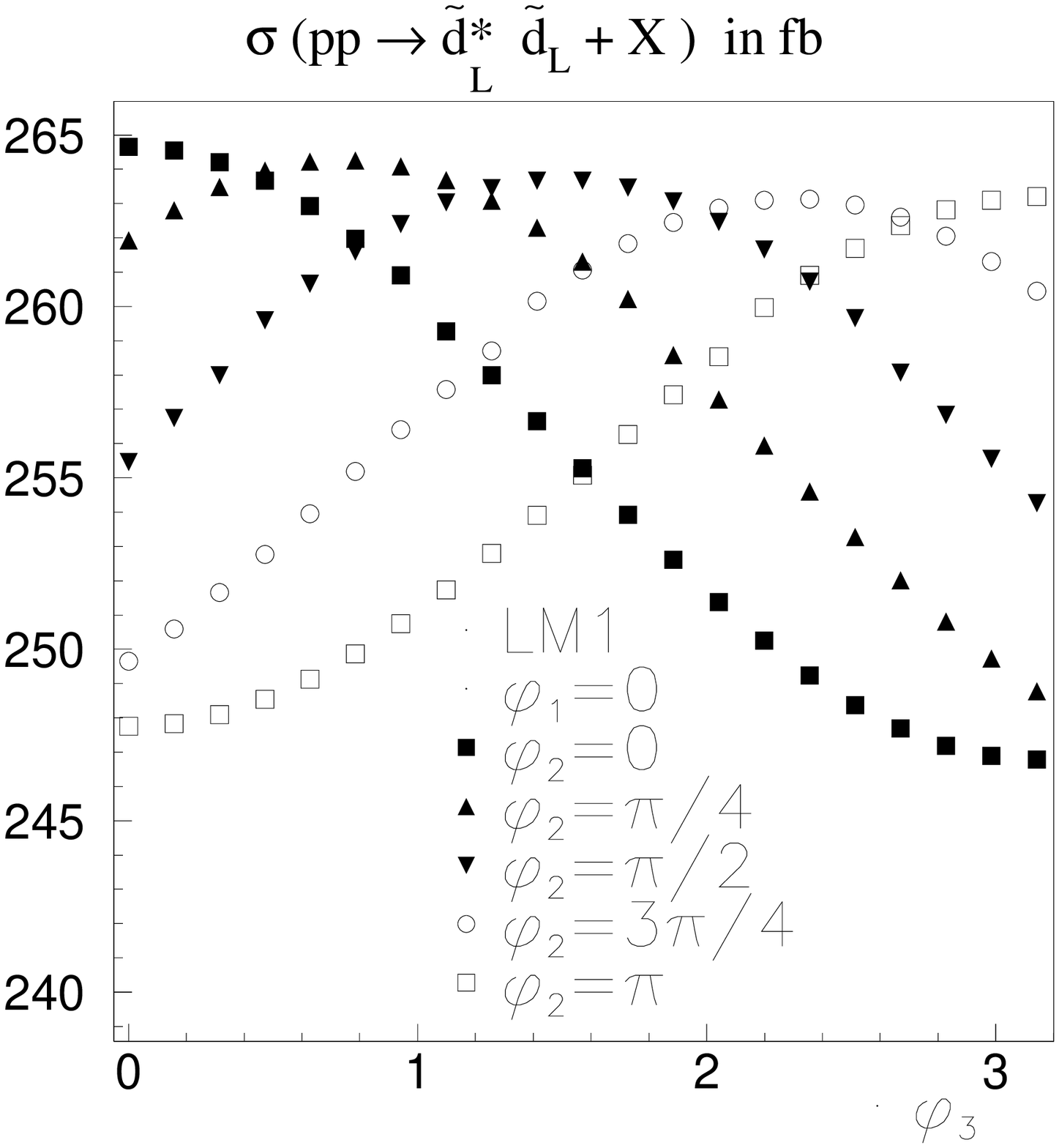}
\includegraphics[width=5cm]{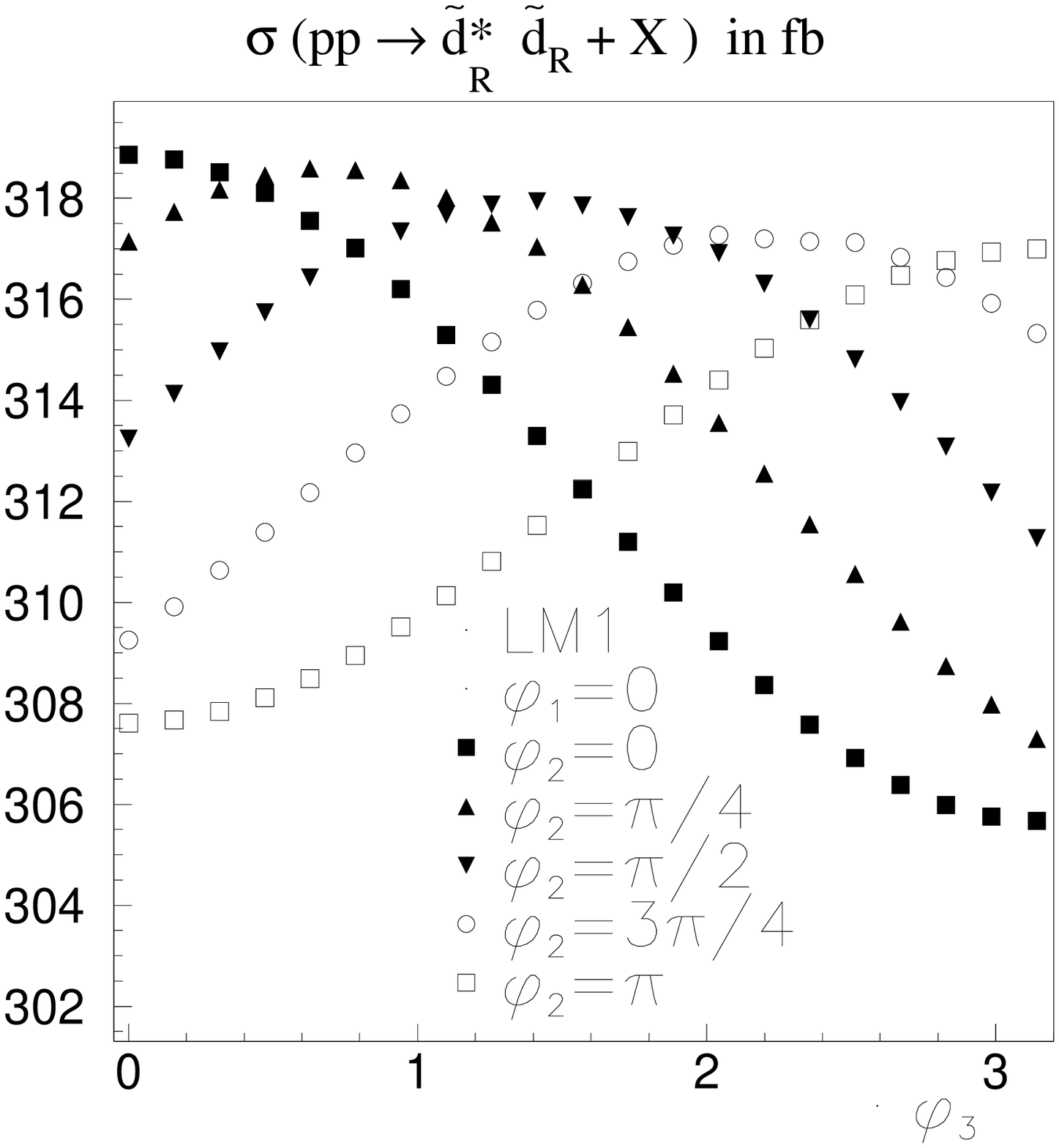}
\includegraphics[width=5cm]{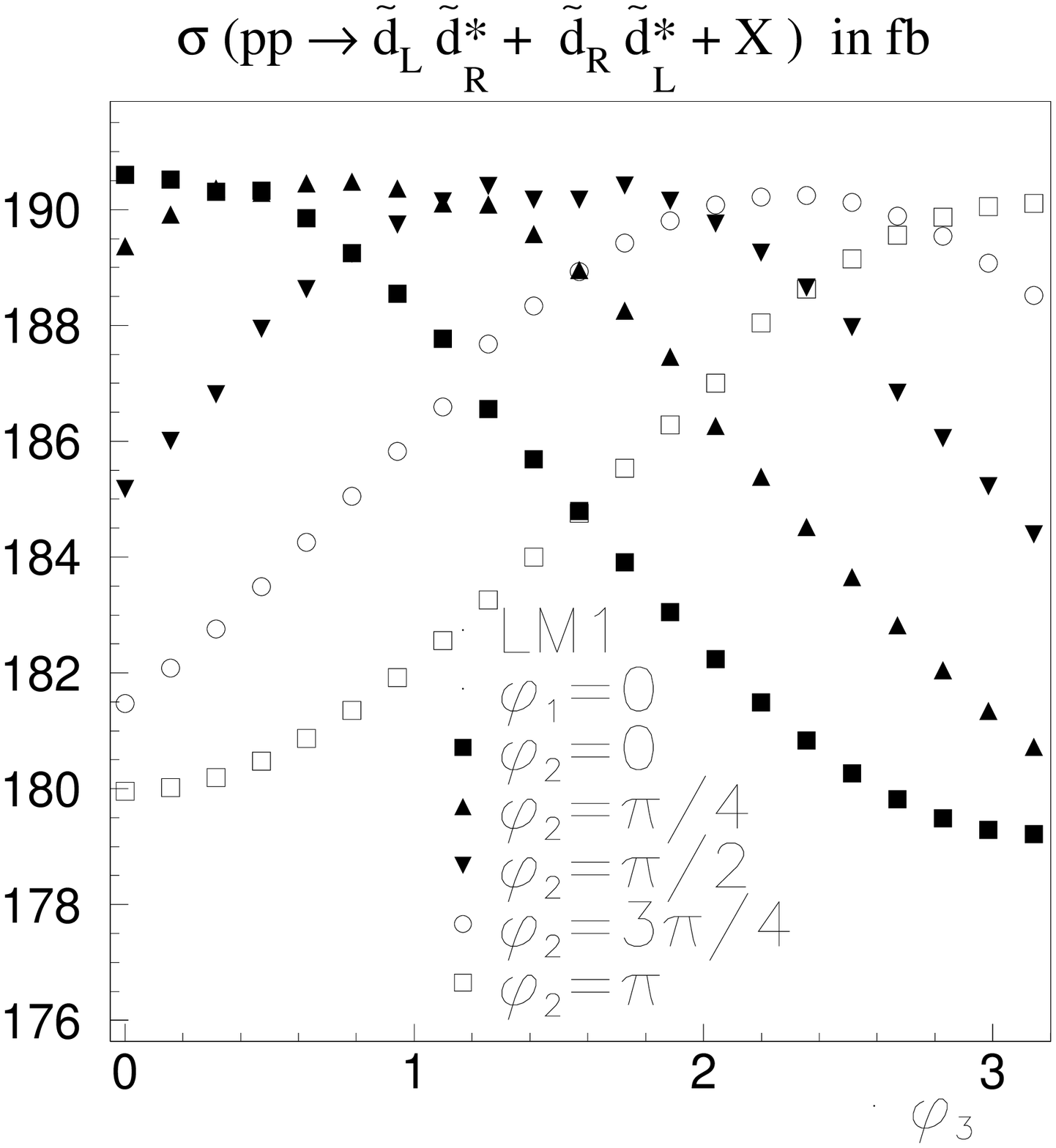}
\includegraphics[width=5cm]{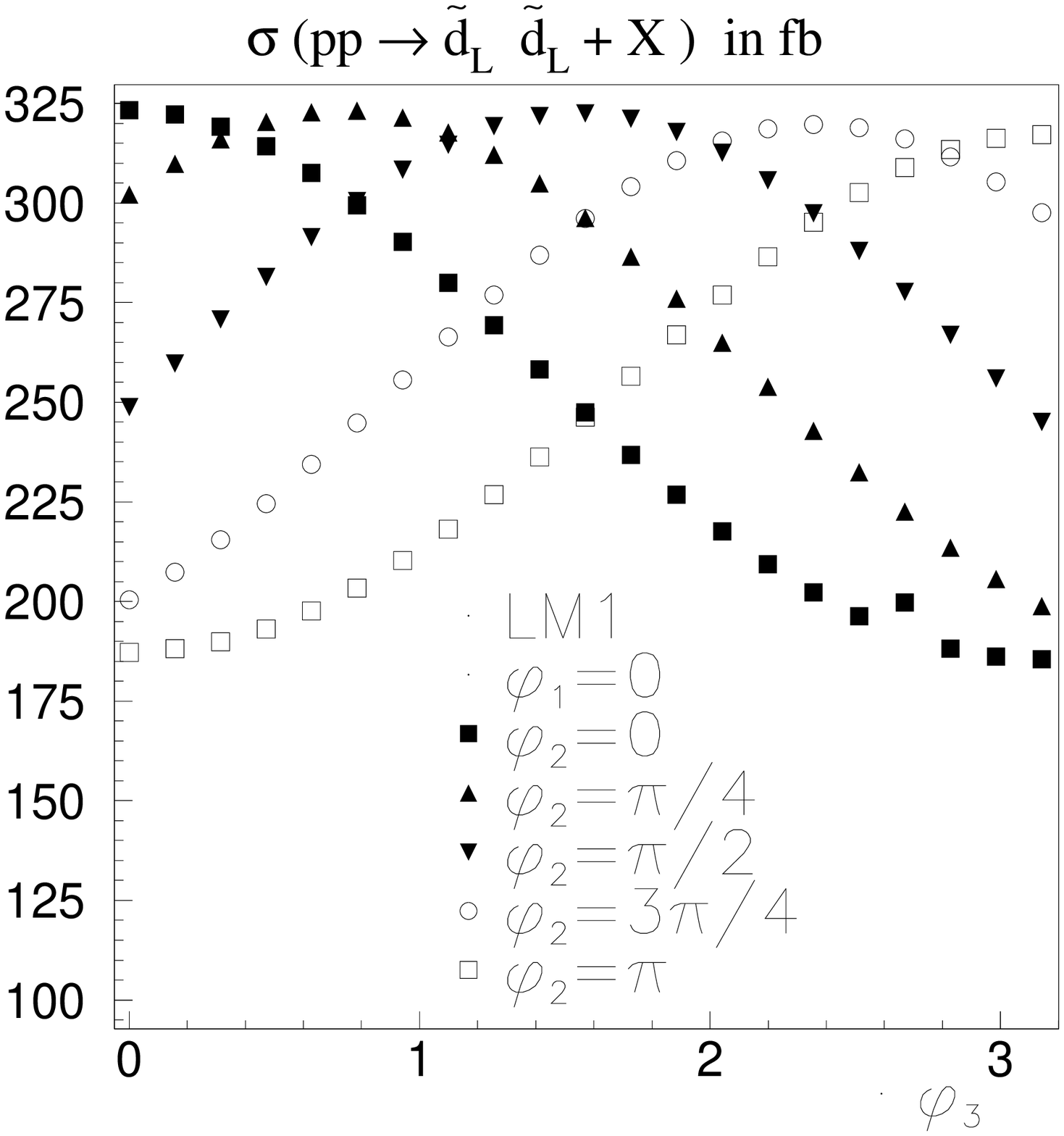}
\includegraphics[width=5cm]{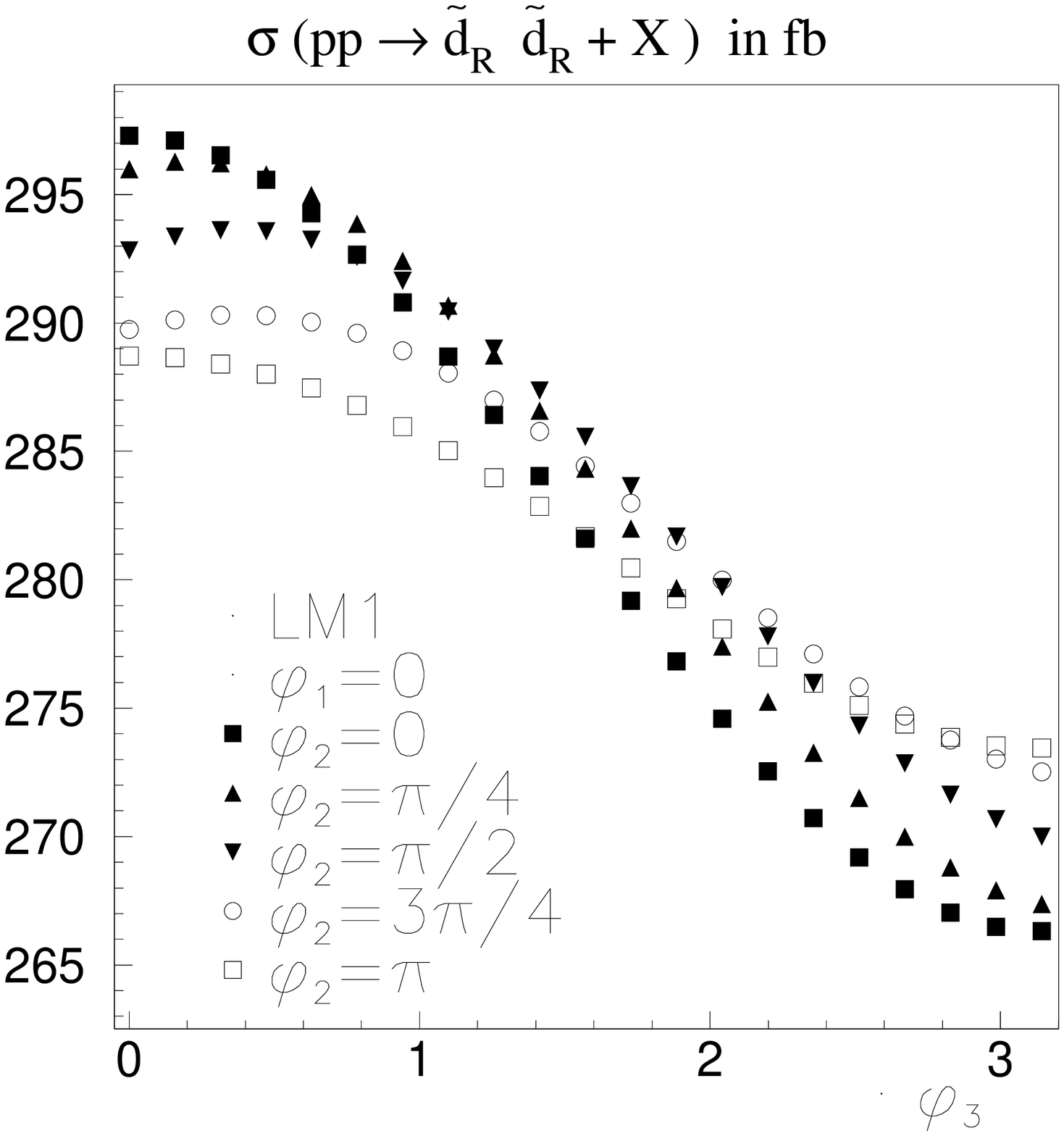}
\includegraphics[width=5cm]{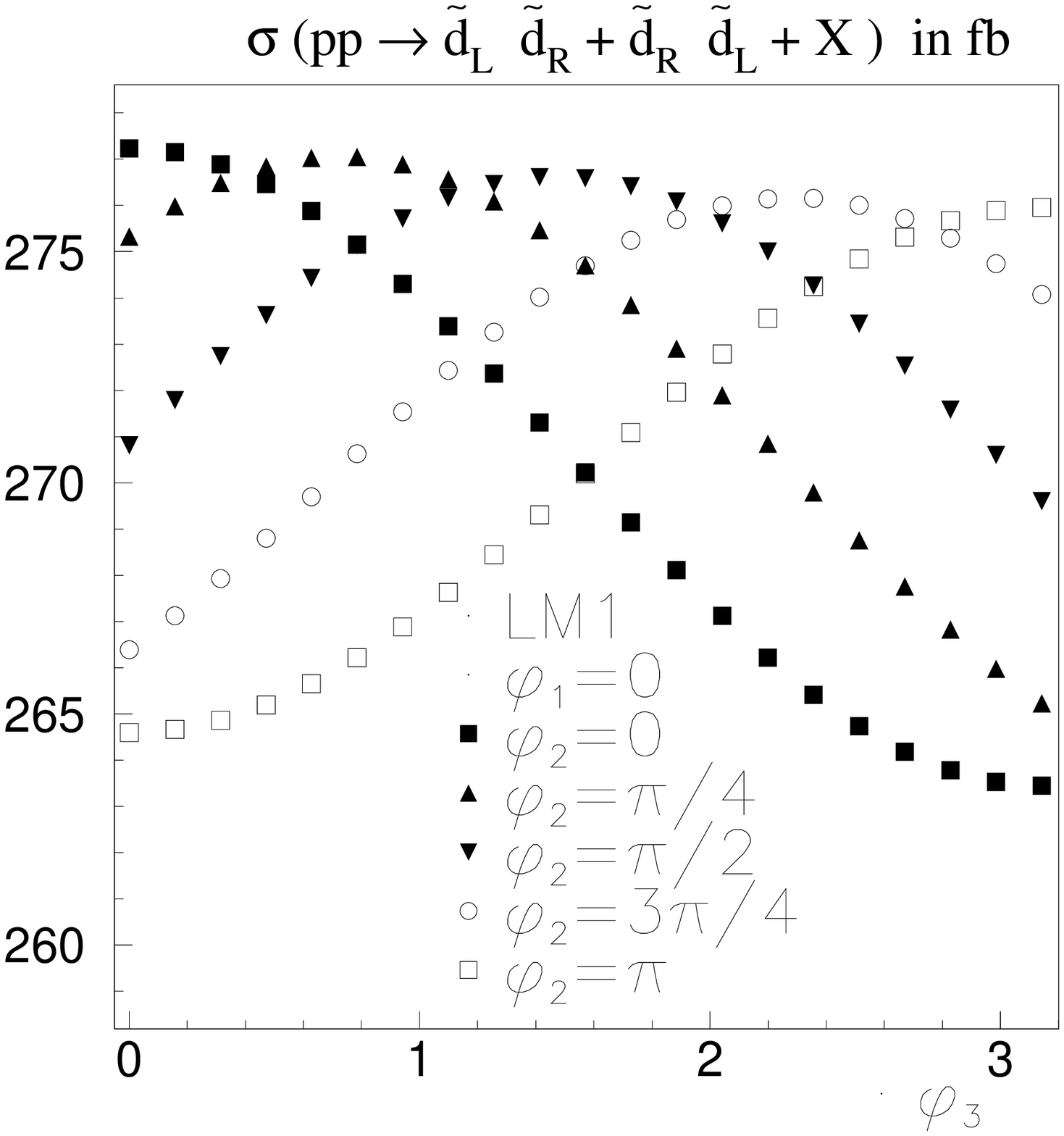}
\caption{The same as in Fig. \ref{fig-sig-up_0} but for pair
production of down or strange squarks. }\label{fig-sig-down_0}
\end{figure}

\begin{figure}
\includegraphics[width=5cm]{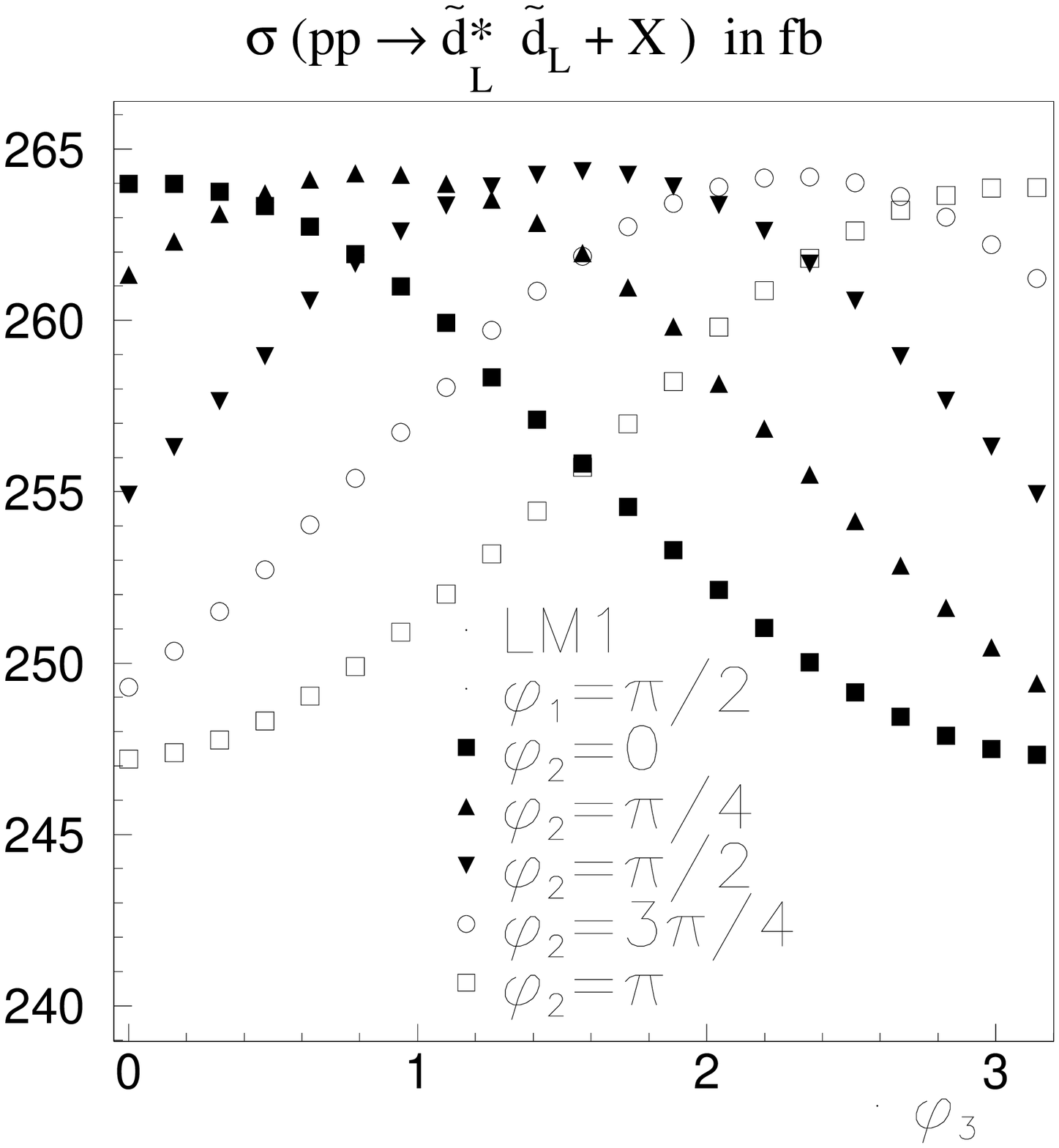}
\includegraphics[width=5cm]{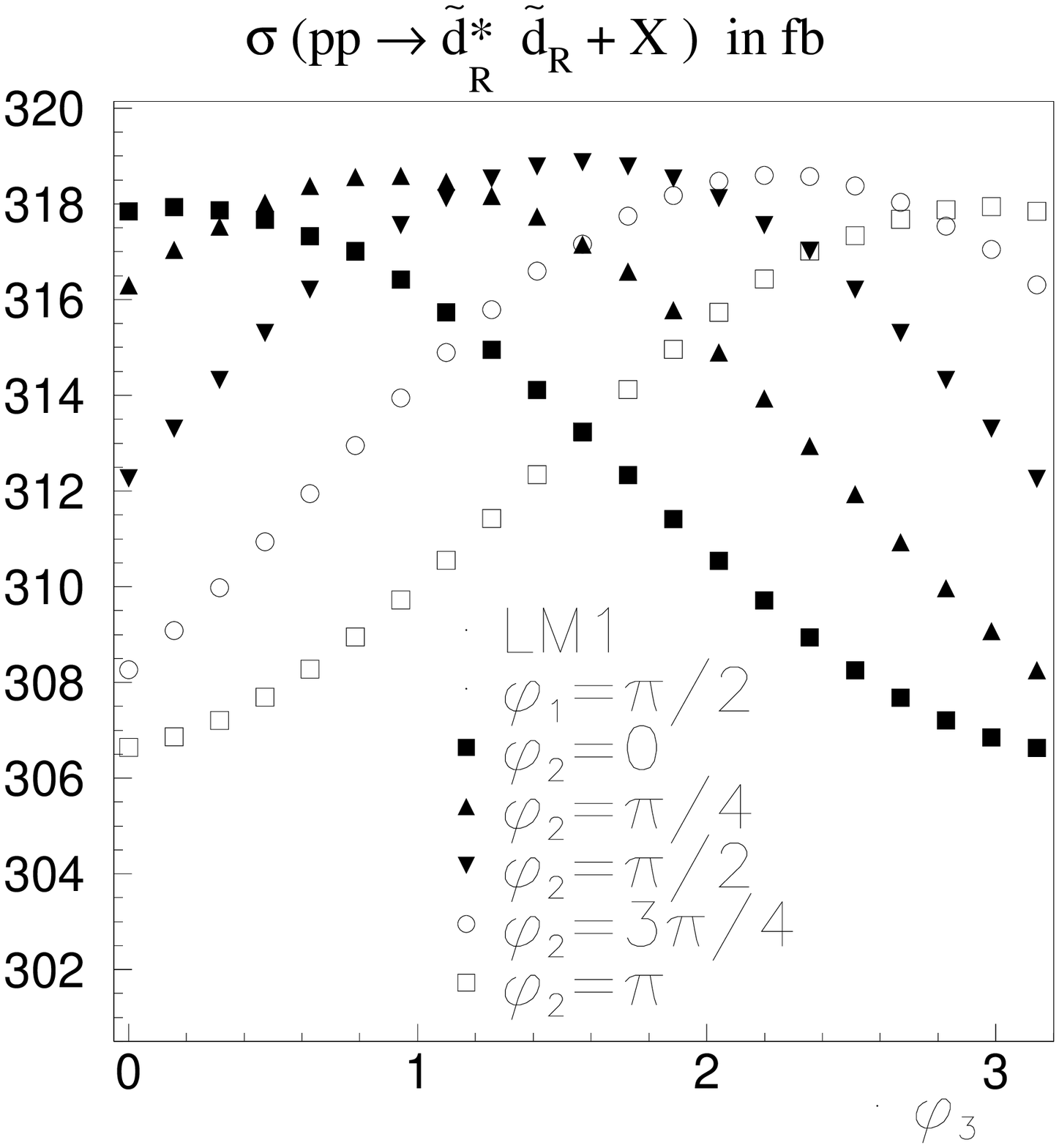}
\includegraphics[width=5cm]{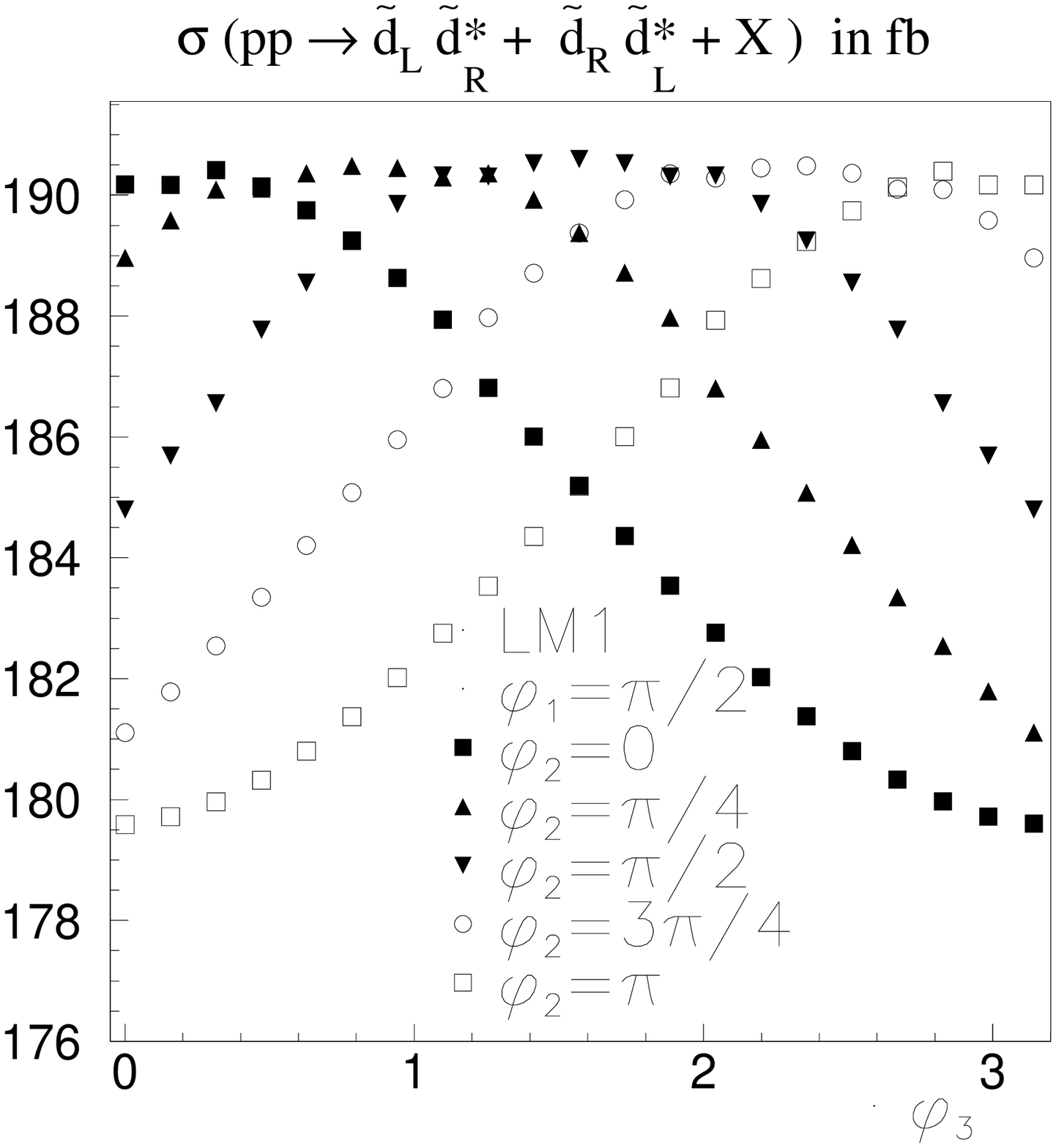}
\includegraphics[width=5cm]{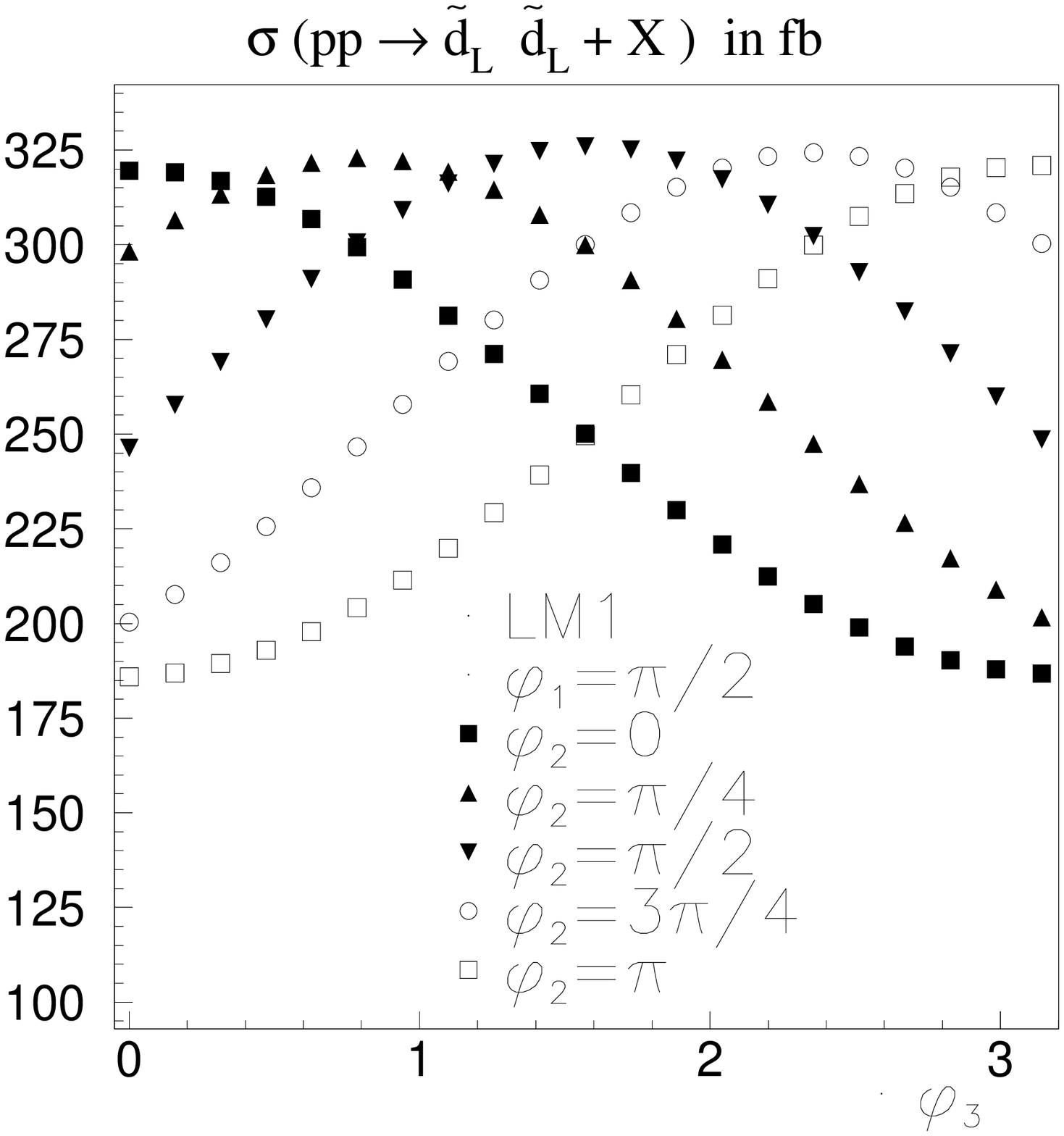}
\includegraphics[width=5cm]{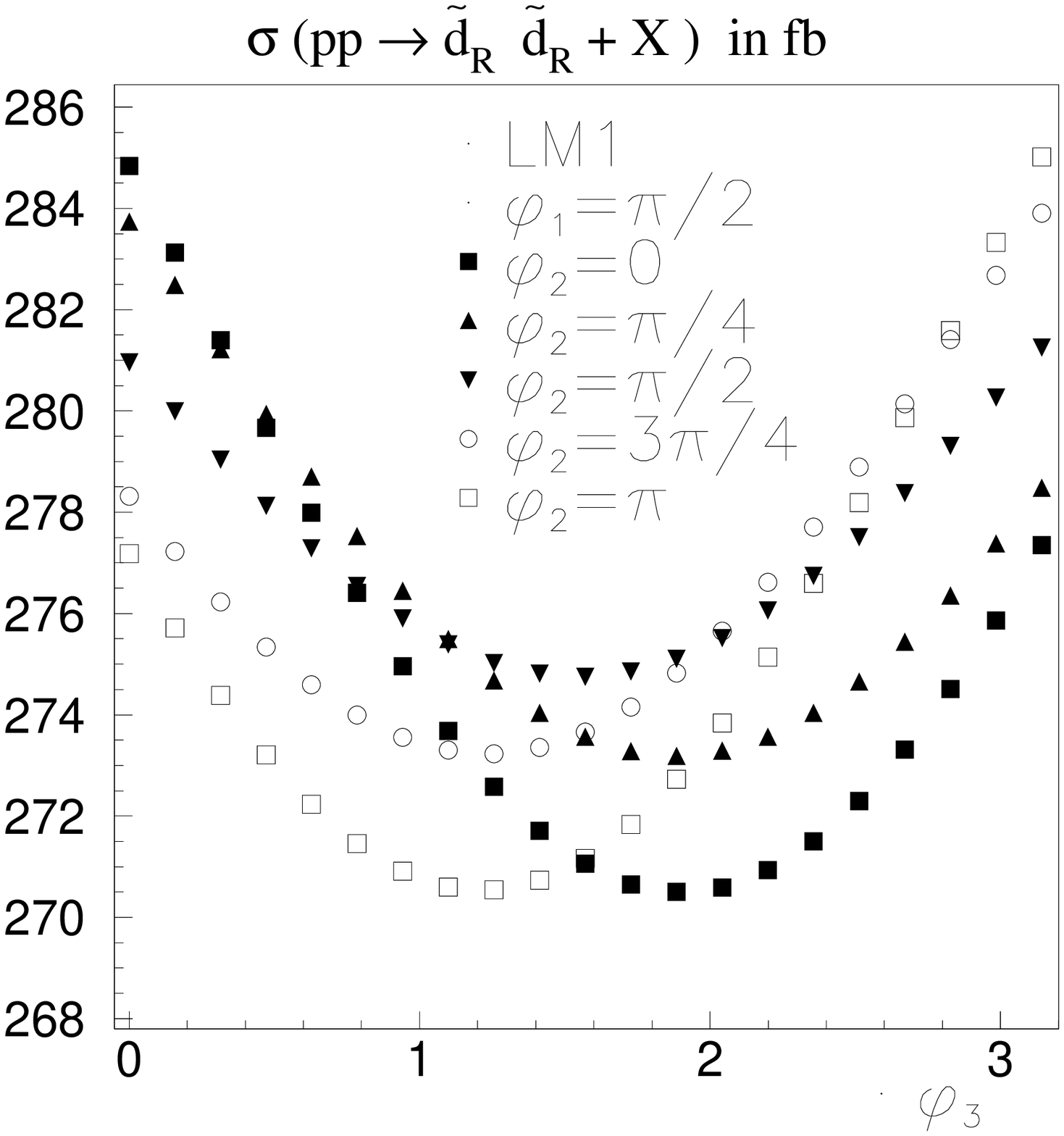}
\includegraphics[width=5cm]{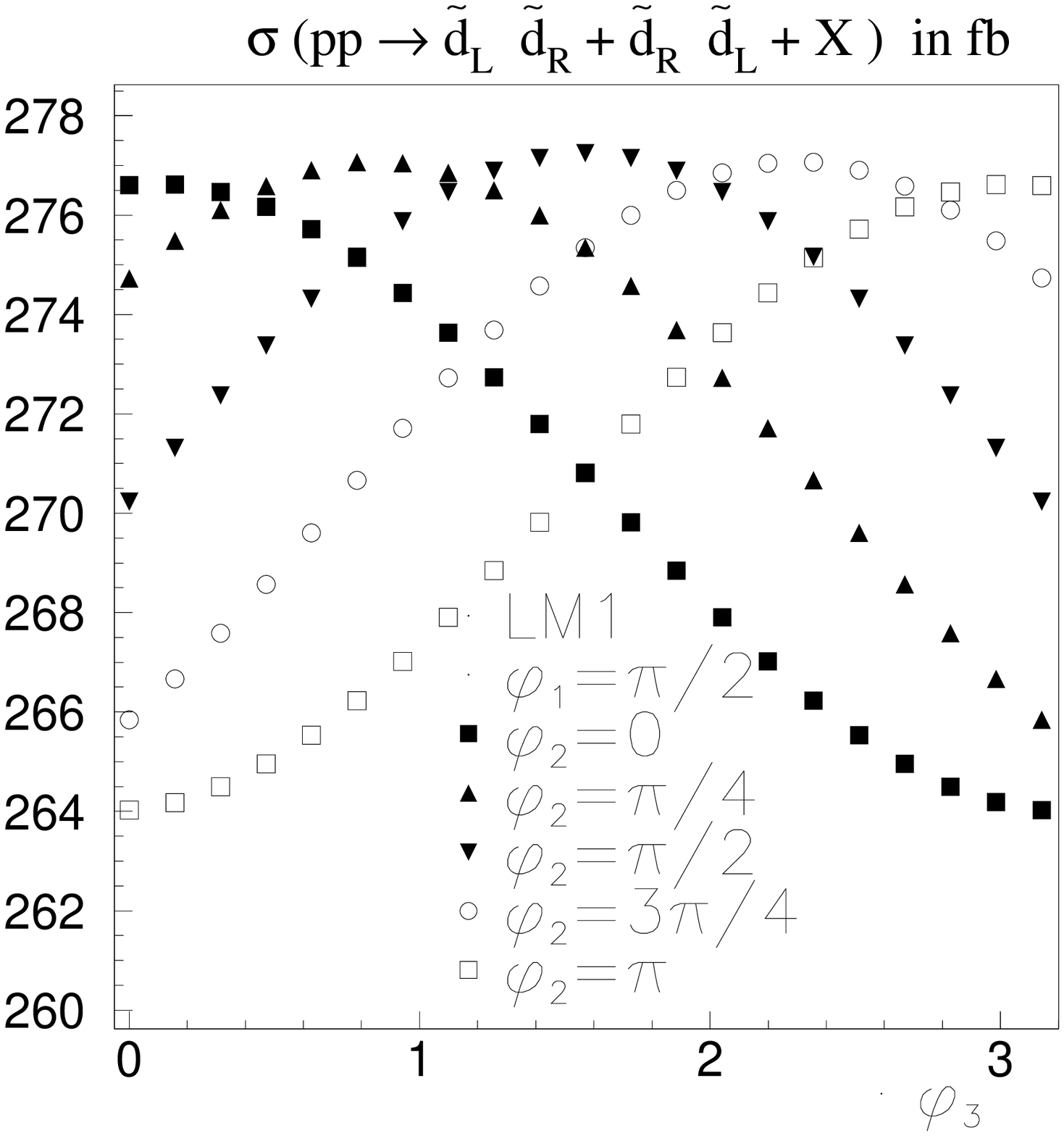}
\caption{ The same as in Fig. \ref{fig-sig-up_pi2}  but for pair
production of down or strange squarks.}\label{fig-sig-down_pi2}
\end{figure}

For the sake of completeness, we plot in Figs.
\ref{fig-sig-down_0} and \ref{fig-sig-down_pi2} the same
production modes for down-type (scalar down or strange) squarks.
Their dependencies on $\varphi_2$ and $\varphi_3$ are similar to
what we found on up-type squark production. Their variation with
$\varphi_1$ is also similar. Clearly, difference between pair
productions of up-type and down-type squarks follow from
differences in squark masses (down-type squarks weigh relatively
heavier than up-type ones after taking into account the $D$-term
effects) and from changes in the couplings $i.e.$ $\left(A_{u_L
i}, A_{u_R i}\right) \rightarrow \left(A_{d_L i}, A_{d_R
i}\right)$ and $\left(C_{u LL}, C_{u RR}\right) \rightarrow
\left(C_{d LL}, C_{d RR}\right)$. For the LM1 point under concern,
down-type squark pair-production turns out to be significantly
smaller than that of the up-type squarks. This is eventually tied
up to difference between the squark masses and to various
couplings.

\begin{figure}
\includegraphics[width=5cm]{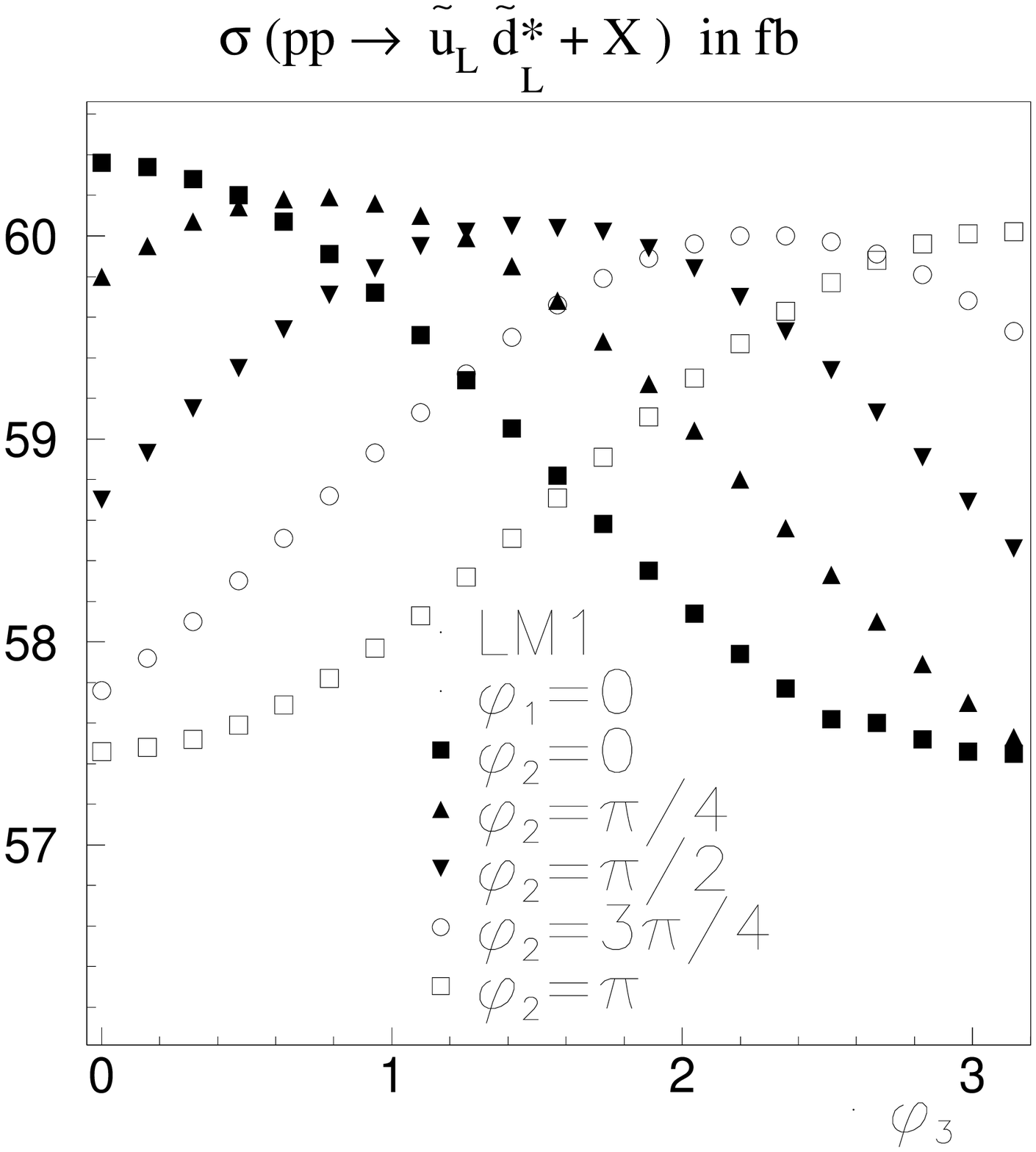}
\includegraphics[width=5cm]{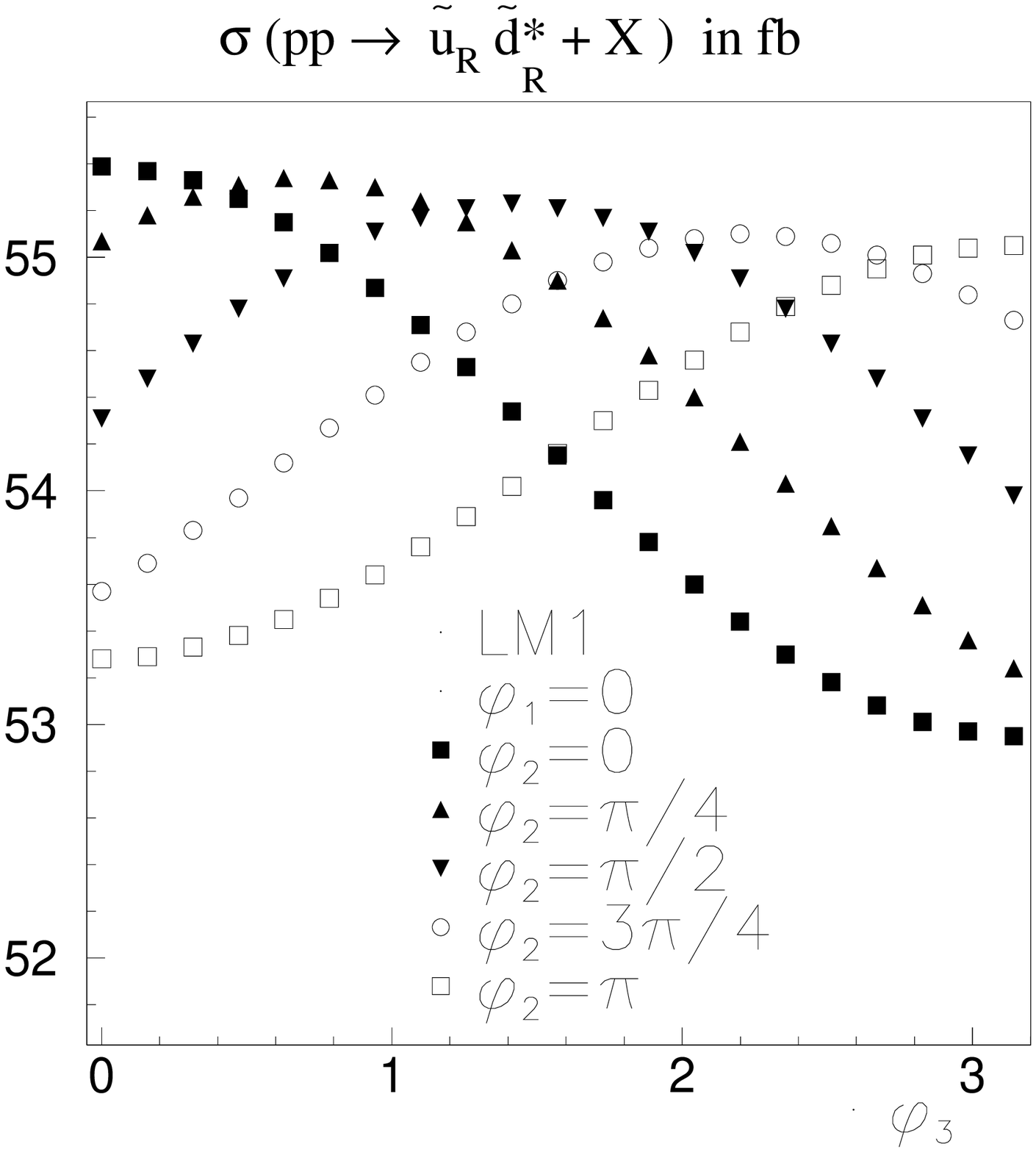}
\includegraphics[width=5cm]{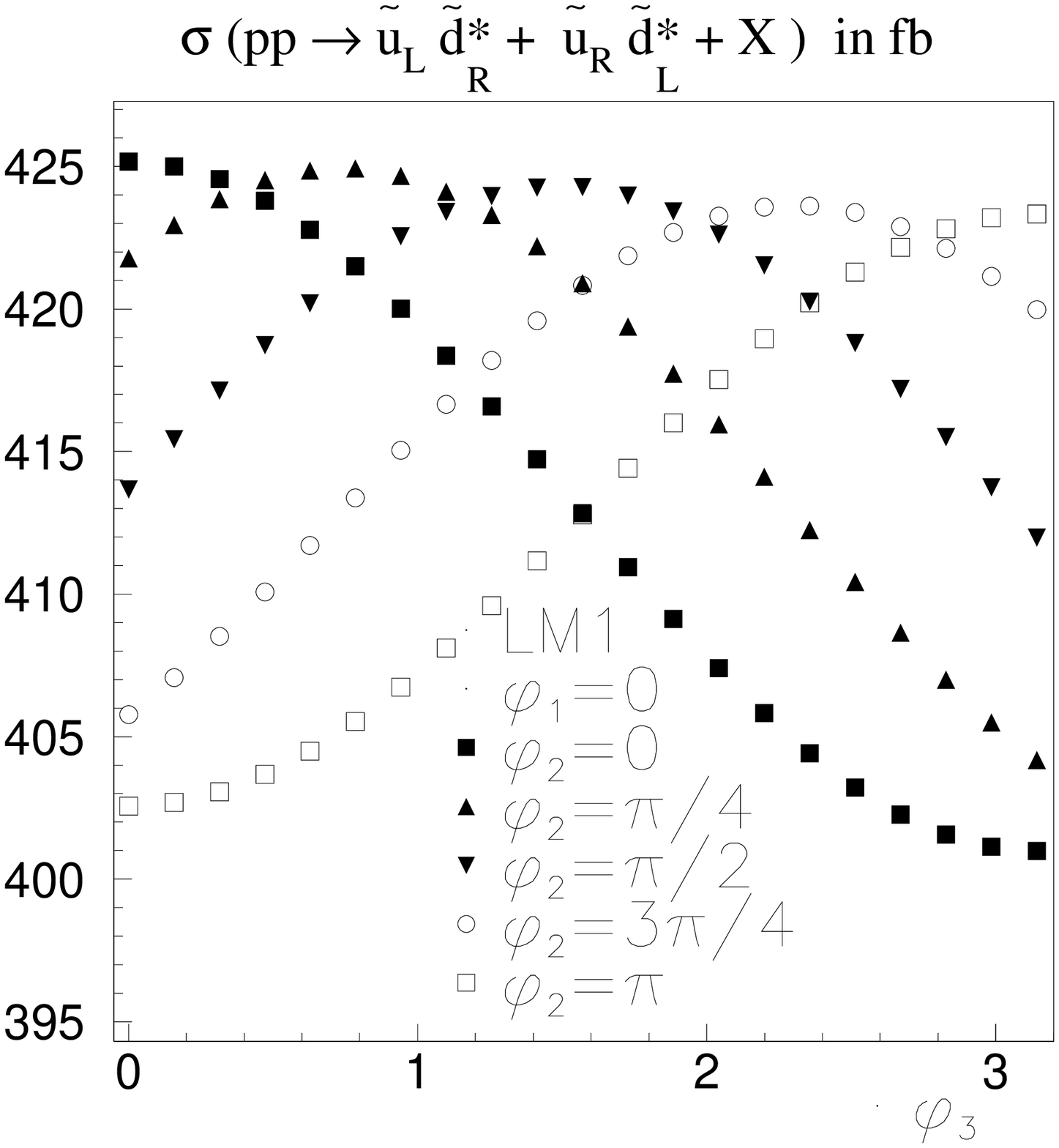}
\includegraphics[width=5cm]{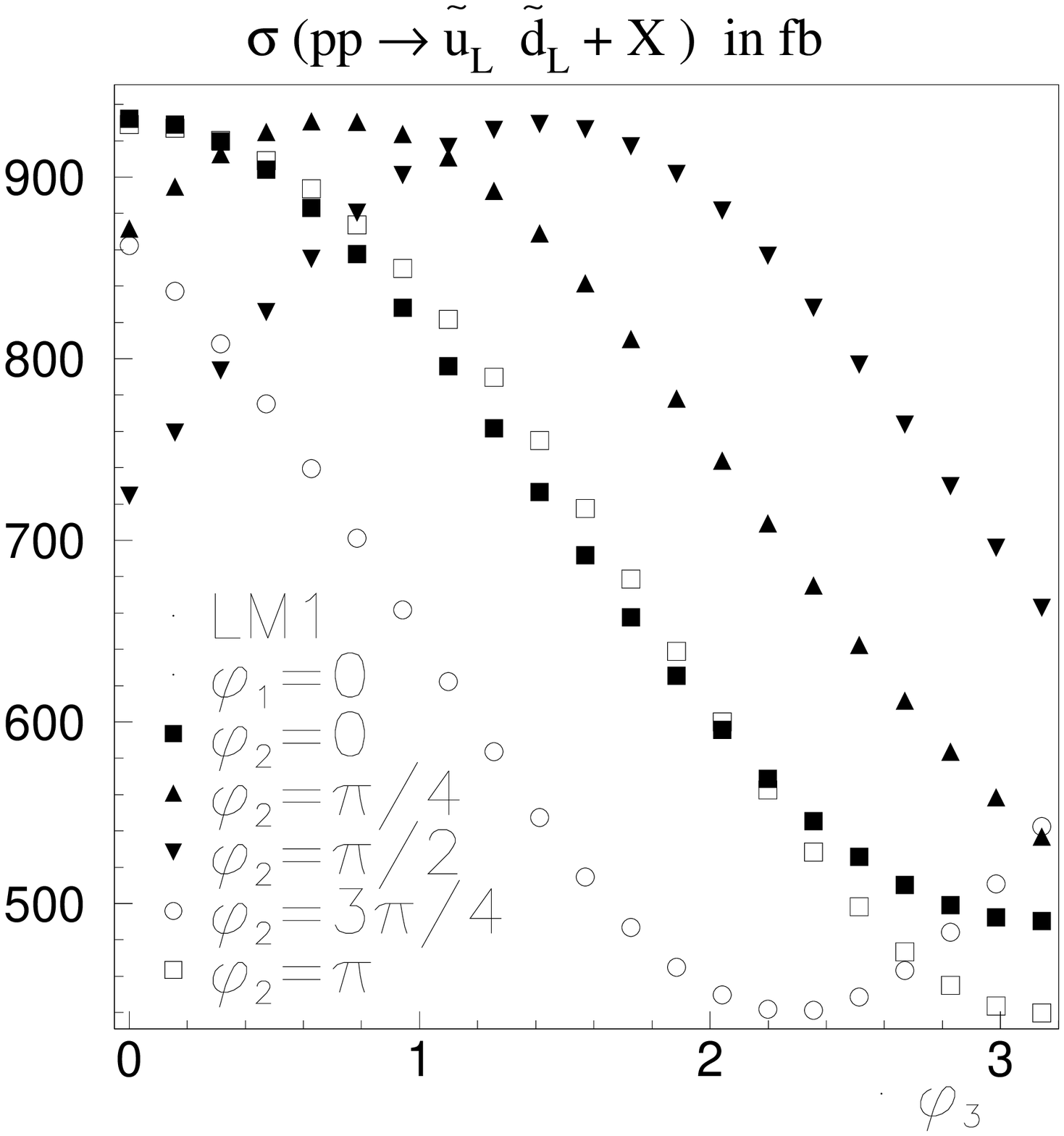}
\includegraphics[width=5cm]{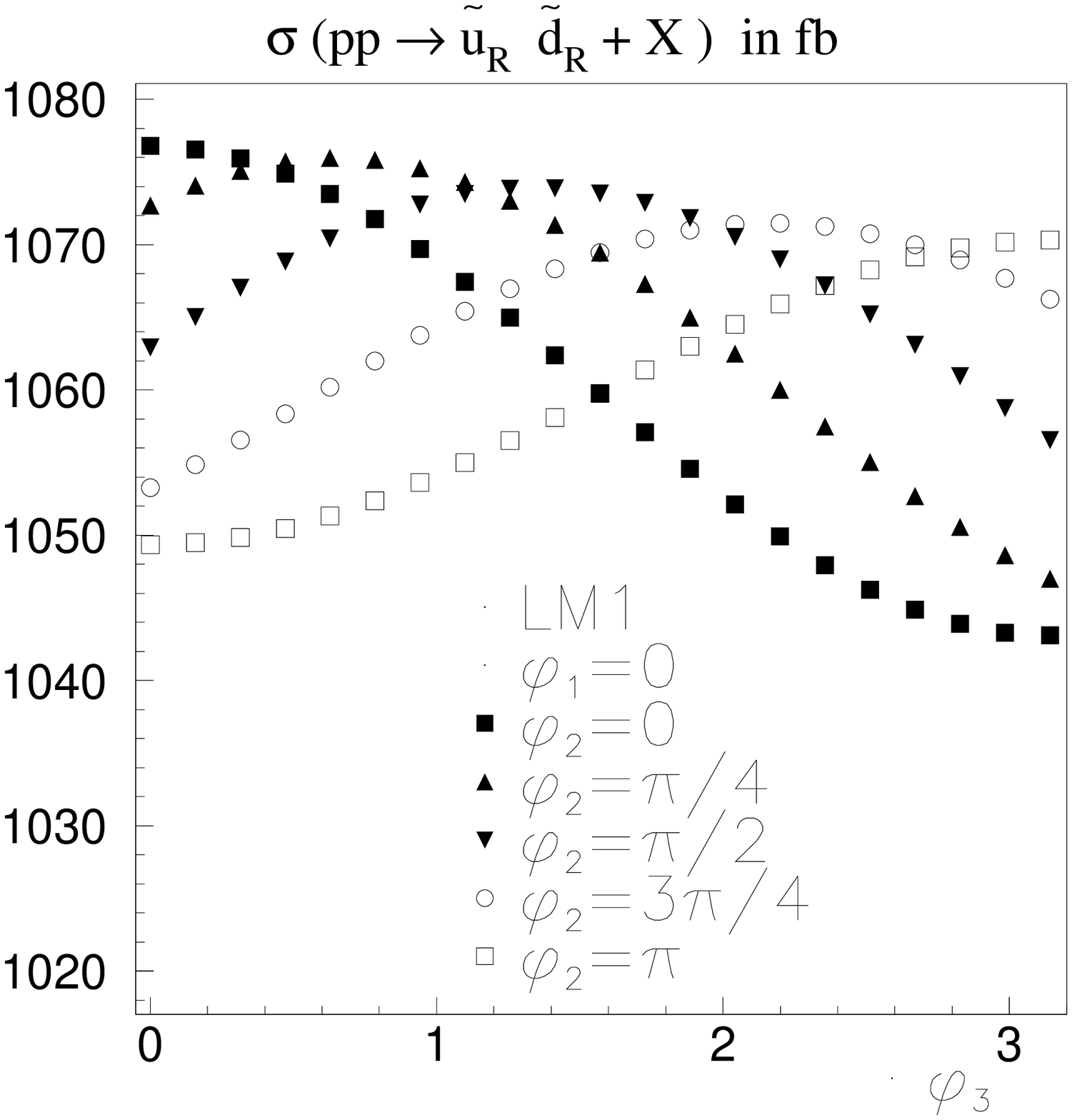}
\includegraphics[width=5cm]{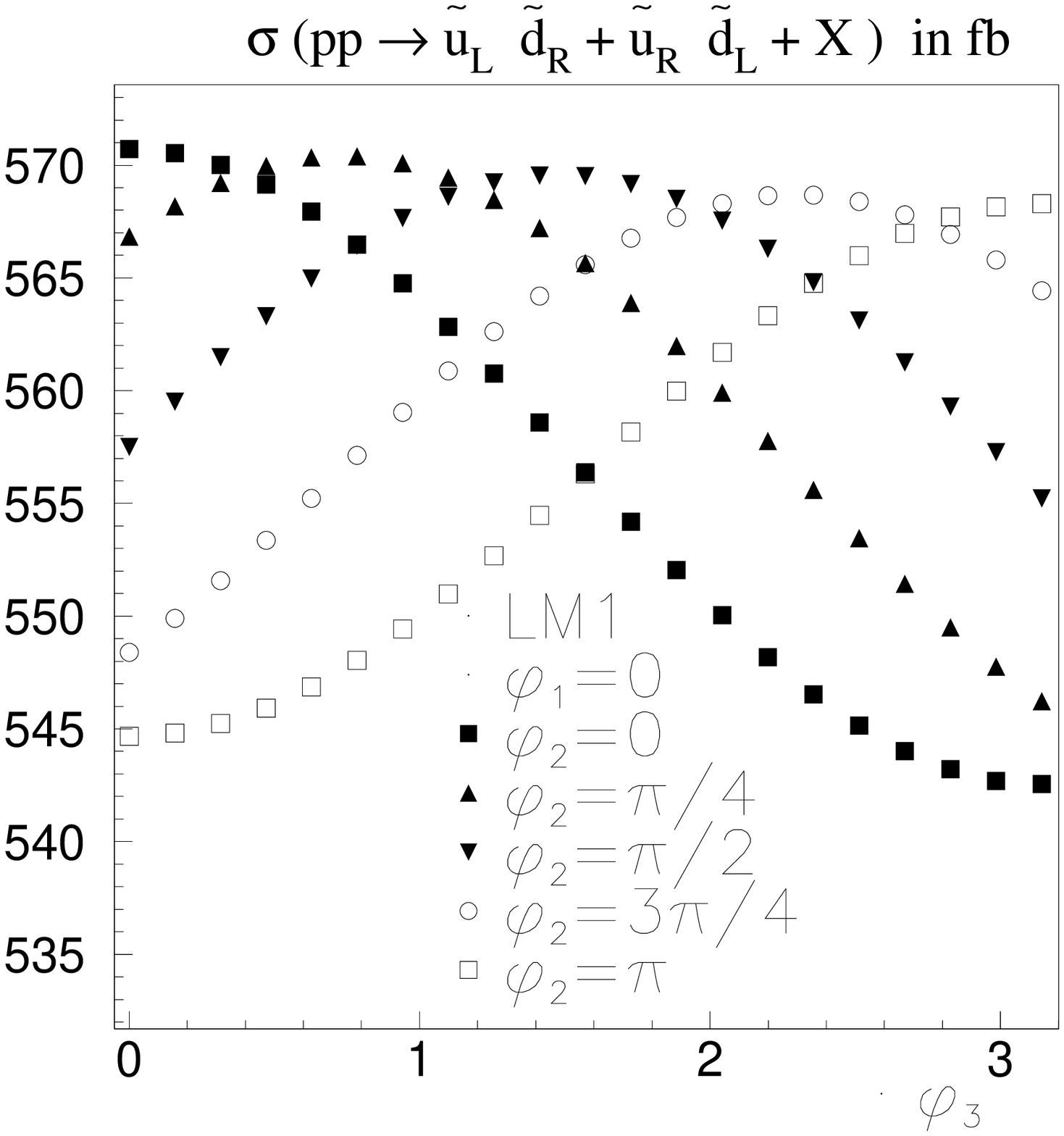}
\caption{Up-down squark pair-production cross sections (in fb) at the
LHC as functions of $\varphi_3$ for $\varphi_1=0$ and several
values of $\varphi_2$. Left: $\sigma(p\, p \rightarrow
\widetilde{u}_L \widetilde{d}_{L}^{\star})$ (top panel) and
$\sigma(p\, p \rightarrow \widetilde{u}_L \widetilde{d}_{L})$
(bottom panel). Middle: $\sigma(p\, p \rightarrow \widetilde{u}_R
\widetilde{d}_{R}^{\star})$ (top panel) and $\sigma(p\, p
\rightarrow \widetilde{u}_R \widetilde{d}_{R})$ (bottom panel).
Right: $\sigma(p\, p \rightarrow \widetilde{u}_L
\widetilde{d}_{R}^{\star})$ (top panel) and $\sigma(p\, p
\rightarrow \widetilde{u}_L \widetilde{d}_{R})$ (bottom
panel).}\label{fig-sig-updown_0}
\end{figure}

\begin{figure}
\includegraphics[width=5cm]{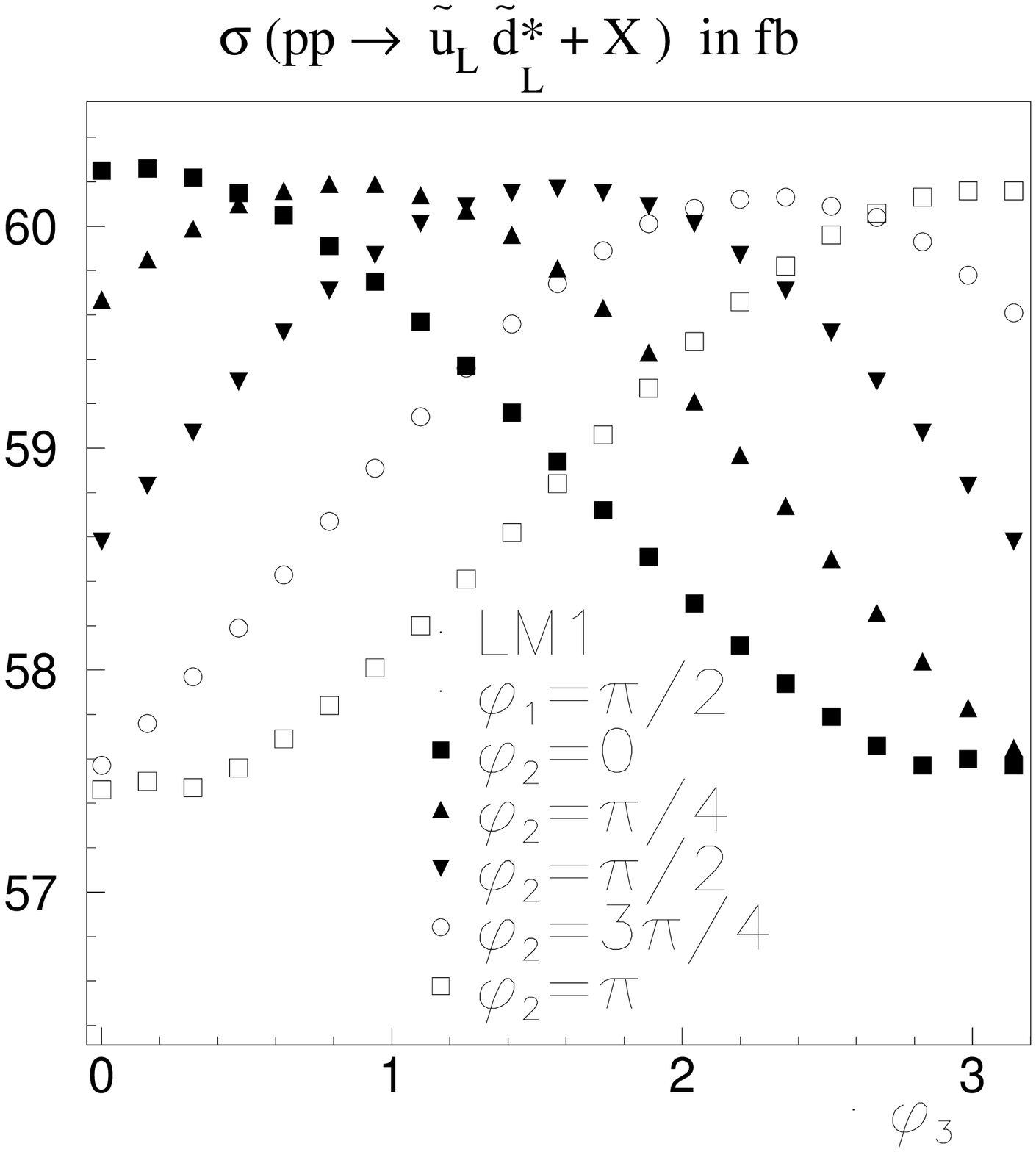}
\includegraphics[width=5cm]{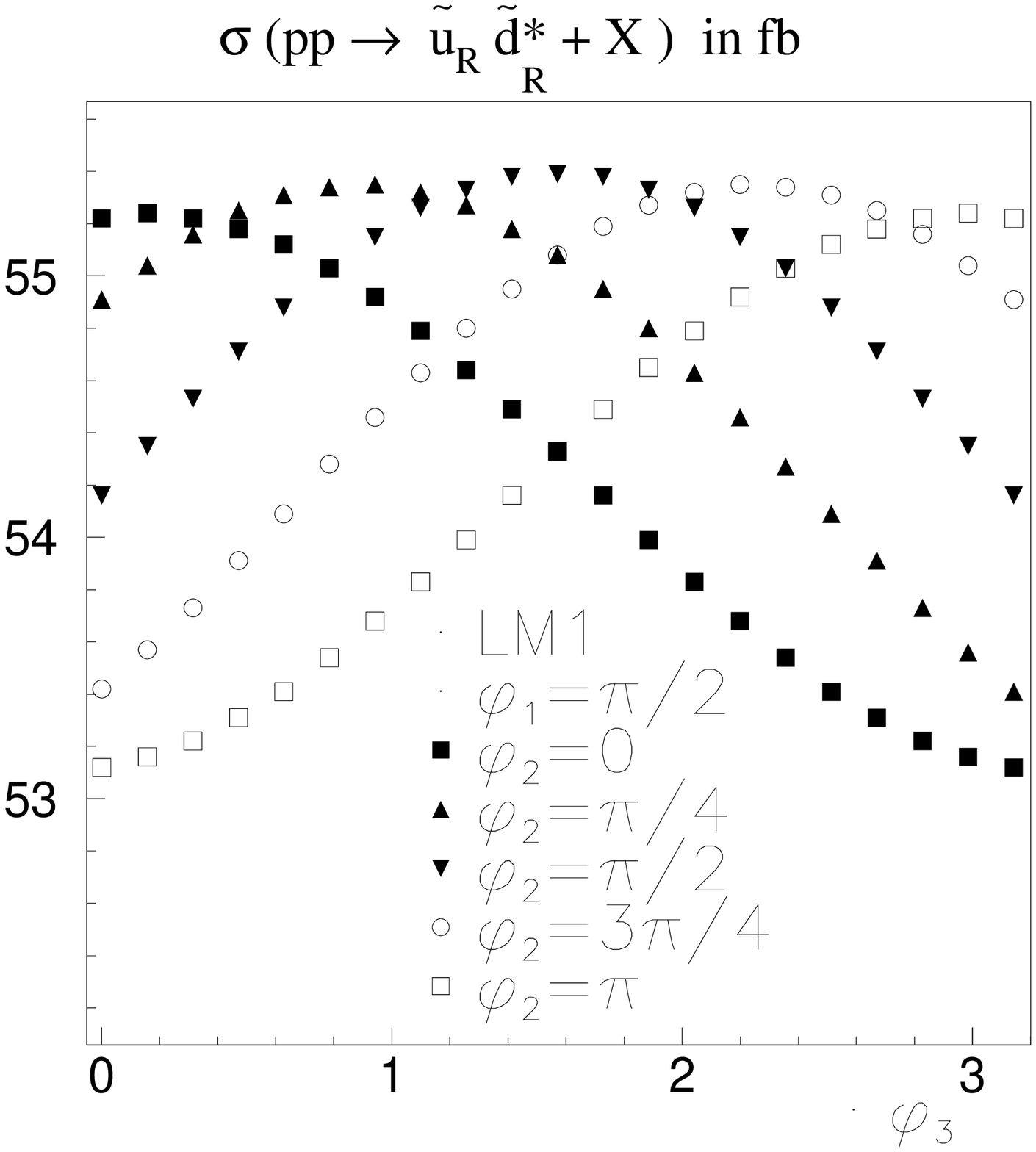}
\includegraphics[width=5cm]{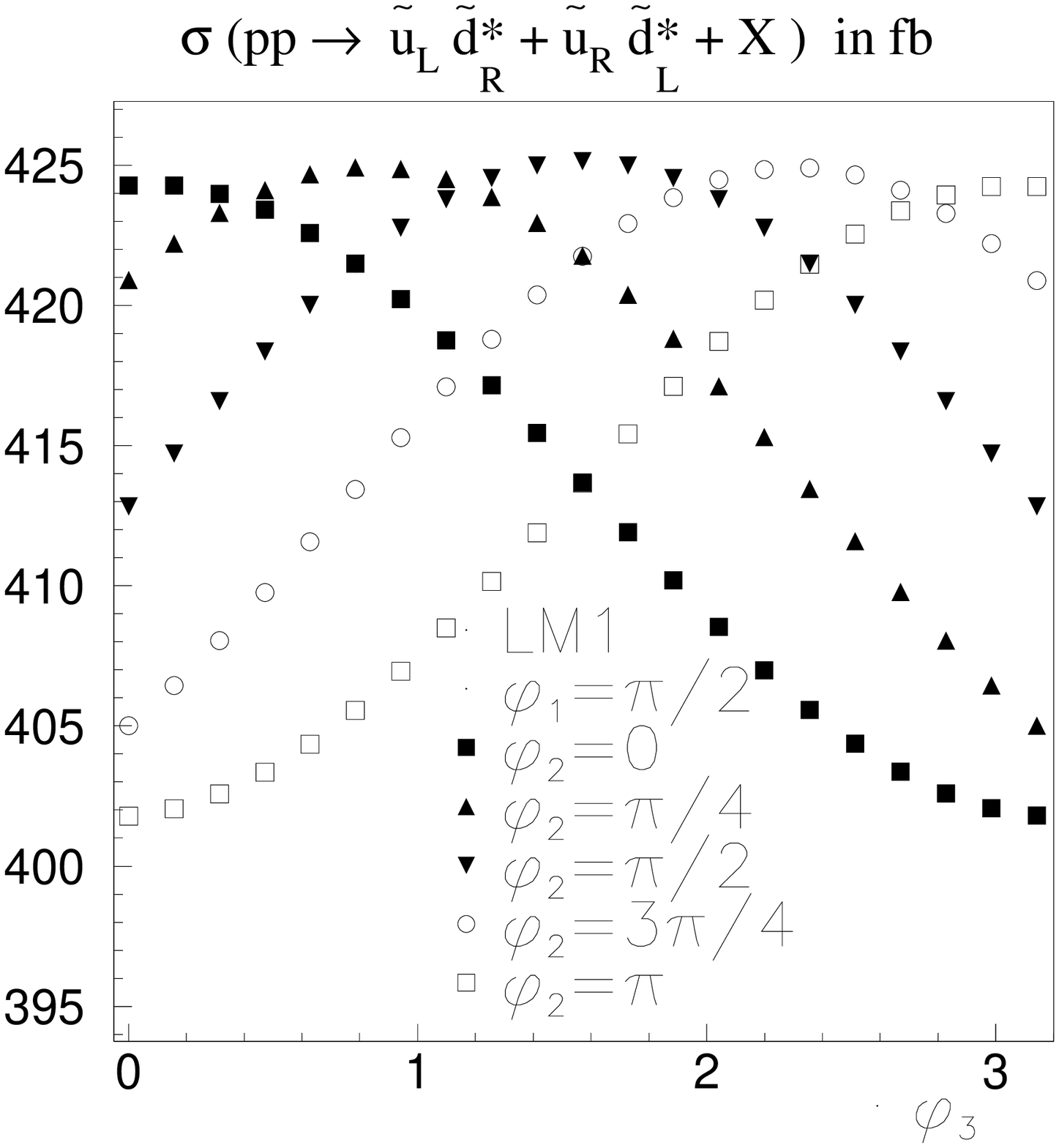}
\includegraphics[width=5cm]{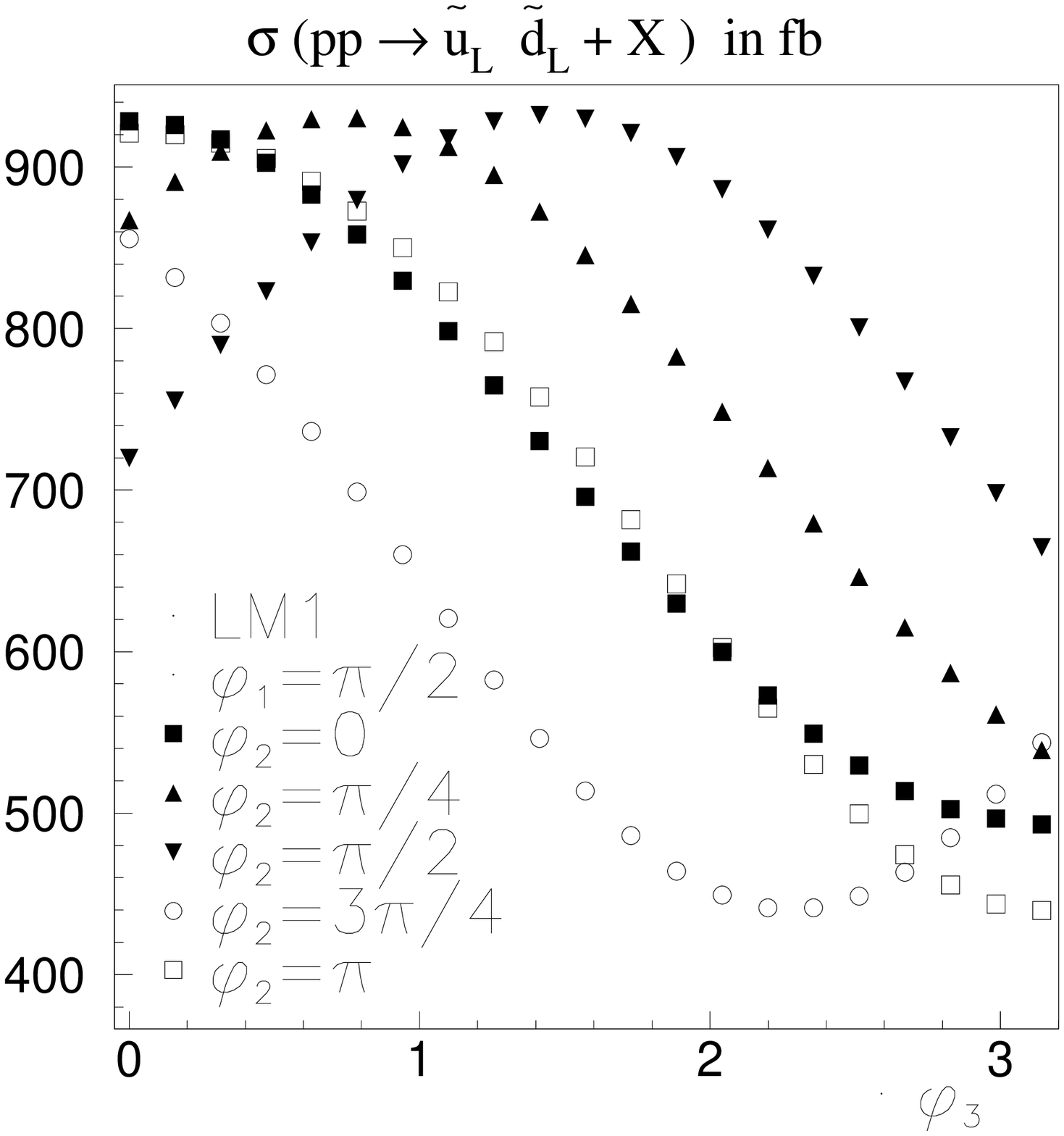}
\includegraphics[width=5cm]{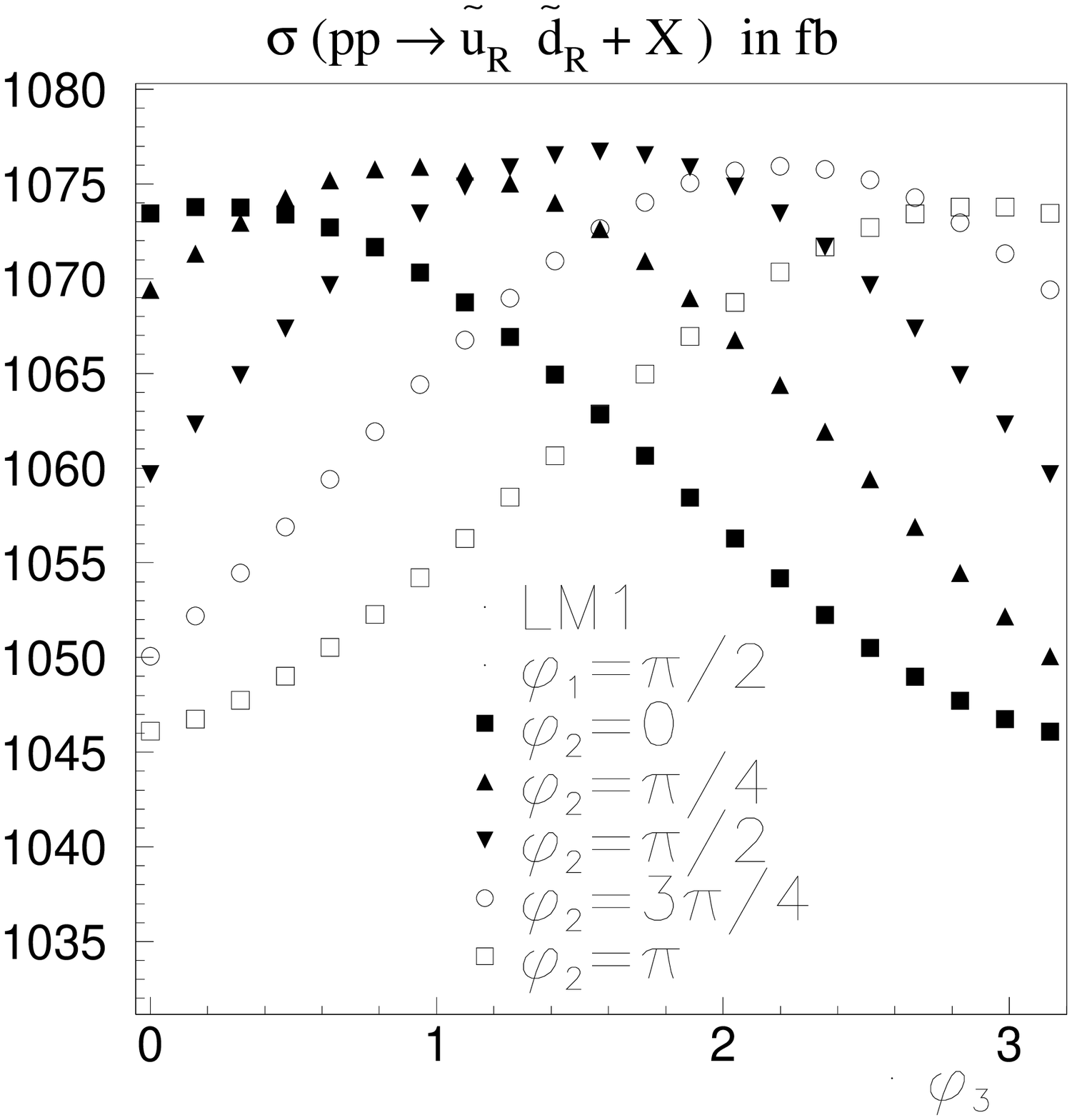}
\includegraphics[width=5cm]{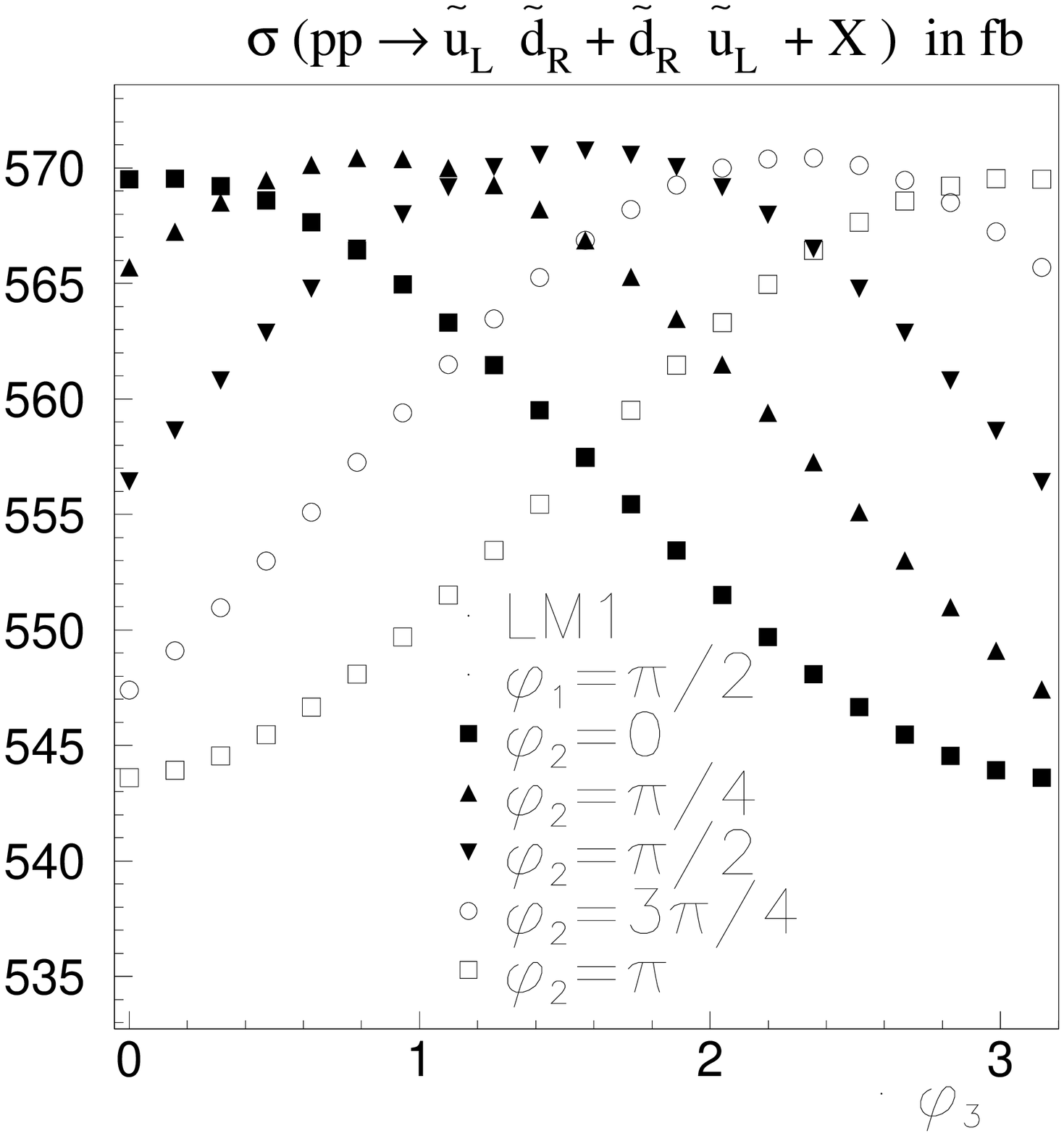}
\caption{The same as in Fig. \ref{fig-sig-updown_0} but for
$\varphi_1=\pi/2$.}\label{fig-sig-updown_2}
\end{figure}

Depicted in Figs. \ref{fig-sig-updown_0} and
\ref{fig-sig-updown_2} are cross sections for associated
production of up-type and down-type quarks for $\varphi_1 =0$ and
$\varphi_1= \pi/2$, respectively. Typically, cross sections are
seen to exhibit significant variations with $\varphi_2$. This is
not surprising at all: chargino sector plays an essential role in
these production modes. This is also confirmed by the manifest
$\varphi_1$ independence of the cross sections. There is one
exception here: $\sigma(p\, p \rightarrow \widetilde{u}_R
\widetilde{d}_{R})$ which exhibits a relatively stronger
dependence on $\varphi_1$ because of the fact that $ p\, p
\rightarrow \widetilde{u}_R \widetilde{d}_{R}$ proceeds with
gluino and bino mediations, only. The largest swing occurs in
$\sigma(p\, p \rightarrow \widetilde{u}_L \widetilde{d}_{L})$
which changes by $\sim 500\ {\rm fb}$ as $\varphi_3$ changes from
$0$ to $\pi$.


\begin{figure}
\includegraphics[width=5cm]{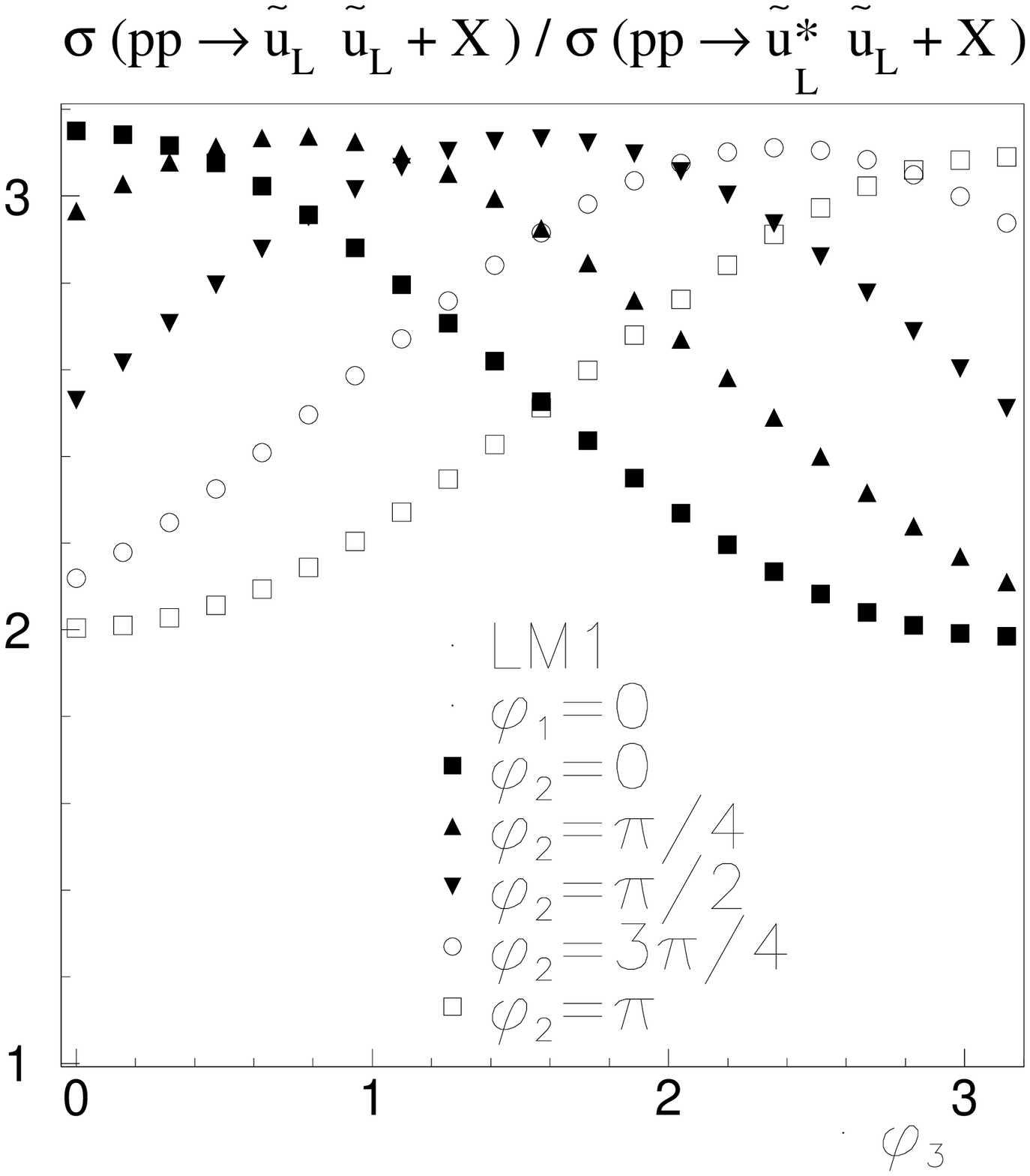}
\includegraphics[width=5cm]{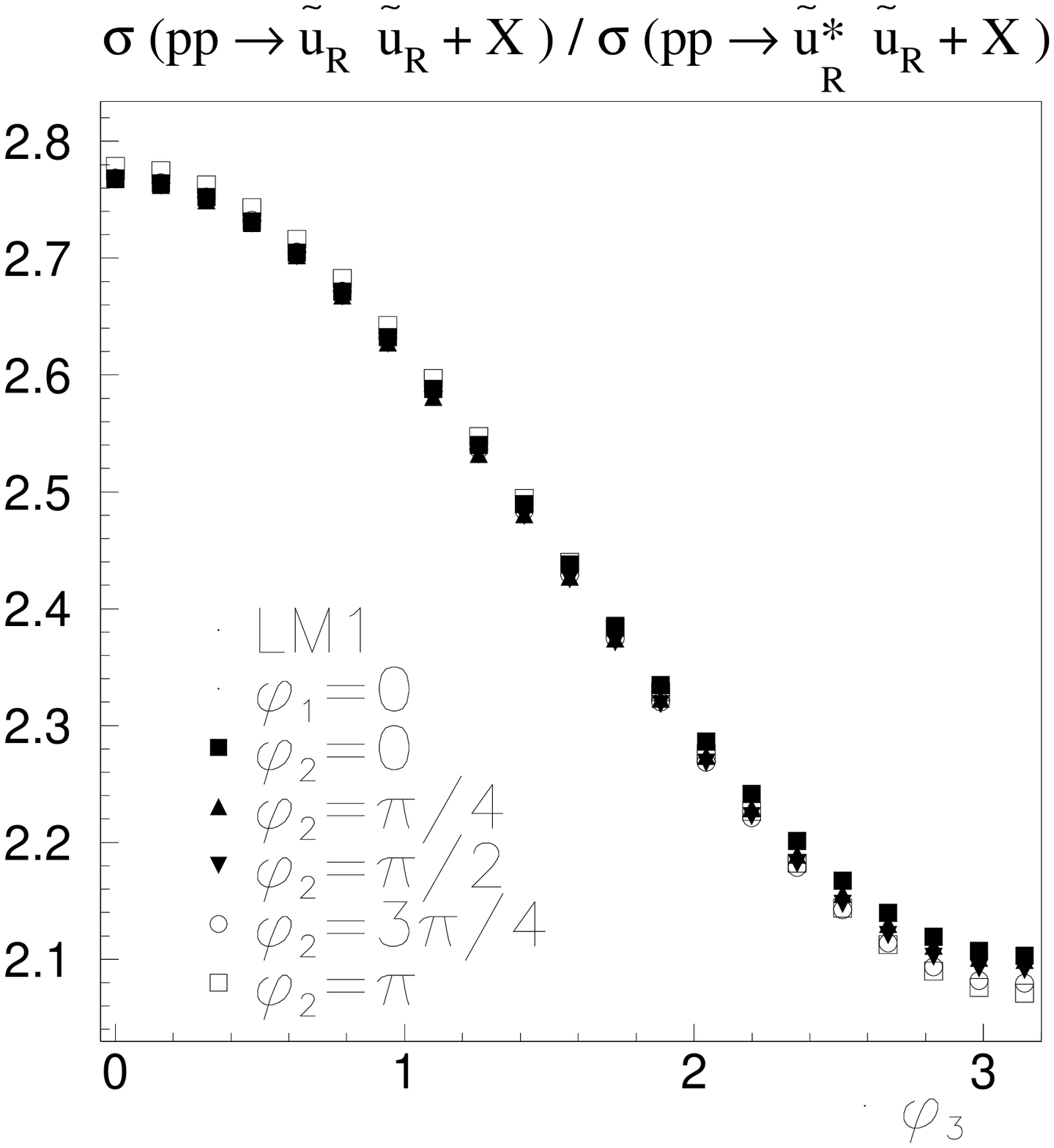}
\includegraphics[width=5cm]{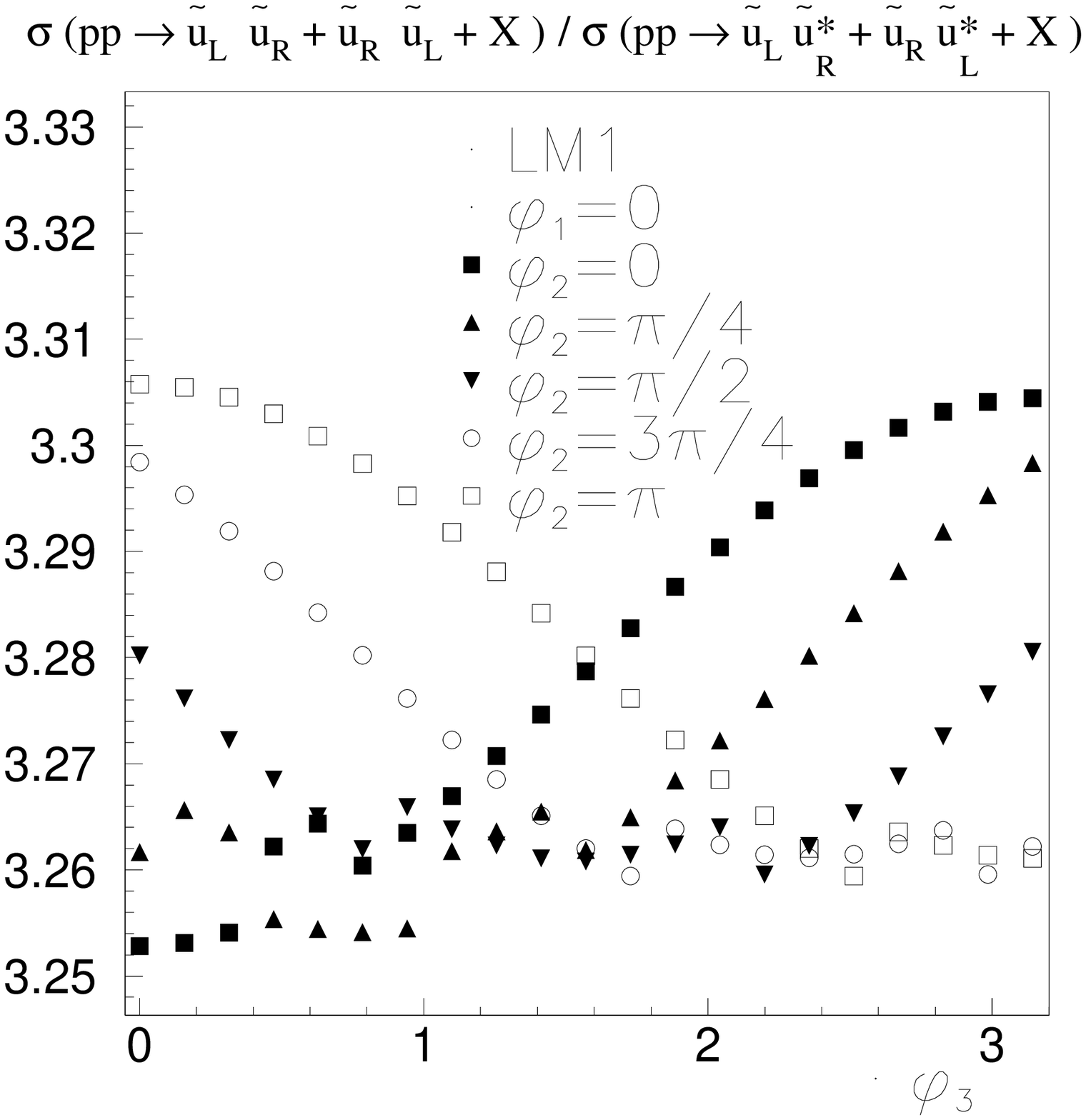}
\includegraphics[width=5cm]{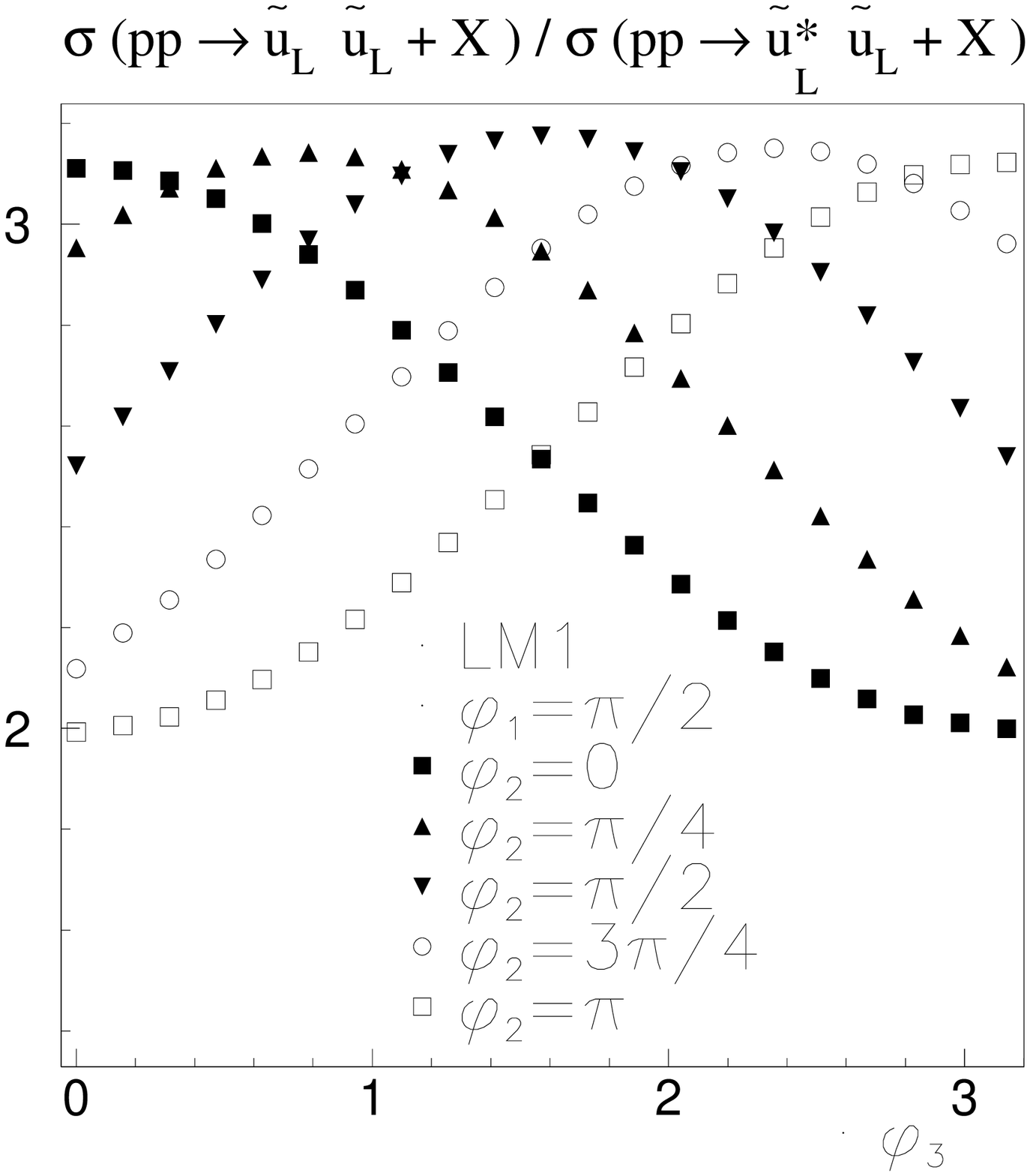}
\includegraphics[width=5cm]{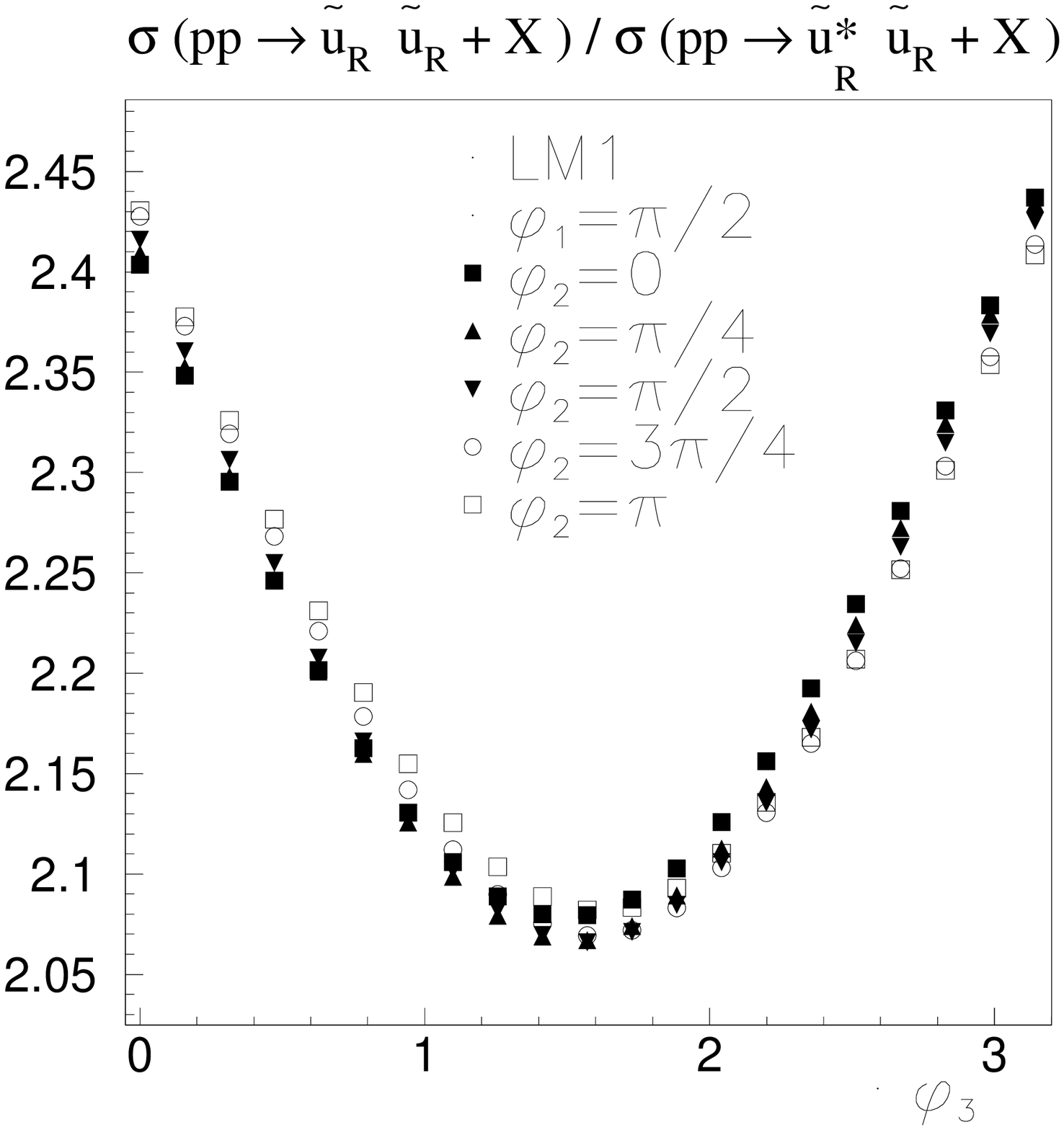}
\includegraphics[width=5cm]{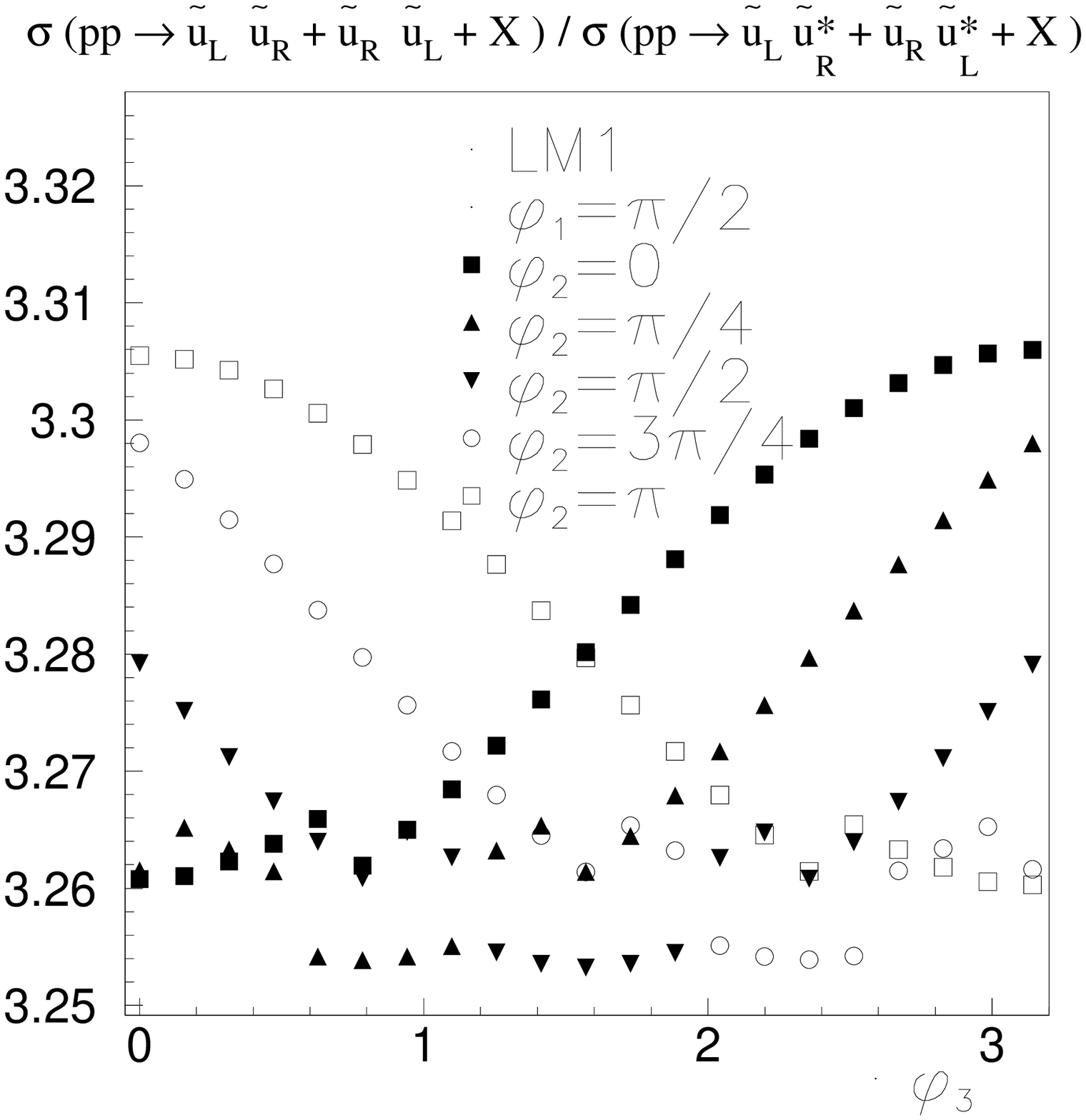}

\caption{ Ratios of up squark pair-production cross sections  at the
LHC as functions of $\varphi_3$ for $\varphi_1=0$ (top panel), $\varphi_1=\pi/2$ (bottom panel) and several values of $\varphi_2$. Left: $\sigma(p\, p \rightarrow \widetilde{u}_L \widetilde{u}_{L})$ / $\sigma(p\, p \rightarrow
\widetilde{u}_L \widetilde{u}_{L}^{\star})$  for $\varphi_1=0$ (top panel) and
 $\sigma(p\, p \rightarrow \widetilde{u}_L \widetilde{u}_{L})$ / $\sigma(p\, p \rightarrow
\widetilde{u}_L \widetilde{u}_{L}^{\star})$ for $\varphi_1=\pi/2$ (bottom panel). Middle:  $\sigma(p\, p \rightarrow \widetilde{u}_R \widetilde{u}_{R})$ / $\sigma(p\, p \rightarrow
\widetilde{u}_R \widetilde{u}_{R}^{\star})$  for $\varphi_1=0$ (top panel) and
 $\sigma(p\, p \rightarrow \widetilde{u}_R \widetilde{u}_{R})$ / $\sigma(p\, p \rightarrow
\widetilde{u}_R \widetilde{u}_{R}^{\star})$  for $\varphi_1=\pi/2$ (bottom panel). Right: $\sigma(p\, p \rightarrow \widetilde{u}_L \widetilde{u}_{R})$ / $\sigma(p\, p \rightarrow
\widetilde{u}_L \widetilde{u}_{R}^{\star})$  for $\varphi_1=0$ (top panel) and
 $\sigma(p\, p \rightarrow \widetilde{u}_L \widetilde{u}_{R})$ / $\sigma(p\, p \rightarrow
\widetilde{u}_L \widetilde{u}_{R}^{\star})$  for $\varphi_1=\pi/2$ (bottom panel).}\label{fig-upratio}
\end{figure}

\begin{figure}
\includegraphics[width=5cm]{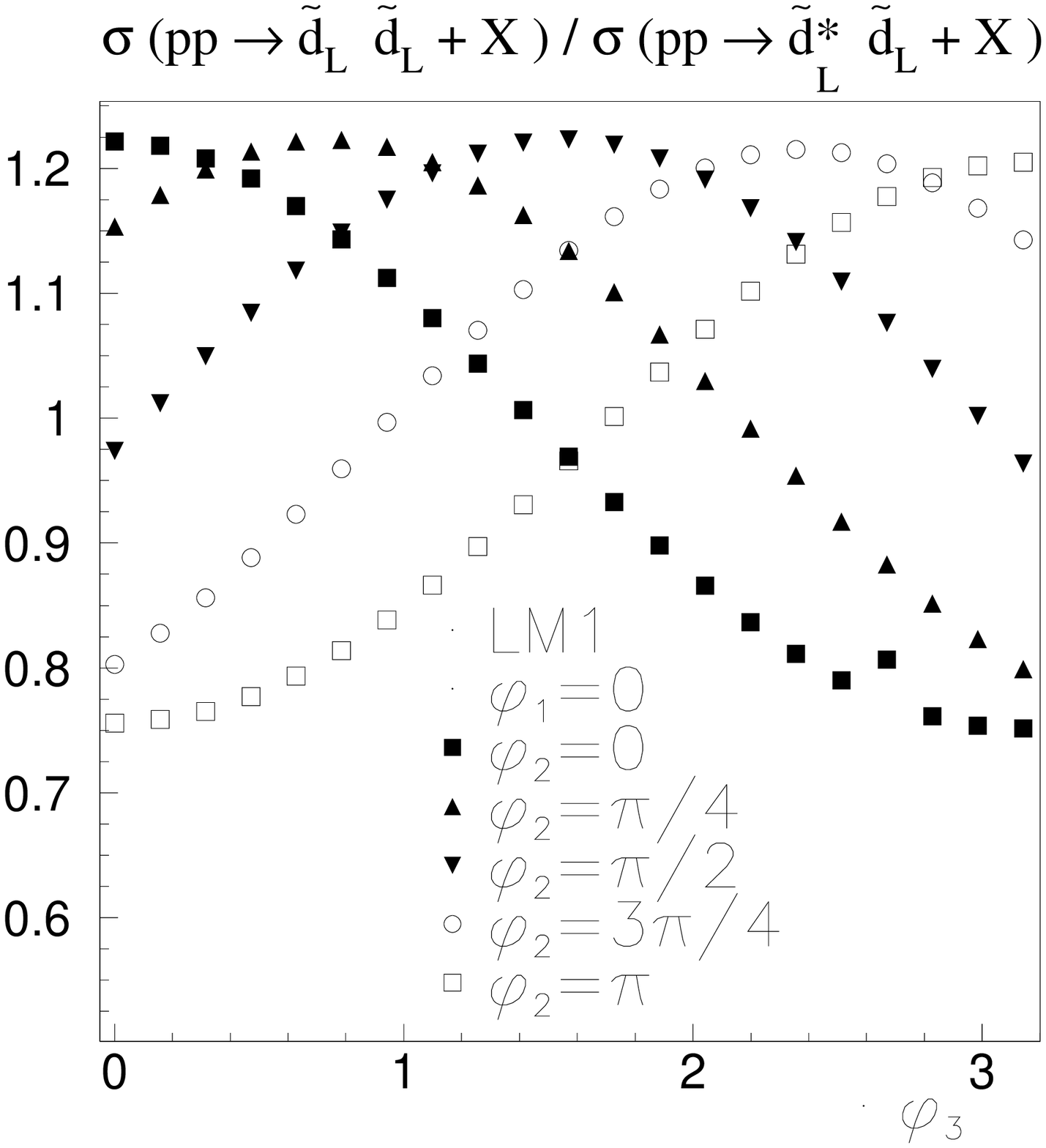}
\includegraphics[width=5cm]{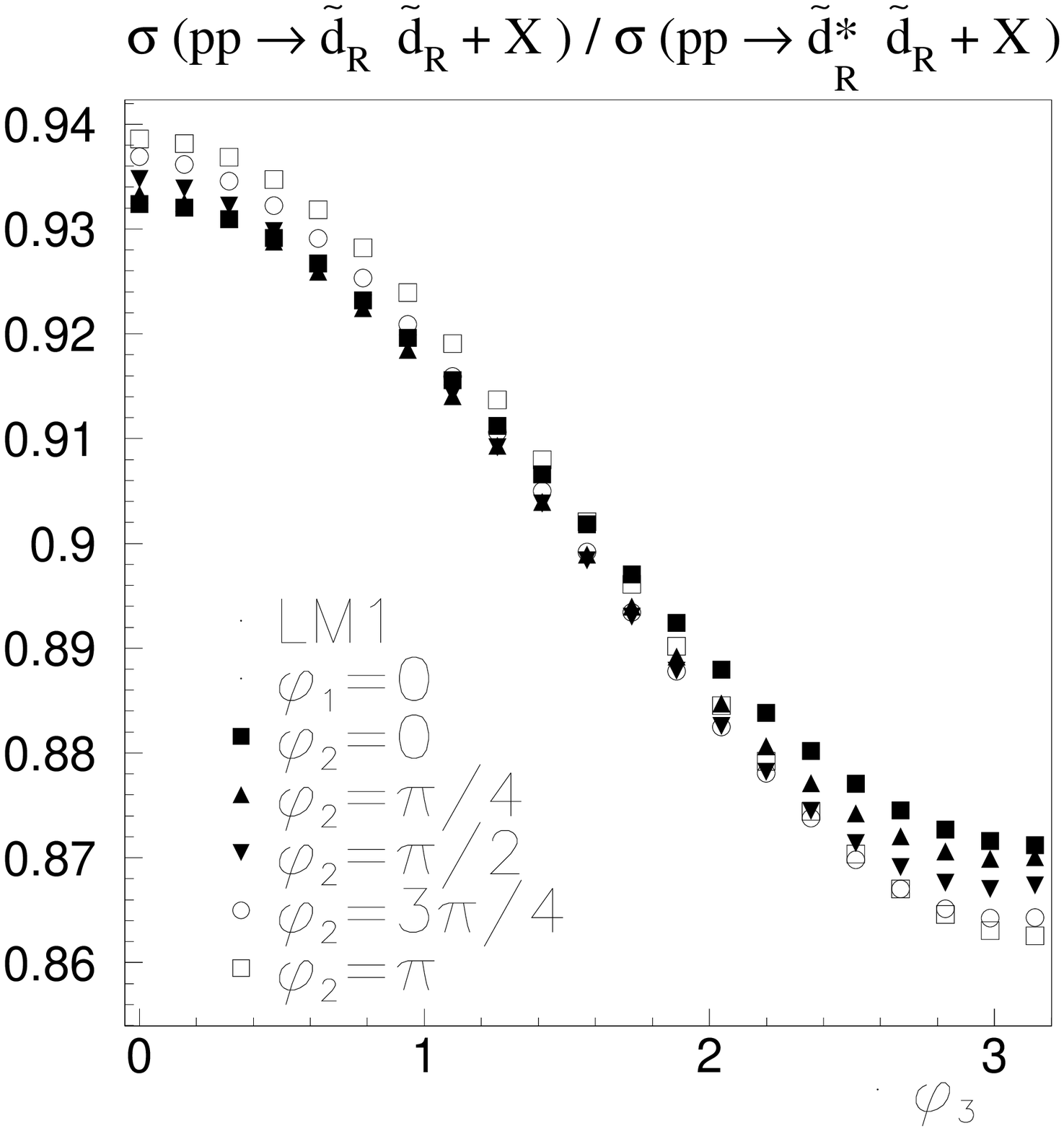}
\includegraphics[width=5cm]{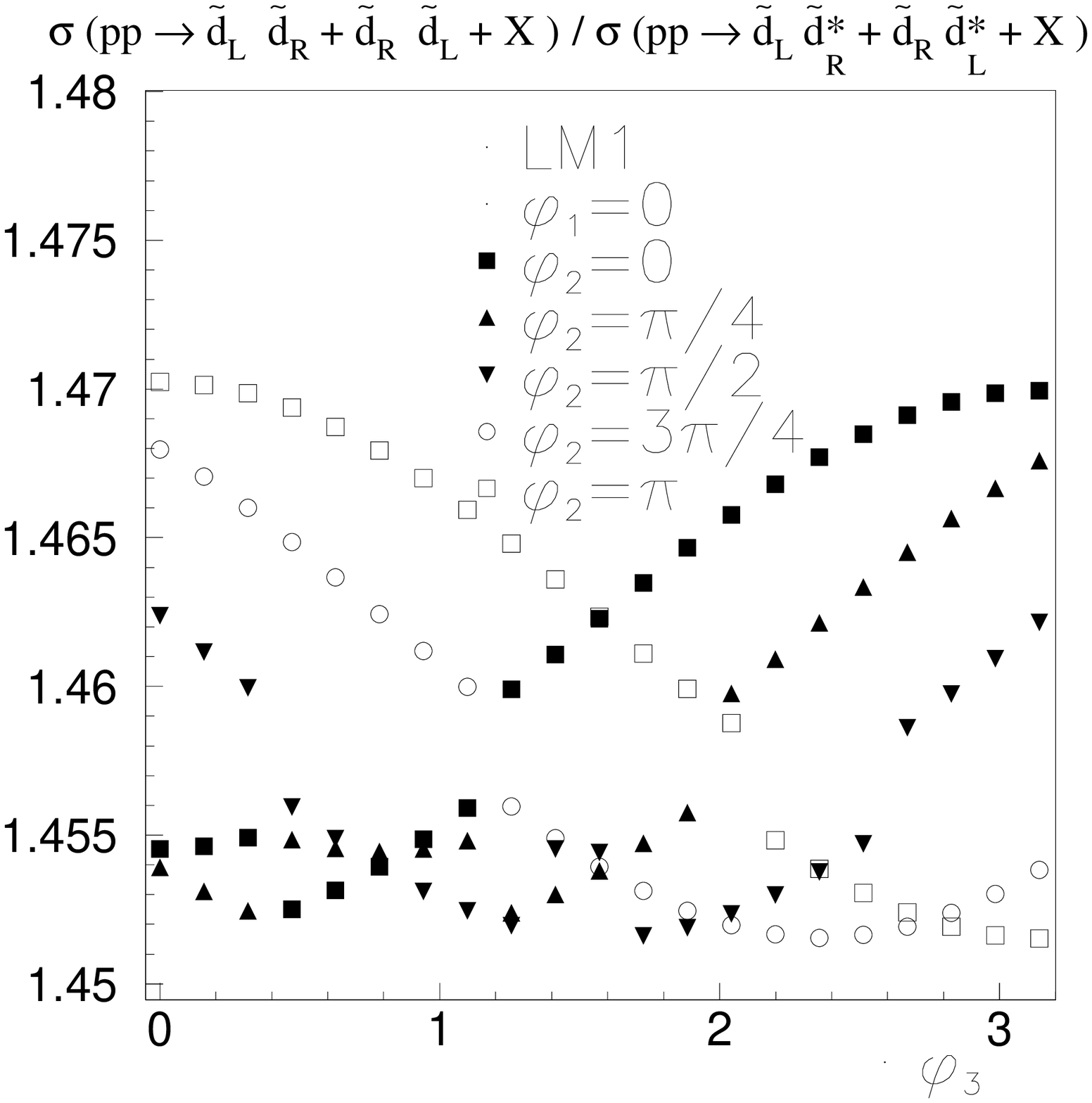}
\includegraphics[width=5cm]{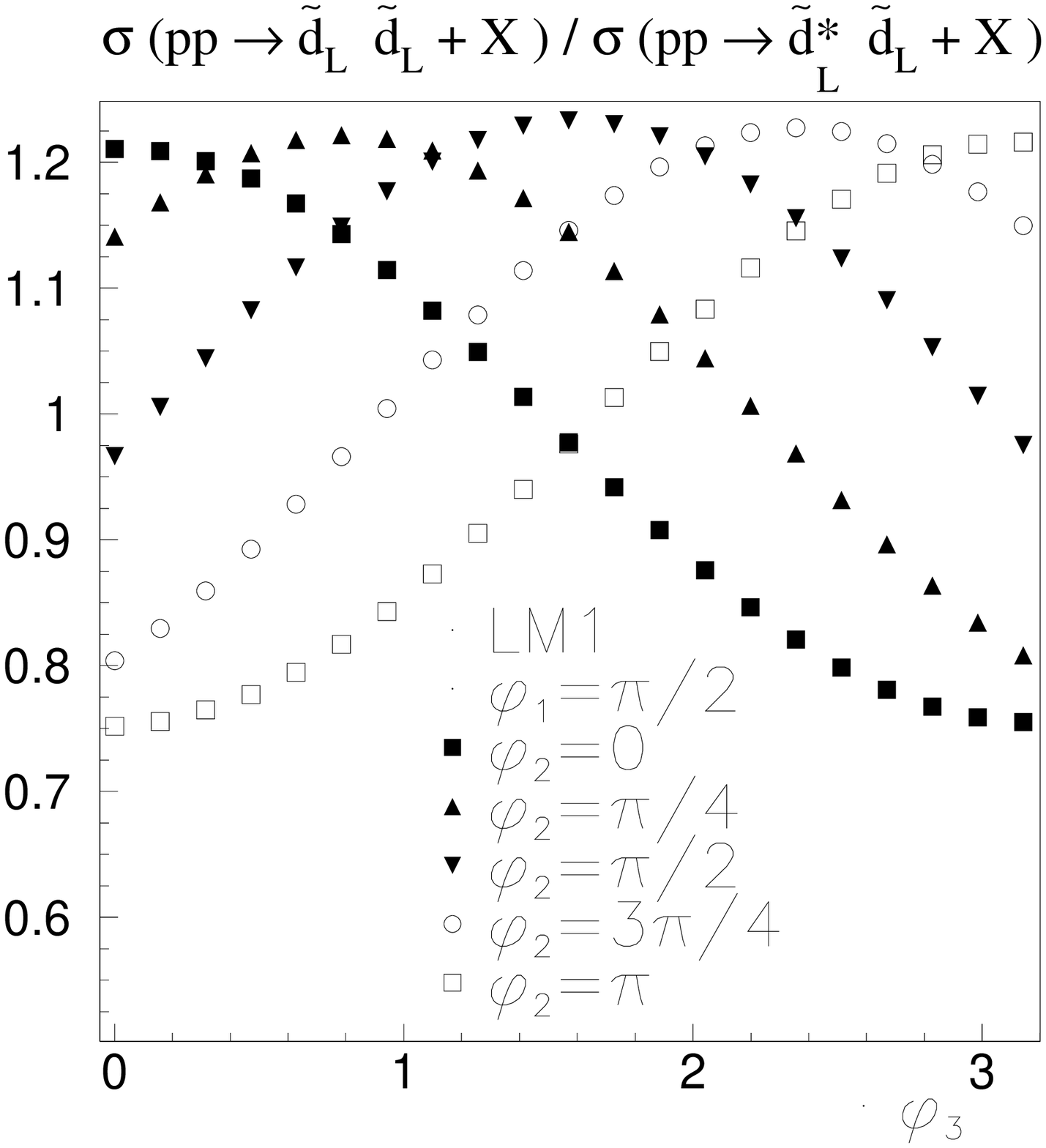}
\includegraphics[width=5cm]{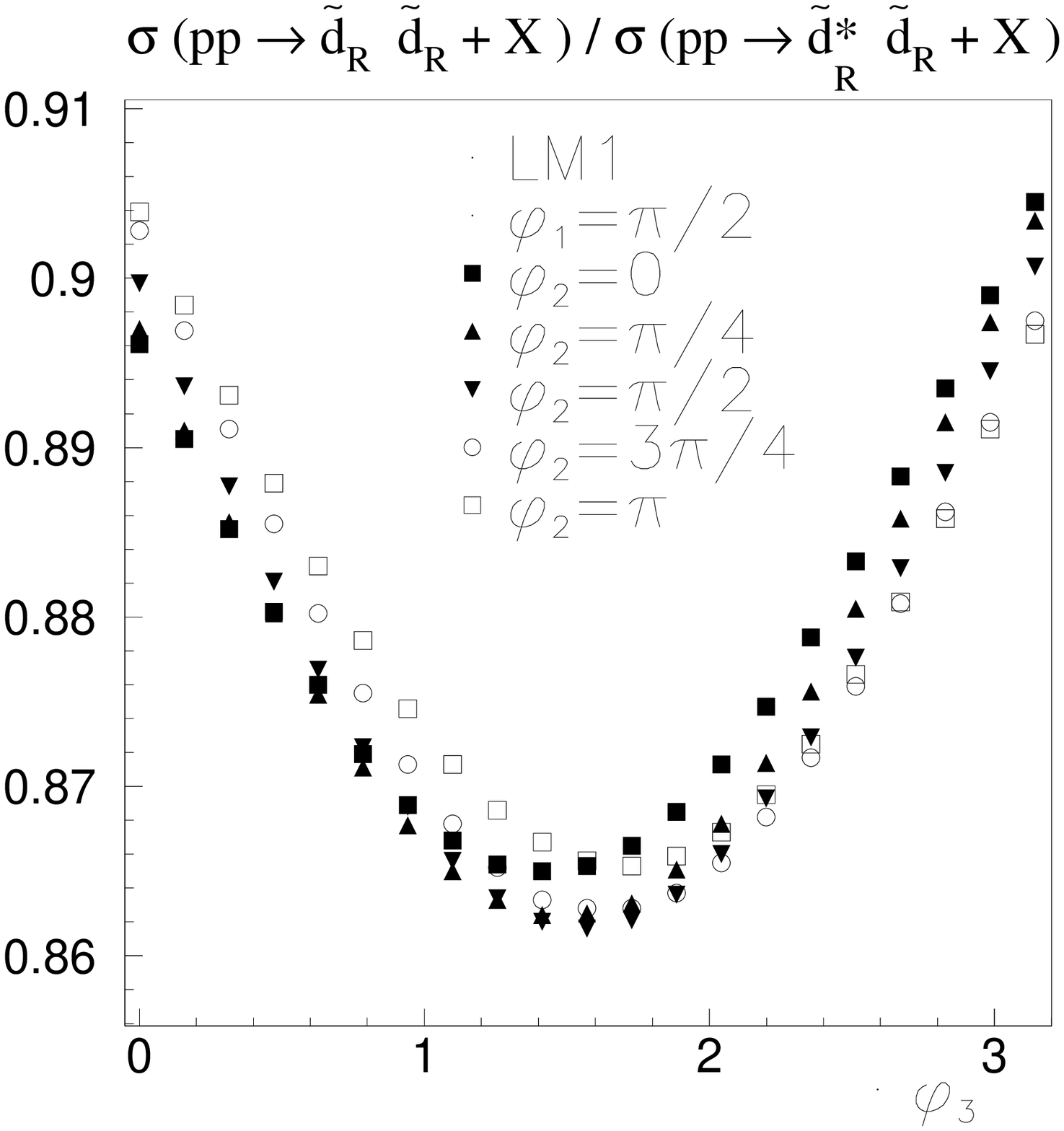}
\includegraphics[width=5cm]{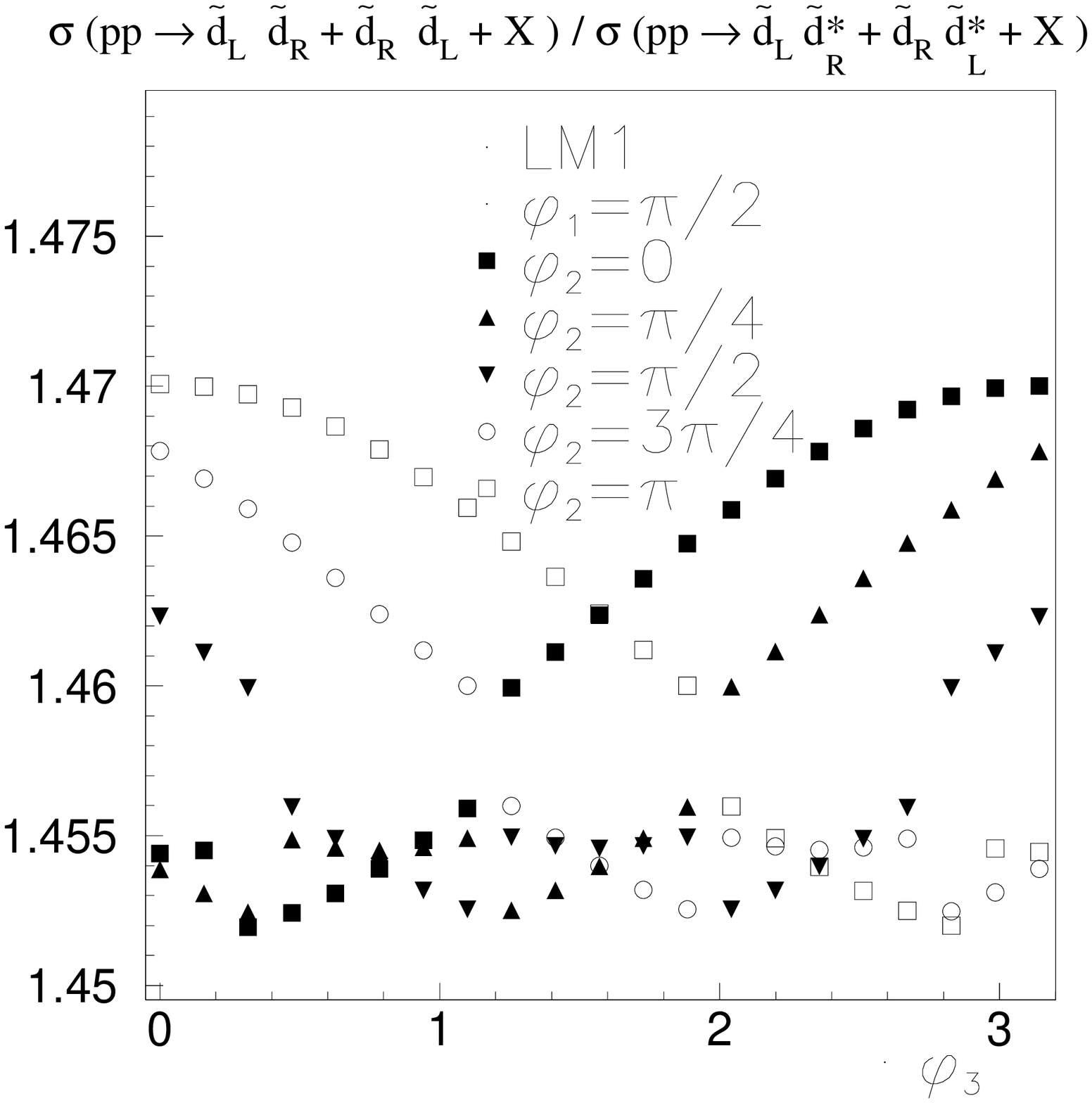}

\caption{ The same as in Fig. \ref{fig-upratio} but for down
squark production.}\label{fig-downratio}
\end{figure}

\begin{figure}
\includegraphics[width=5cm]{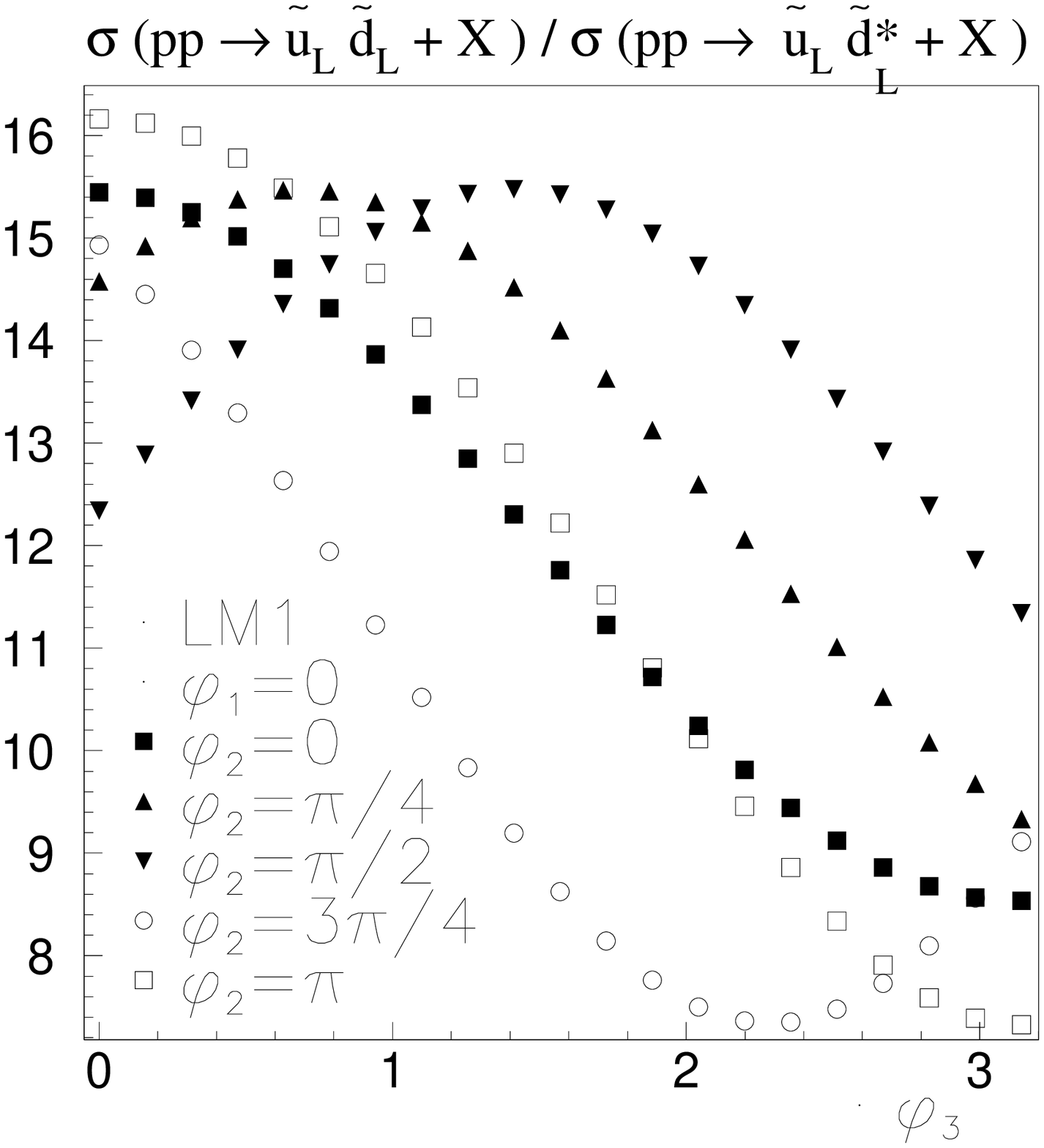}
\includegraphics[width=5cm]{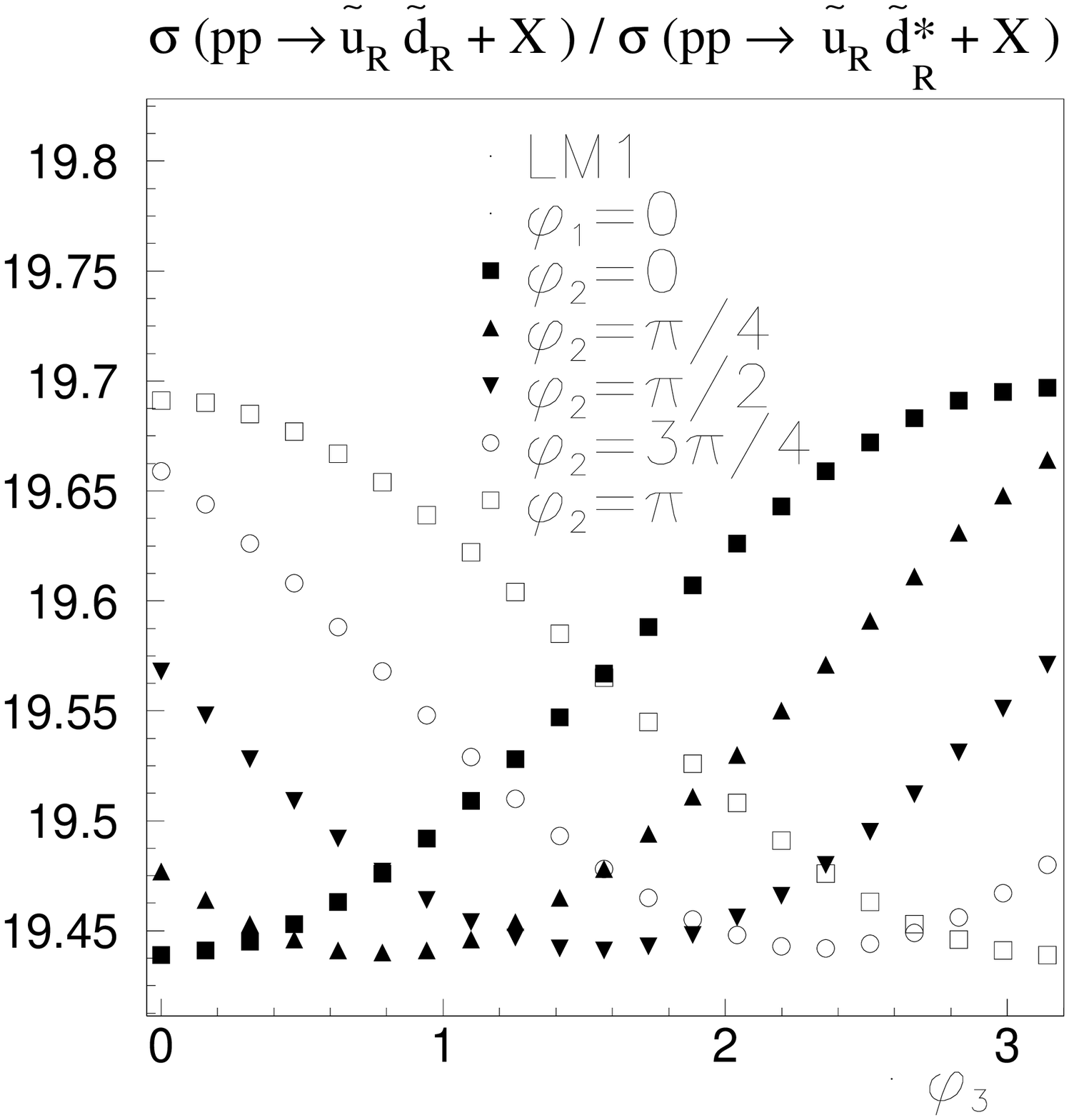}
\includegraphics[width=5cm]{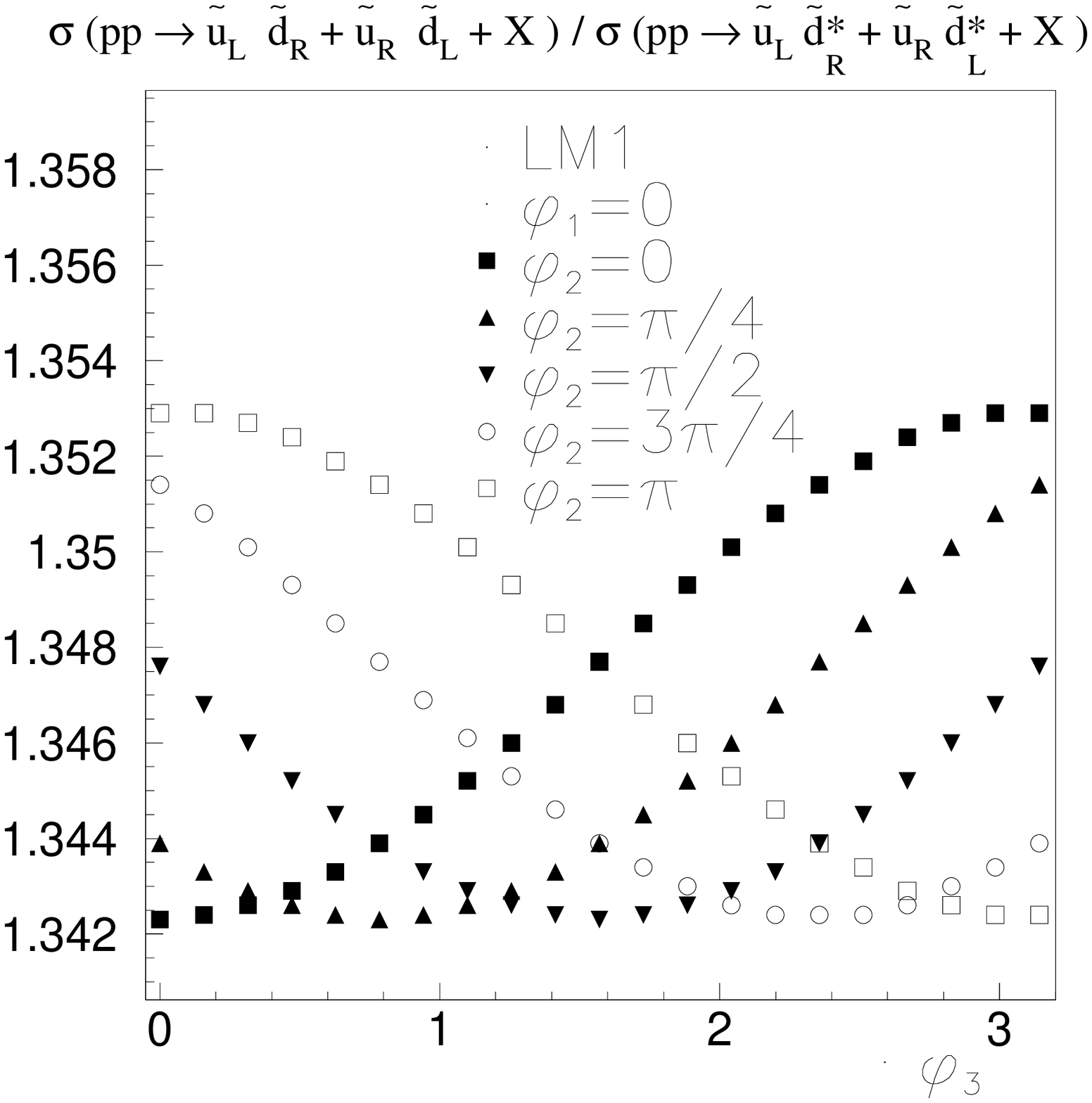}
\includegraphics[width=5cm]{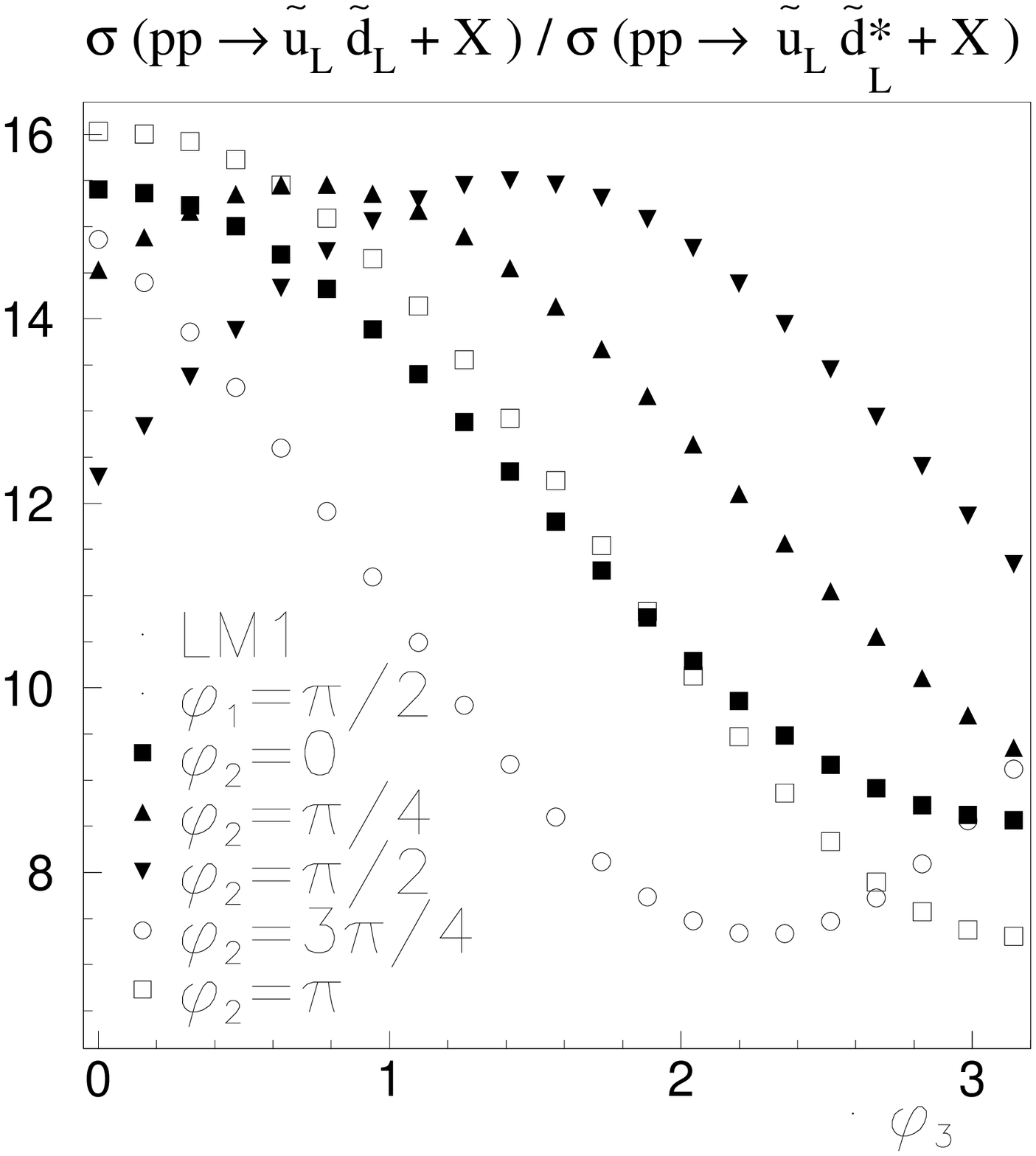}
\includegraphics[width=5cm]{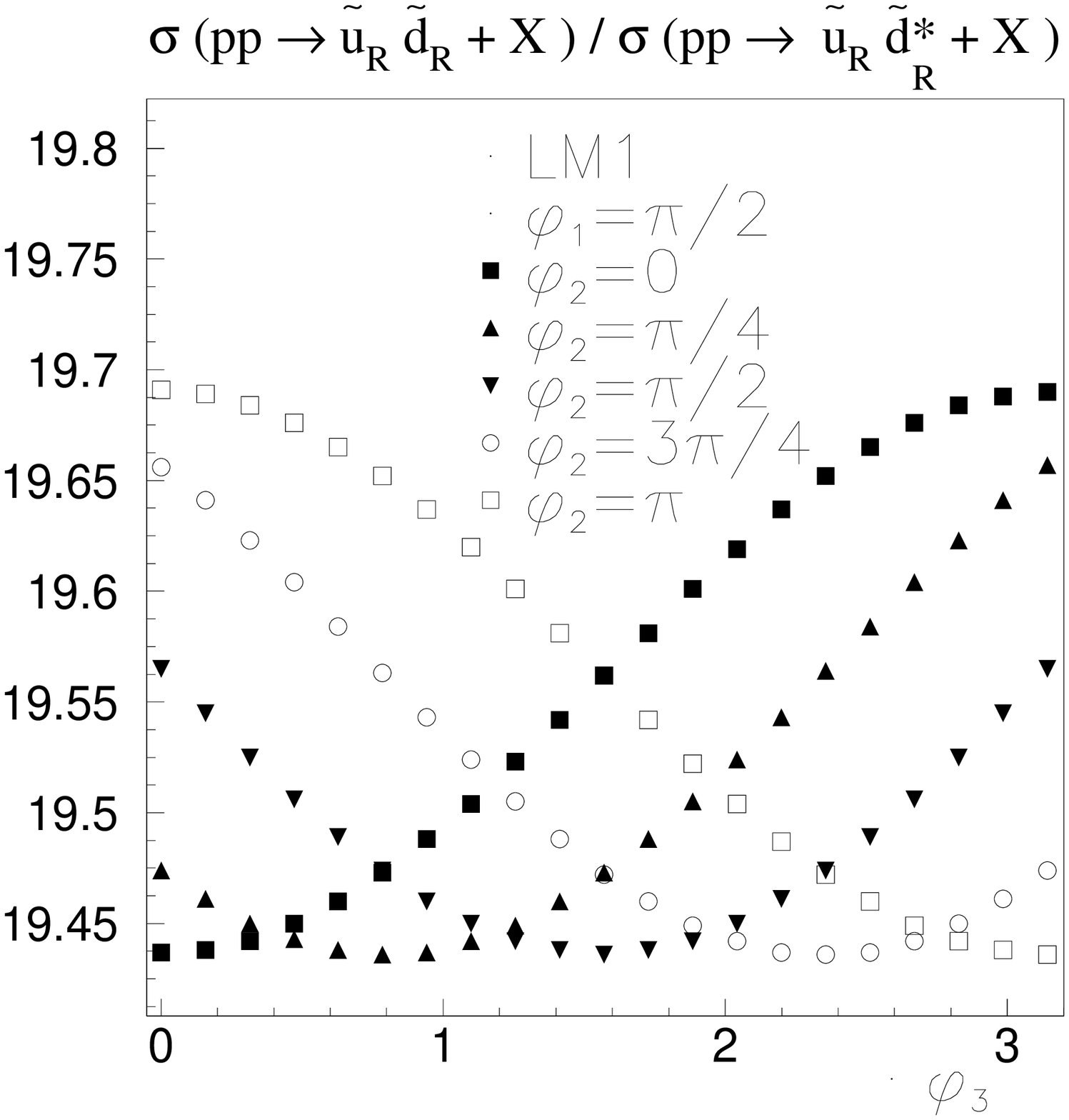}
\includegraphics[width=5cm]{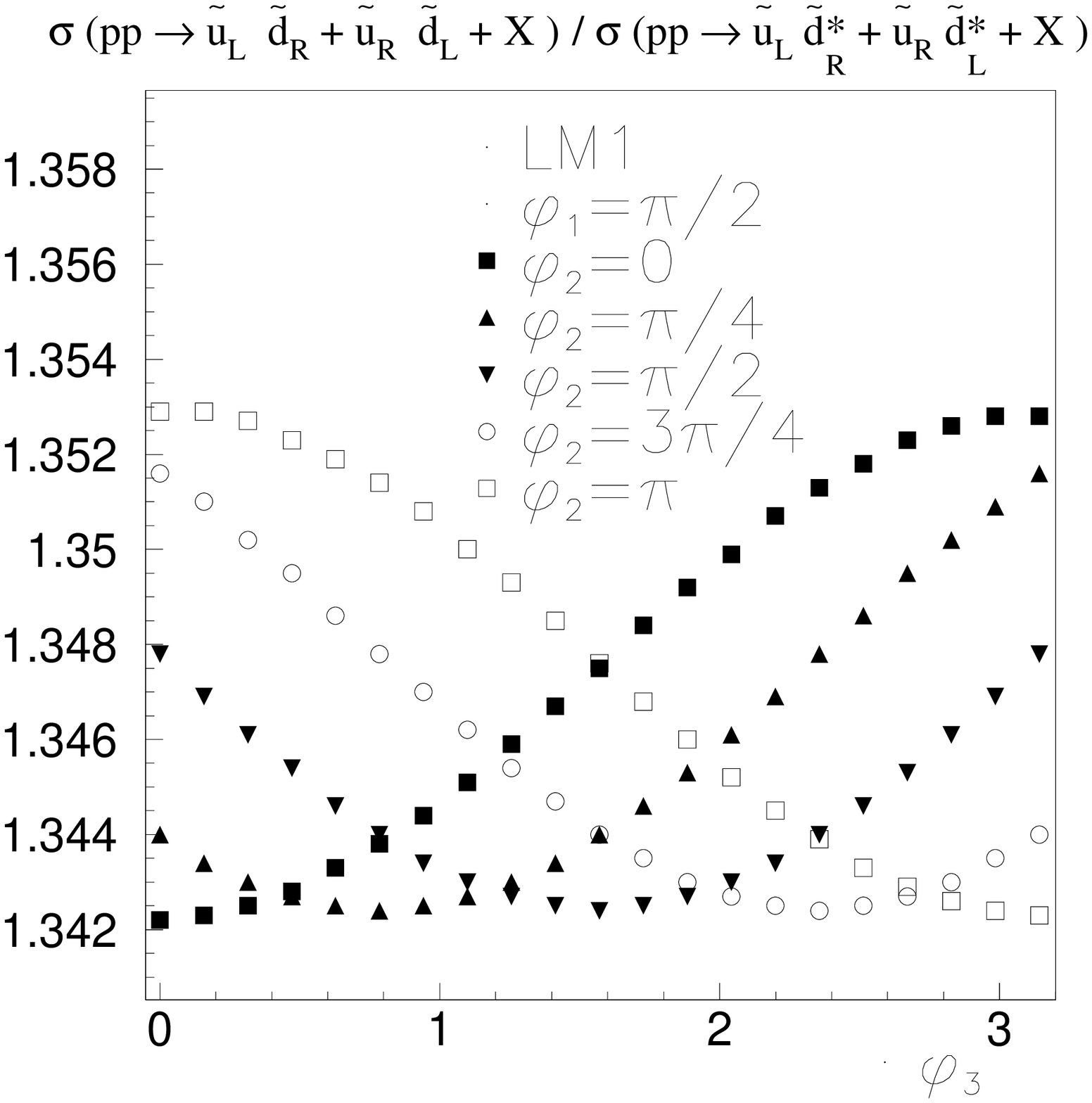}

\caption{ Ratios of up-down squark pair-production cross sections  at the
LHC as functions of $\varphi_3$ for $\varphi_1=0$ (top panel), $\varphi_1=\pi/2$ (bottom panel) and several values of $\varphi_2$.
Left: $\sigma(p\, p \rightarrow \widetilde{u}_L \widetilde{d}_{L})$ / $\sigma(p\, p \rightarrow
\widetilde{u}_L \widetilde{d}_{L}^{\star})$  for $\varphi_1=0$ (top panel) and
 $\sigma(p\, p \rightarrow \widetilde{u}_L \widetilde{d}_{L})$ / $\sigma(p\, p \rightarrow
\widetilde{u}_L \widetilde{d}_{L}^{\star})$ for $\varphi_1=\pi/2$ (bottom panel). Middle:  $\sigma(p\, p \rightarrow \widetilde{u}_R \widetilde{d}_{R})$ / $\sigma(p\, p \rightarrow
\widetilde{u}_R \widetilde{d}_{R}^{\star})$  for $\varphi_1=0$ (top panel) and
 $\sigma(p\, p \rightarrow \widetilde{u}_R \widetilde{d}_{R})$ / $\sigma(p\, p \rightarrow
\widetilde{u}_R \widetilde{d}_{R}^{\star})$  for $\varphi_1=\pi/2$ (bottom panel). Right: $\sigma(p\, p \rightarrow \widetilde{u}_L \widetilde{d}_{R})$ / $\sigma(p\, p \rightarrow
\widetilde{u}_L \widetilde{d}_{R}^{\star})$  for $\varphi_1=0$ (top panel) and
 $\sigma(p\, p \rightarrow \widetilde{u}_L \widetilde{d}_{R})$ / $\sigma(p\, p \rightarrow
\widetilde{u}_L \widetilde{d}_{R}^{\star})$  for $\varphi_1=\pi/2$ (bottom panel).}\label{fig-updownratio}
\end{figure}

Shown in Figs. \ref{fig-upratio}, \ref{fig-downratio} and
\ref{fig-updownratio} are ratios of the cross sections for
$\widetilde{q} \widetilde{\hat q}$ production to those for
$\widetilde{q} \widetilde{\hat q}^{\star}$. These figures are
intended for determining the relative population of
squark--anti-squark and squark-squark pairs in collider
environment at the LHC. Fig. \ref{fig-upratio} dictates that
$\widetilde{u} \widetilde{\hat u}$ production cross section is
2--3 times larger than $\widetilde{u} \widetilde{\hat u}^{\star}$
production cross section. Therefore, for a given luminosity, only
30--50$\%$ of up-type squark pairs will be up--anti-up squarks.
Contrary to up squark sector, the corresponding ratios in
down-type squark sector remain ${\cal{O}}(1)$ as is seen in Fig.
\ref{fig-downratio}. Therefore, one expects down-type
squark--squark and squark--anti-squark pairs to be produced
approximately equal in number. This manifest difference between
up-- and down--squark pair production could be useful in collider
searches for squarks (from their decays into certain leptonic
final states, for example).

Perhaps the most interesting is up--down production. Indeed, as is
seen in Fig. \ref{fig-updownratio}, $\sigma(p\, p \rightarrow
\widetilde{u}_L \widetilde{d}_{L})$ / $\sigma(p\, p \rightarrow
\widetilde{u}_L \widetilde{d}_{L}^{\star})$ ranges from 10-20
depending on the phases while $\sigma(p\, p \rightarrow
\widetilde{u}_R \widetilde{d}_{R})$ / $\sigma(p\, p \rightarrow
\widetilde{u}_R \widetilde{d}_{R}^{\star})$ is approximately fixed
to 20. These numbers imply that the number of squark--anti-squark
pairs are only $\sim 5-10\%$ of the squark-squark pairs for these
modes. Unlike these similar-chirality modes, the
dissimilar-chirality mode $\sigma(p\, p \rightarrow
\widetilde{u}_L \widetilde{d}_{R})$ / $\sigma(p\, p \rightarrow
\widetilde{u}_L \widetilde{d}_{R}^{\star})$ is ${\cal{O}}(1)$.
Clearly, none of these ratios exhibit a strong variation with
$\varphi_1$.

 From all these three figures, Figs. \ref{fig-upratio},
\ref{fig-downratio} and \ref{fig-updownratio}, we conclude that
squark--squark production cross sections are, at least for LM1
under consideration, larger or equal to squark--anti-squark
production. The main reason for this, apart from the impact of
squark masses themselves, is the proportionality of squark--squark
production cross sections to the exchanged gaugino mass. Indeed,
squark--anti-squark production involves transferred momentum
rather than the gaugino mass in the $t$-channel (see also
\cite{thomas}).

The analysis in this subsection requires a knowledge of what
squark with what chirality is produced. For numerical results
illustrated in Figs.\ref{fig-sig-up_0}--\ref{fig-updownratio} to
make sense experiments must be able to differentiate among
$\widetilde{q}_L$, $\widetilde{q}_R$, $\widetilde{q}_L^{\star}$
and $\widetilde{q}_R^{\star}$. The chirality information can be
inferred from their decay pattern:
\begin{eqnarray}
\widetilde{q}_R \rightarrow (\mbox{quark jet}) +
\slashchar{P}_T\;,\;\; \widetilde{q}_L \rightarrow (\mbox{quark
jet}) + (\mbox{leptons}) + \slashchar{P}_T
\end{eqnarray}
where a detailed study of such detection modes have been given in
\cite{CMS} and references therein.

Other than chirality there is the question of flavor. Indeed, in a
real experimental situation it could be quite difficult to know if
the 'left-handed squark' produced is $\widetilde{u}$ or
$\widetilde{c}$ or $\widetilde{d}$ or $\widetilde{s}$. From the
scratch we know that $\widetilde{u}$ and $\widetilde{c}$ are
hardly differentiable except for their small mass splitting and
possible flavor-violation effects between $\widetilde{c}$ and
$\widetilde{t}$ squarks. In fact, all the $\widetilde{u}
\widetilde{u}$ or $\widetilde{u} \widetilde{u}^{\star}$ cross
sections plotted above may be regarded as half of the
$\widetilde{u}$ or $\widetilde{c}$ production cross sections.
Similar observations hold also for $\widetilde{d}$ and
$\widetilde{s}$ productions. To this end, there is a degree of
flavor-blindness in cross sections plotted in
Figs.\ref{fig-sig-up_0}--\ref{fig-updownratio}. However, for the
figures above to make sense one has to know if the squark produced
is up-type or down-type or their anti-particles. This, indeed,
could be a quite difficult task since it necessitates a detailed
knowledge of the electric charges of the debris produced by the
collision (which should, in principle, be possible by measuring
curvatures of the particle tracks in the detector).

\subsection{Squark Pair-Production: Definite Flavor and Indefinite Chirality}
\begin{figure}
\includegraphics[width=5cm]{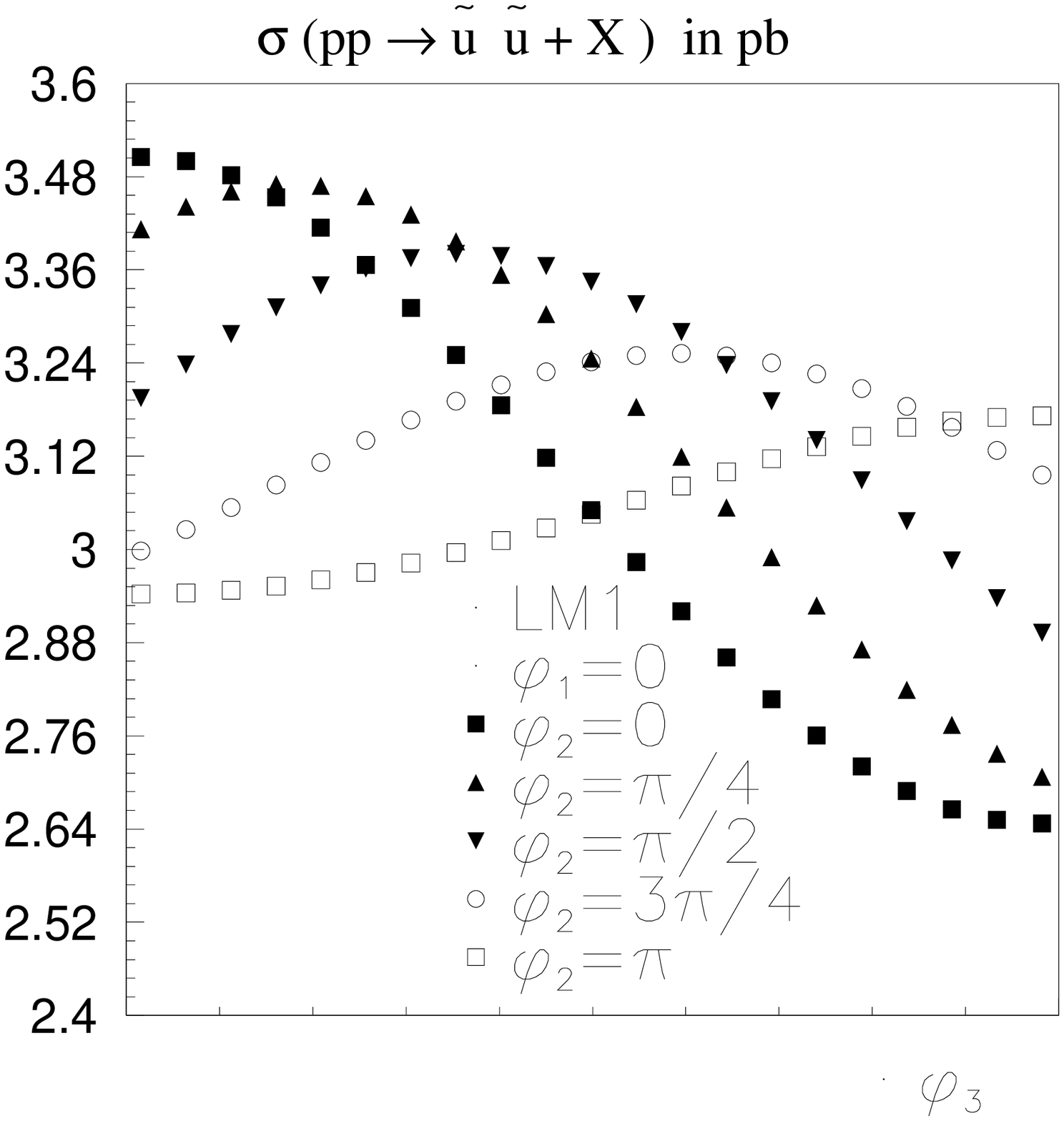}
\includegraphics[width=5cm]{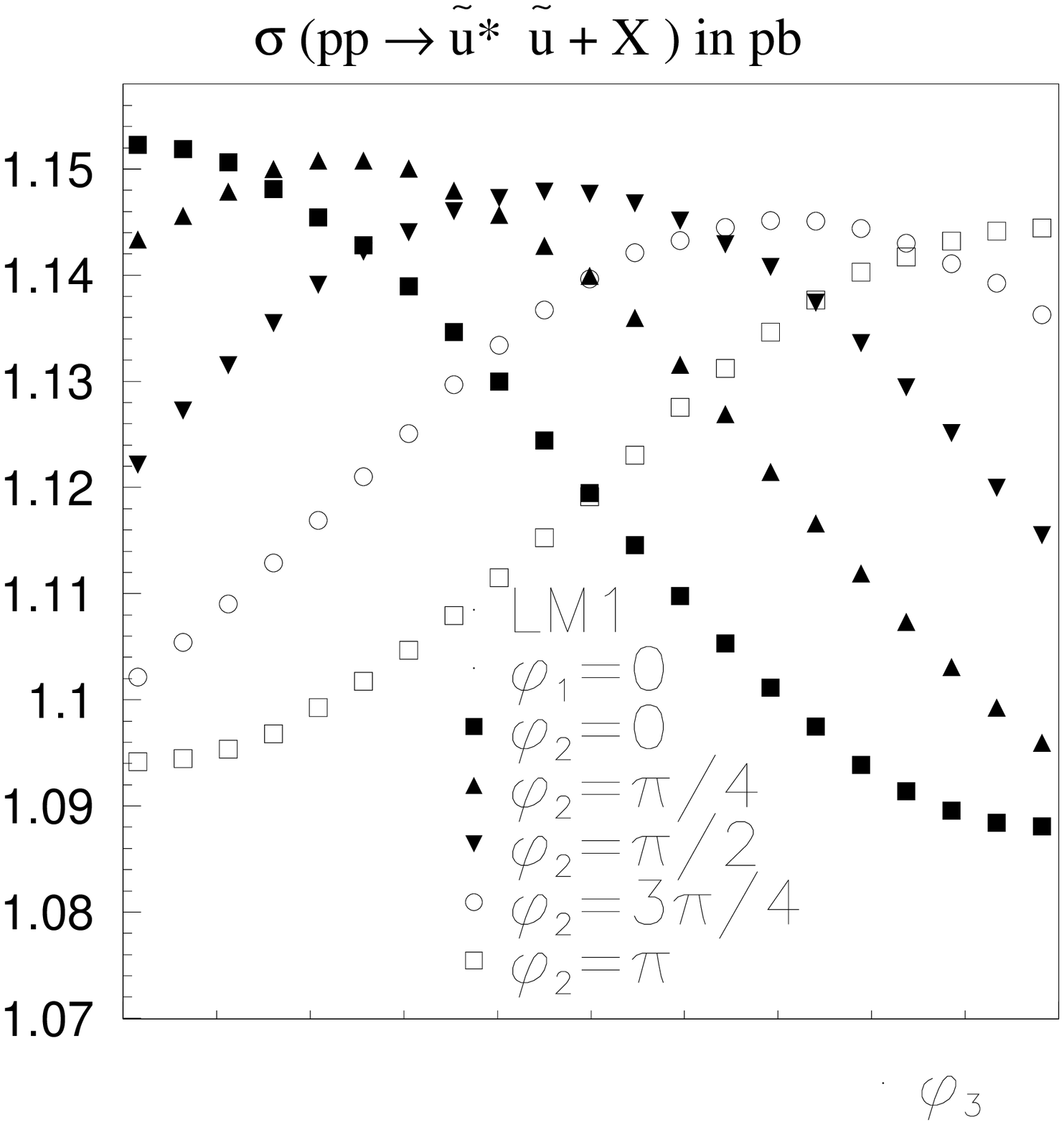}
\includegraphics[width=5cm]{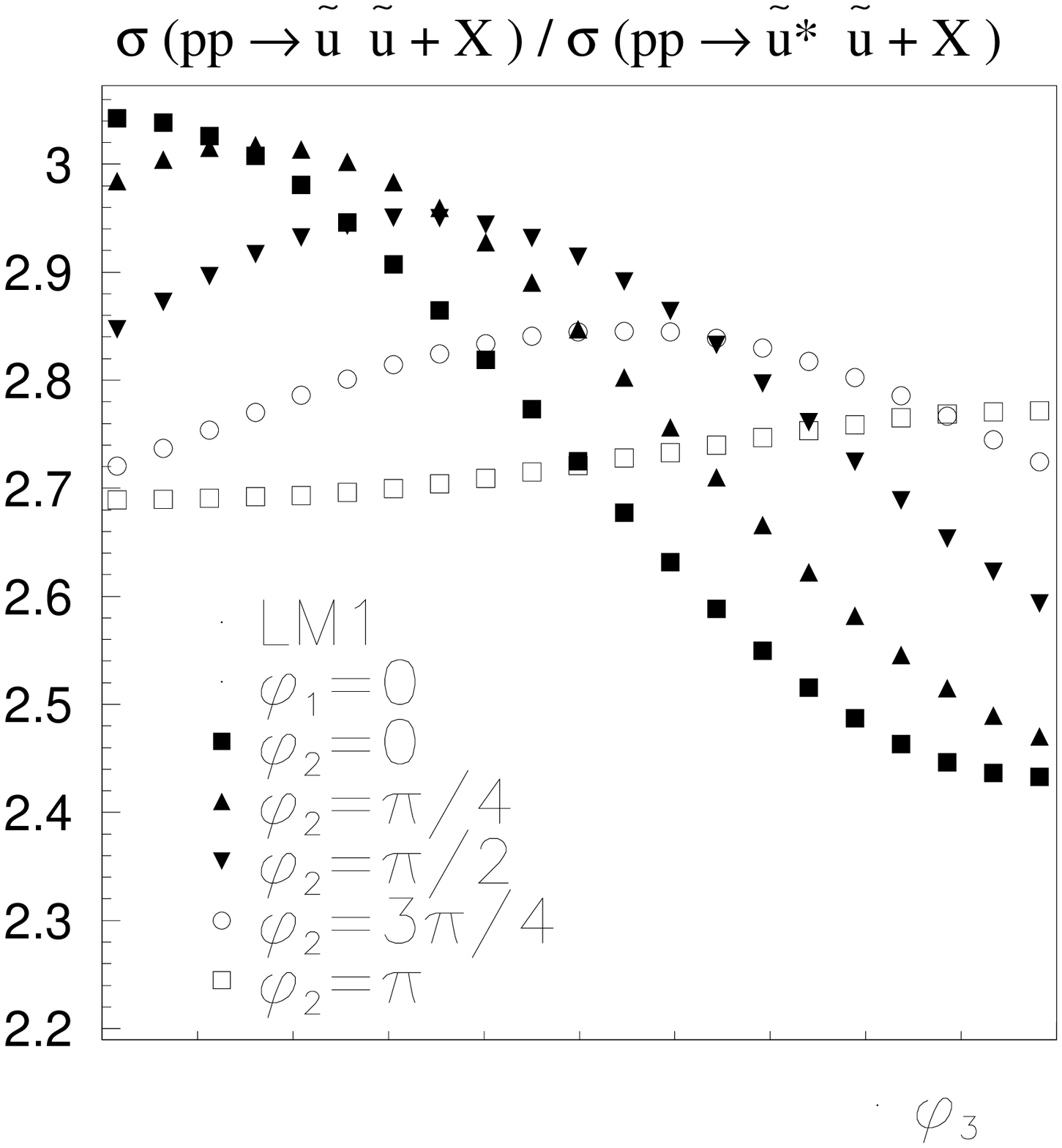}
\includegraphics[width=5cm]{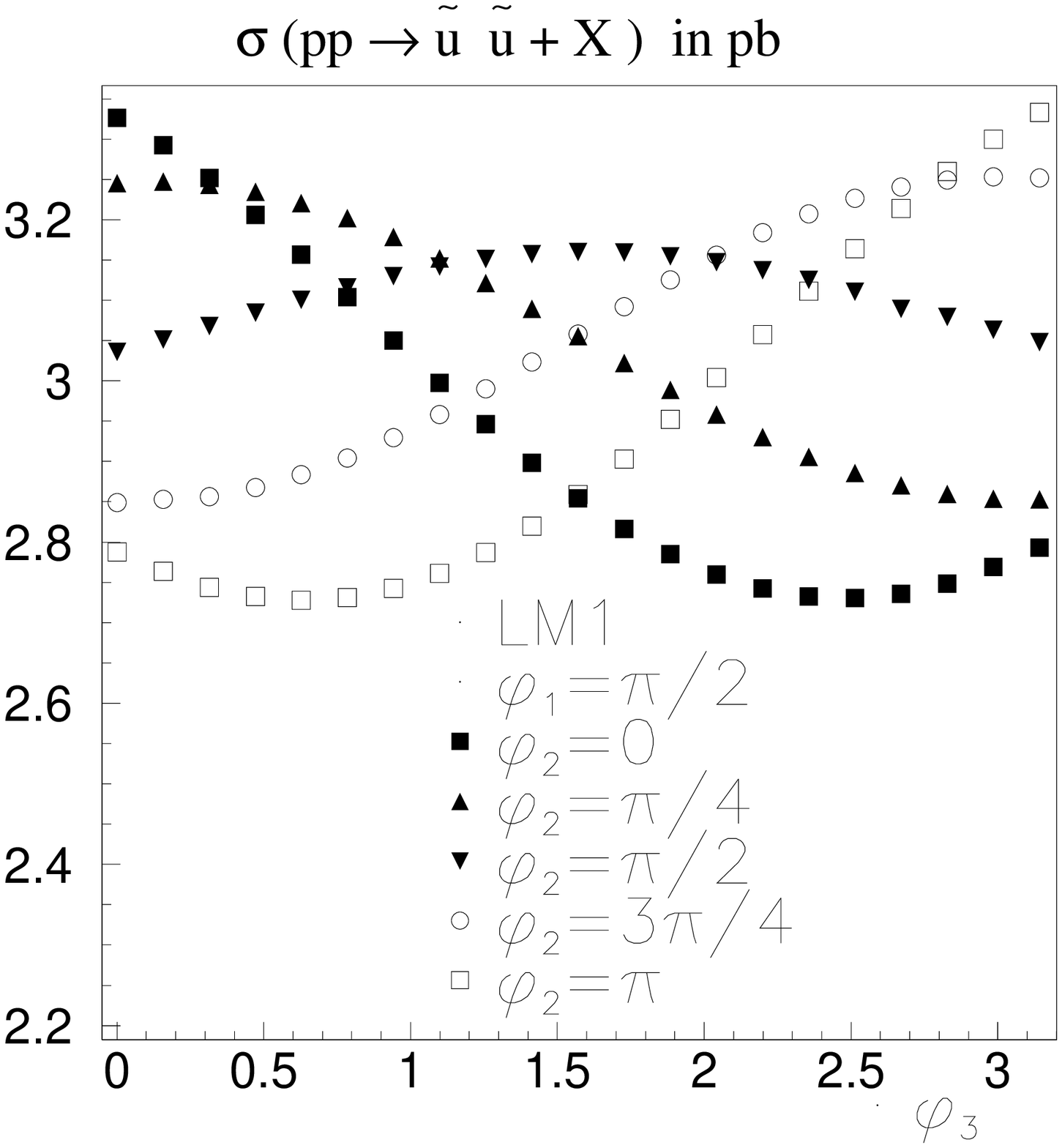}
\includegraphics[width=5cm]{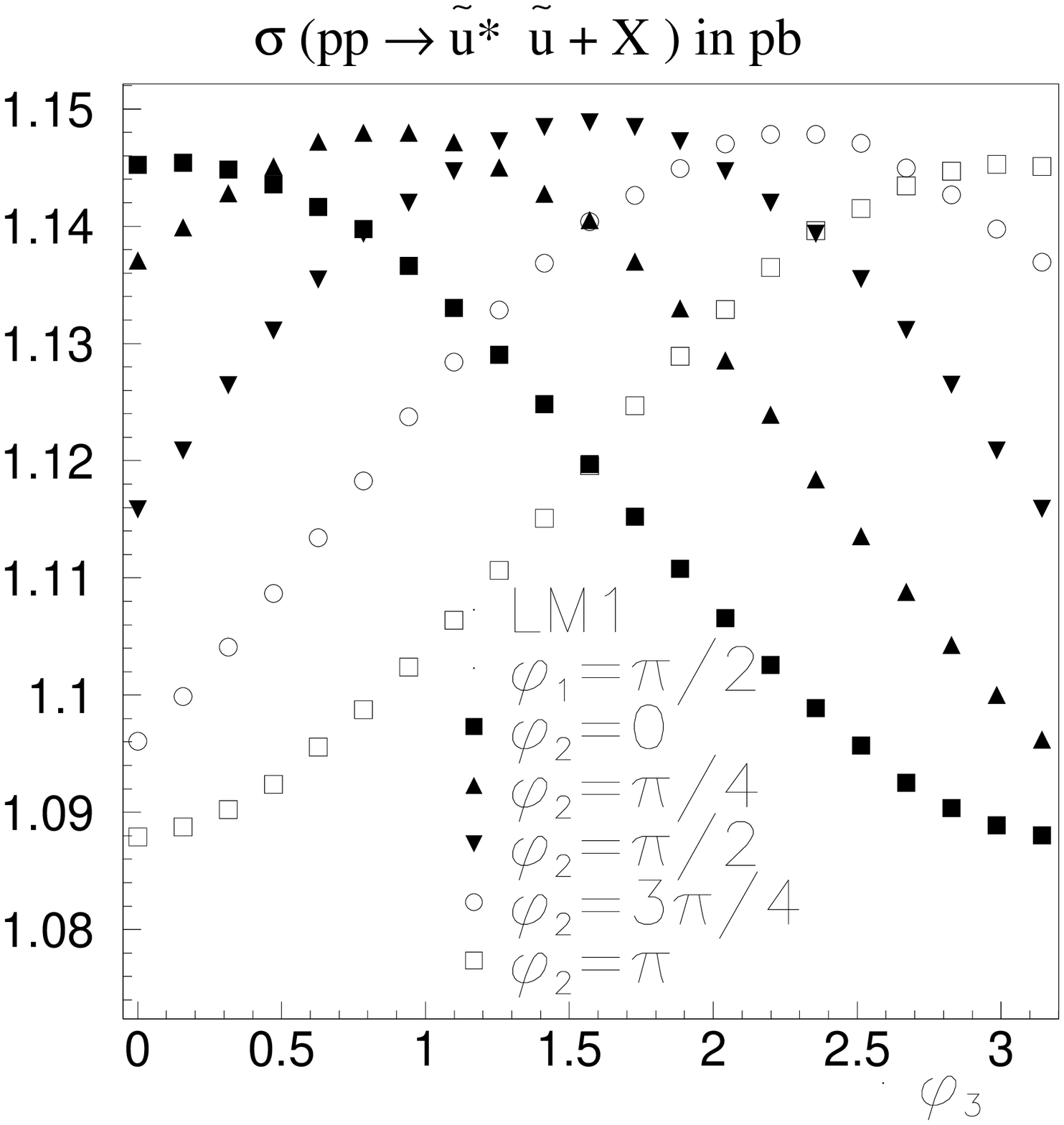}
\includegraphics[width=5cm]{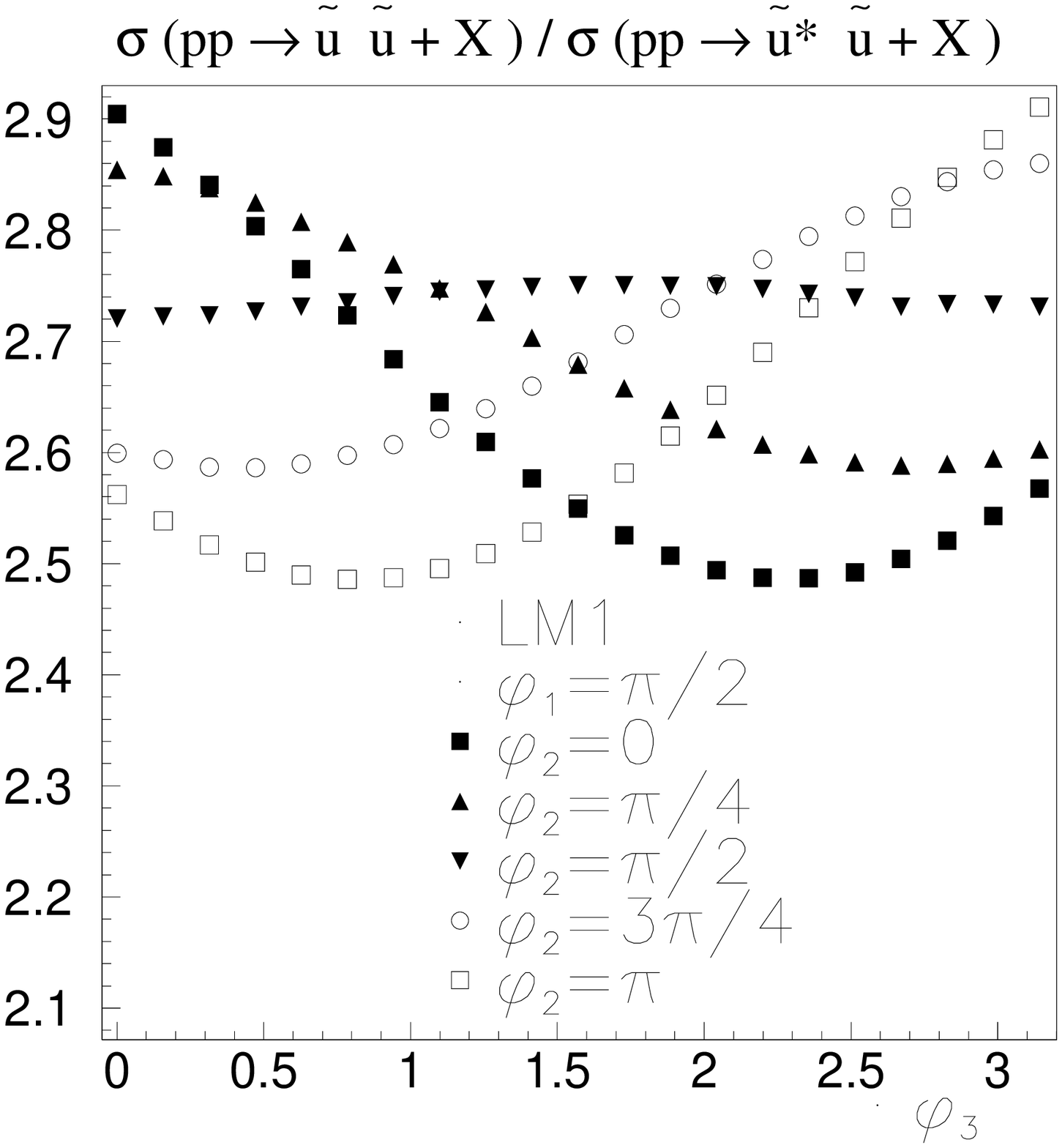}

\caption{The chirality-blind up-squark pair-production cross
sections $\sigma(p\, p \rightarrow \widetilde{u} \widetilde{u}) =
\sum_{ X=L,R; Y=L,R} \sigma(p\, p \rightarrow \widetilde{u}_X
\widetilde{u}_{Y})$ and $\sigma(p\, p \rightarrow \widetilde{u}
\widetilde{u}^{\star}) = \sum_{X=L,R; Y=L,R} \sigma(p\, p
\rightarrow \widetilde{u}_X \widetilde{u}_{Y}^{\star})$ at the LHC
as functions of $\varphi_3$ for $\varphi_1=0$ (top panel),
$\varphi_1=\pi/2$ (bottom panel) and several values of
$\varphi_2$. Left: $\sigma(p\, p \rightarrow \widetilde{u}
\widetilde{u})$   for $\varphi_1=0$ (top panel) and
 $\sigma(p\, p \rightarrow \widetilde{u} \widetilde{u})$ for
 $\varphi_1=\pi/2$ (bottom panel). Middle:
 $\sigma(p\, p \rightarrow \widetilde{u} \widetilde{u}^{\star})$  for $\varphi_1=0$ (top panel) and
 $\sigma(p\, p \rightarrow \widetilde{u} \widetilde{u}^{\star})$
 for $\varphi_1=\pi/2$ (bottom panel). Right: $\sigma(p\, p \rightarrow \widetilde{u} \widetilde{u})$ / $\sigma(p\, p \rightarrow
\widetilde{u} \widetilde{u}^{\star})$  for $\varphi_1=0$ (top panel) and
 $\sigma(p\, p \rightarrow \widetilde{u} \widetilde{u})$ / $\sigma(p\, p \rightarrow
\widetilde{u} \widetilde{u}^{\star})$  for $\varphi_1=\pi/2$ (bottom panel).}\label{fig-uptotal}
\end{figure}

\begin{figure}
\includegraphics[width=5cm]{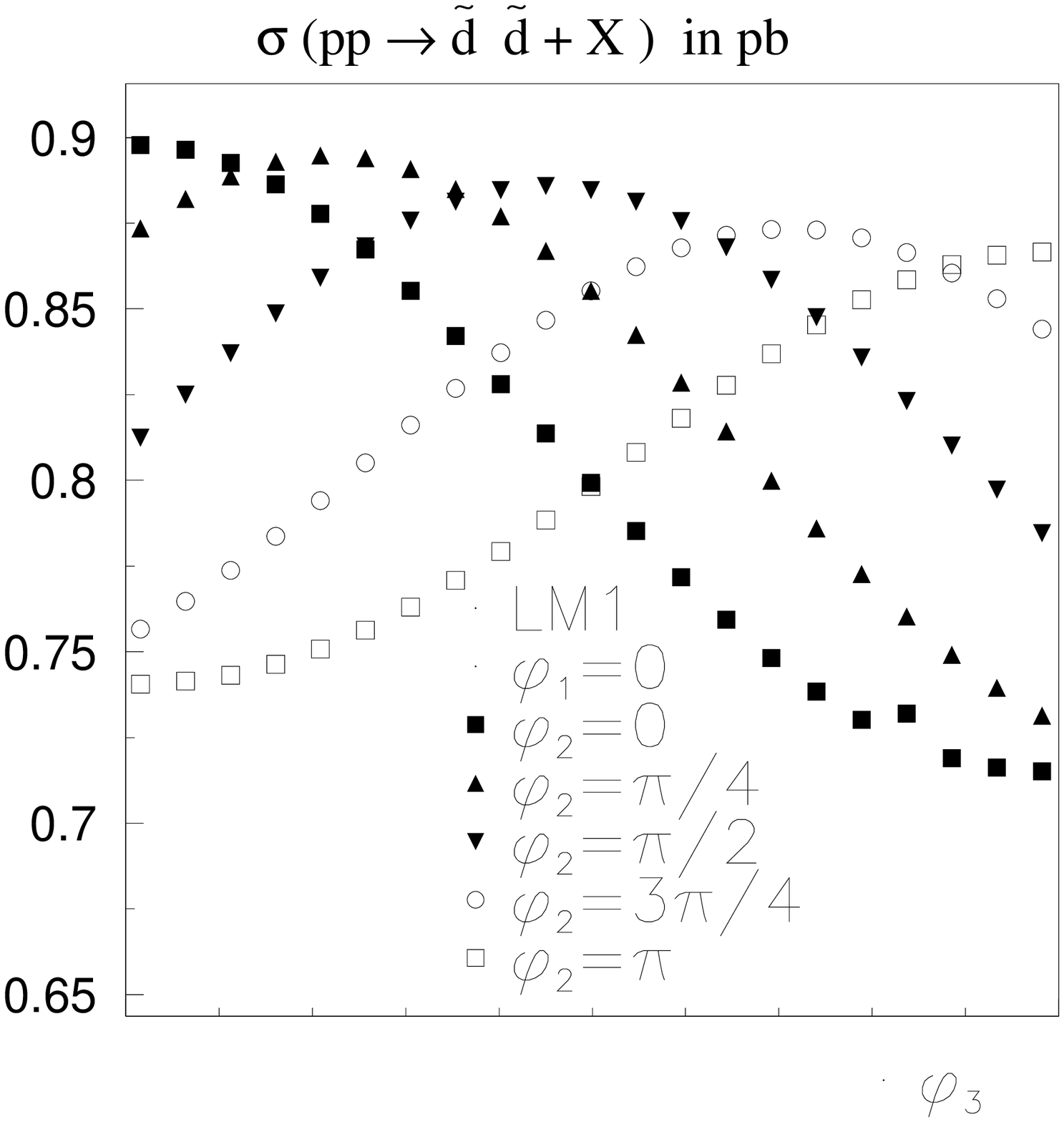}
\includegraphics[width=5cm]{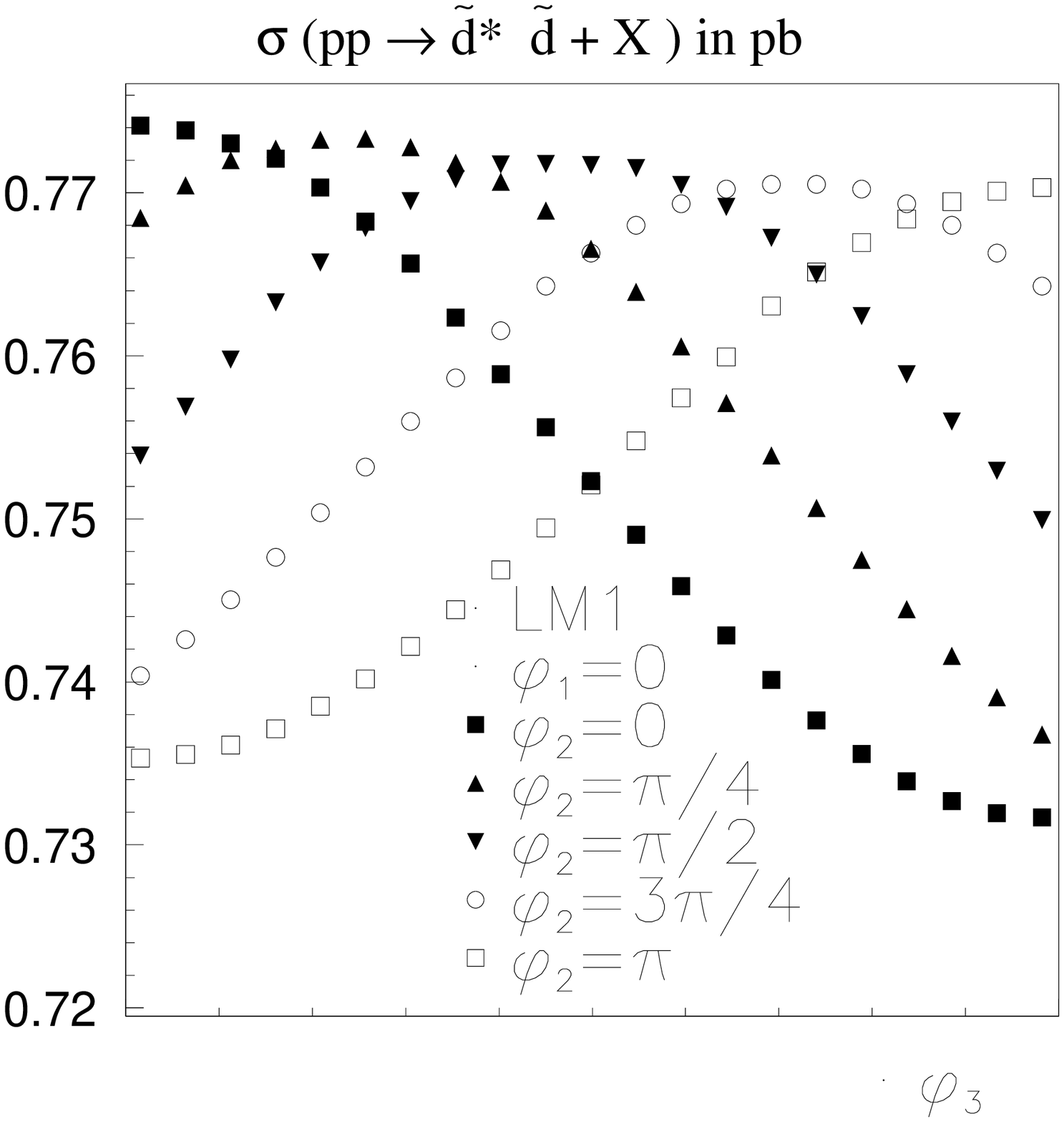}
\includegraphics[width=5cm]{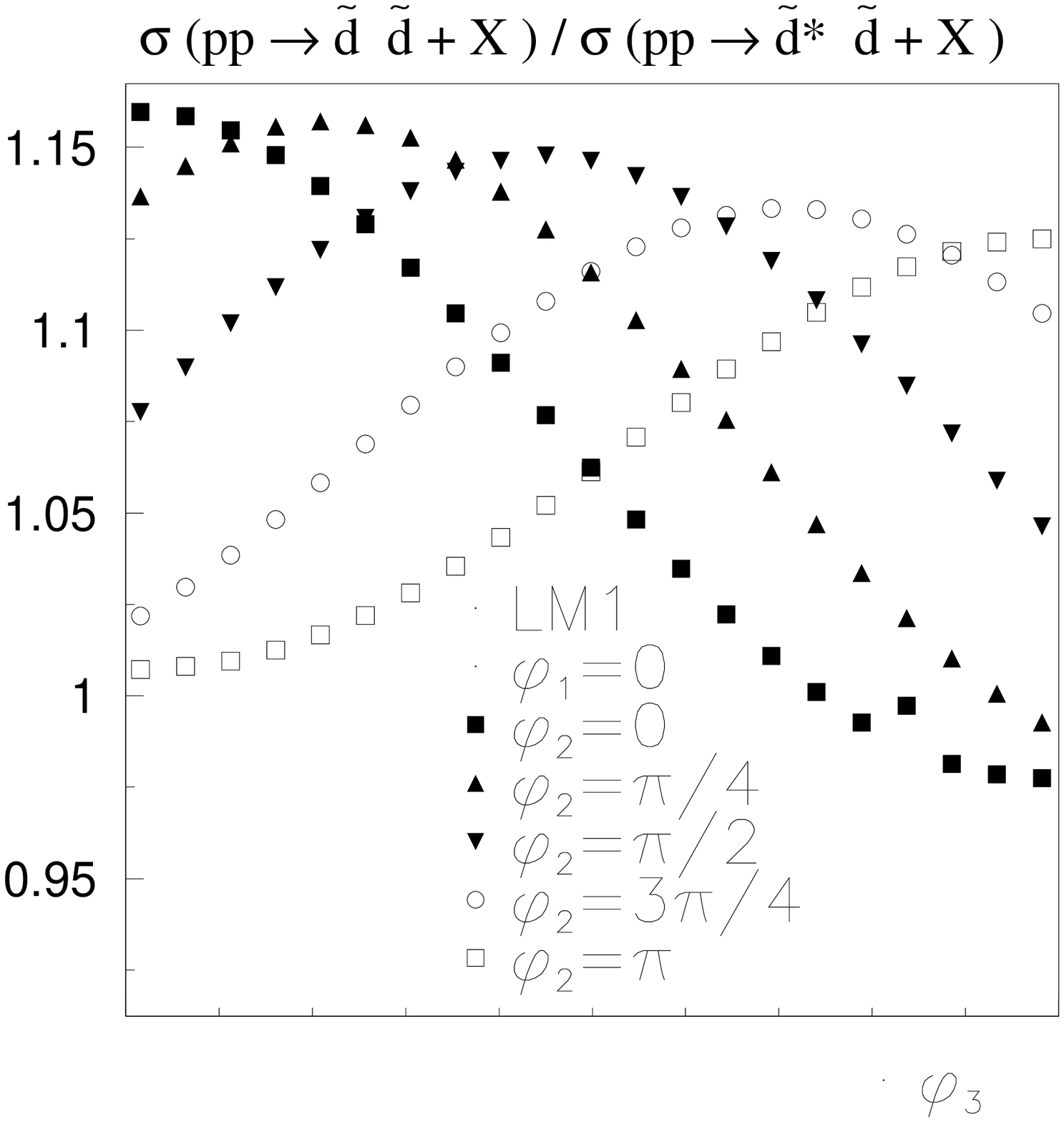}
\includegraphics[width=5cm]{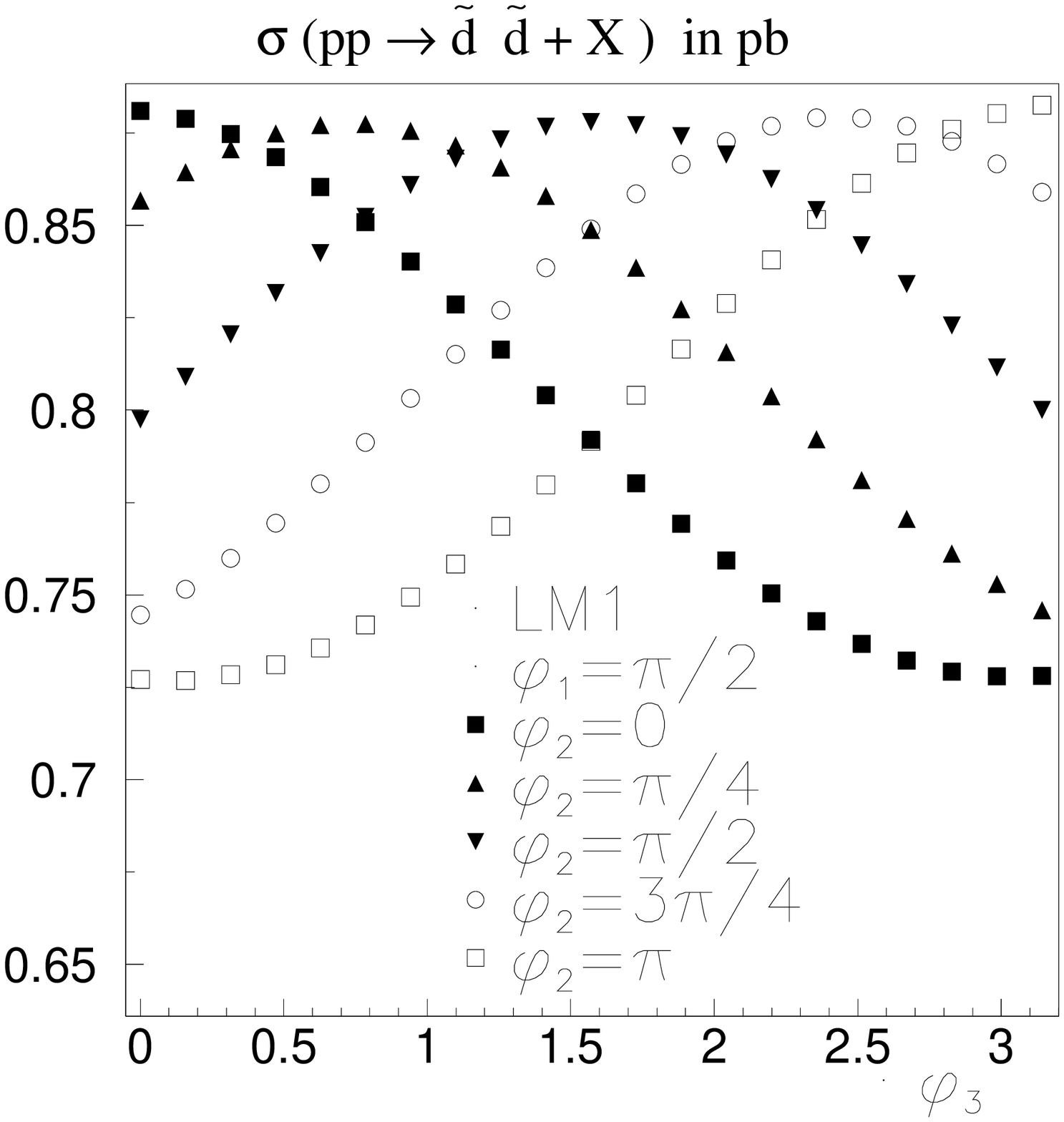}
\includegraphics[width=5cm]{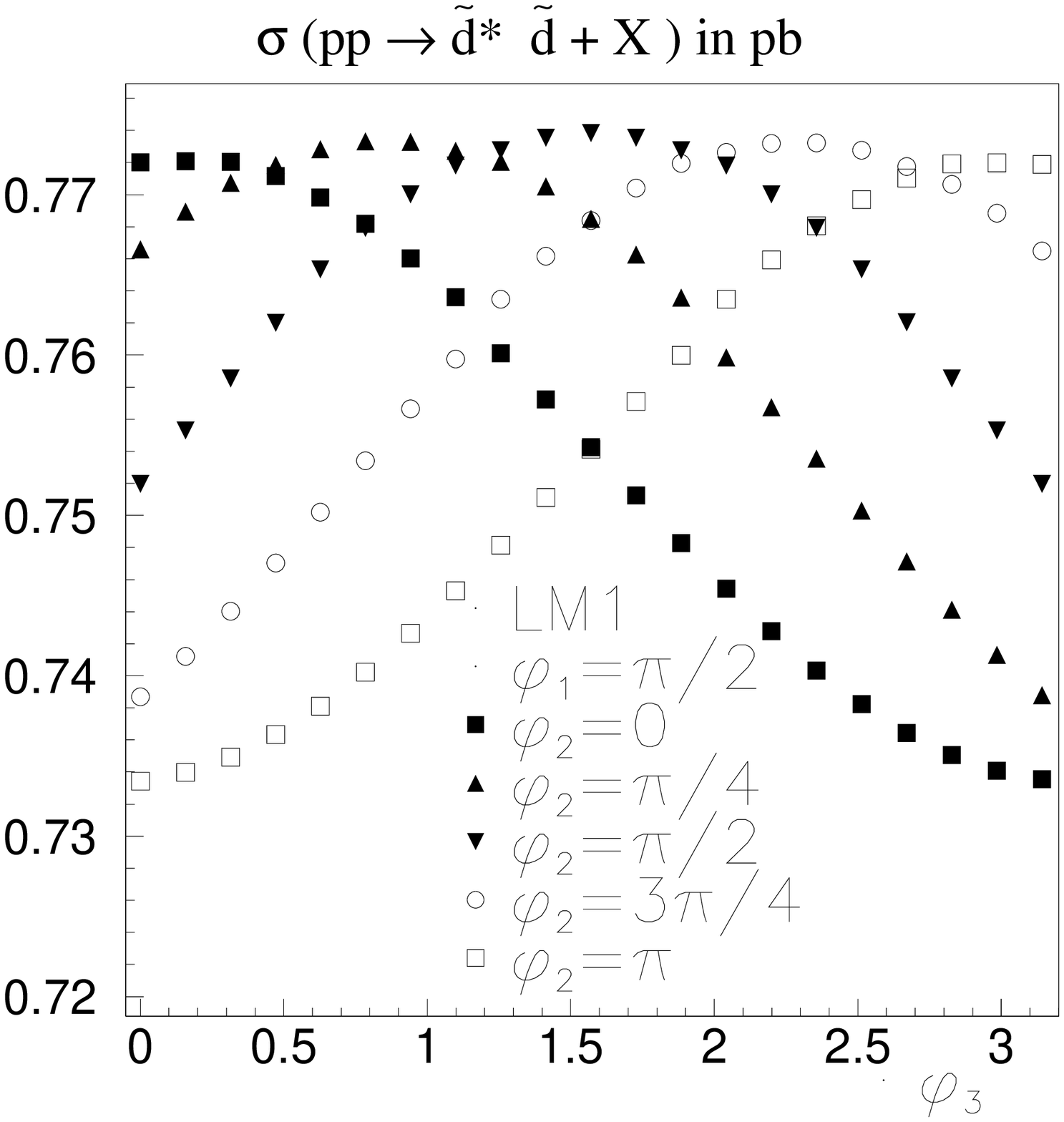}
\includegraphics[width=5cm]{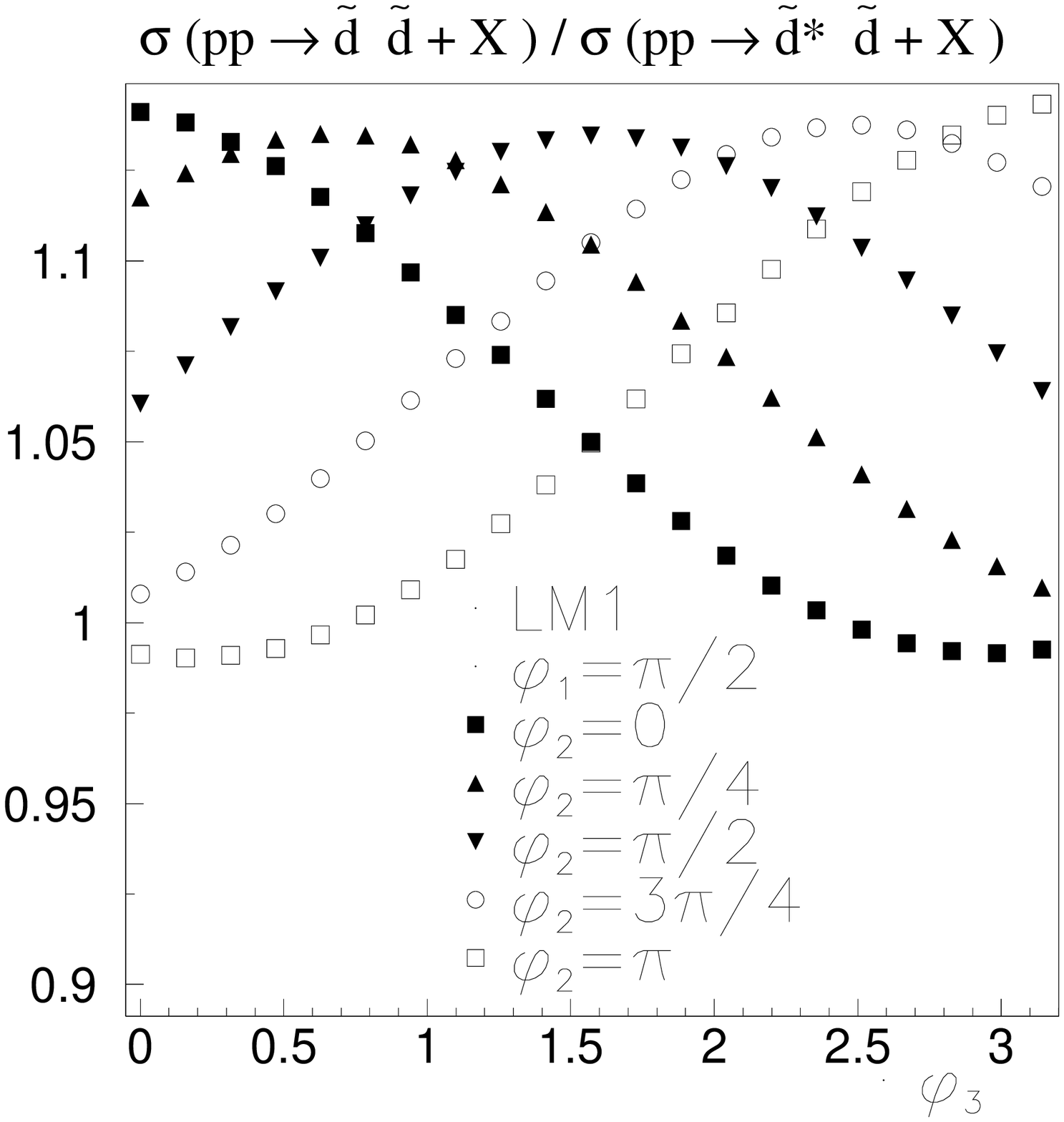}

\caption{ The chirality-blind down-squark production cross
sections with the conventions employed in Fig. \ref{fig-uptotal}.}
\label{fig-downtotal}
\end{figure}

\begin{figure}
\includegraphics[width=5cm]{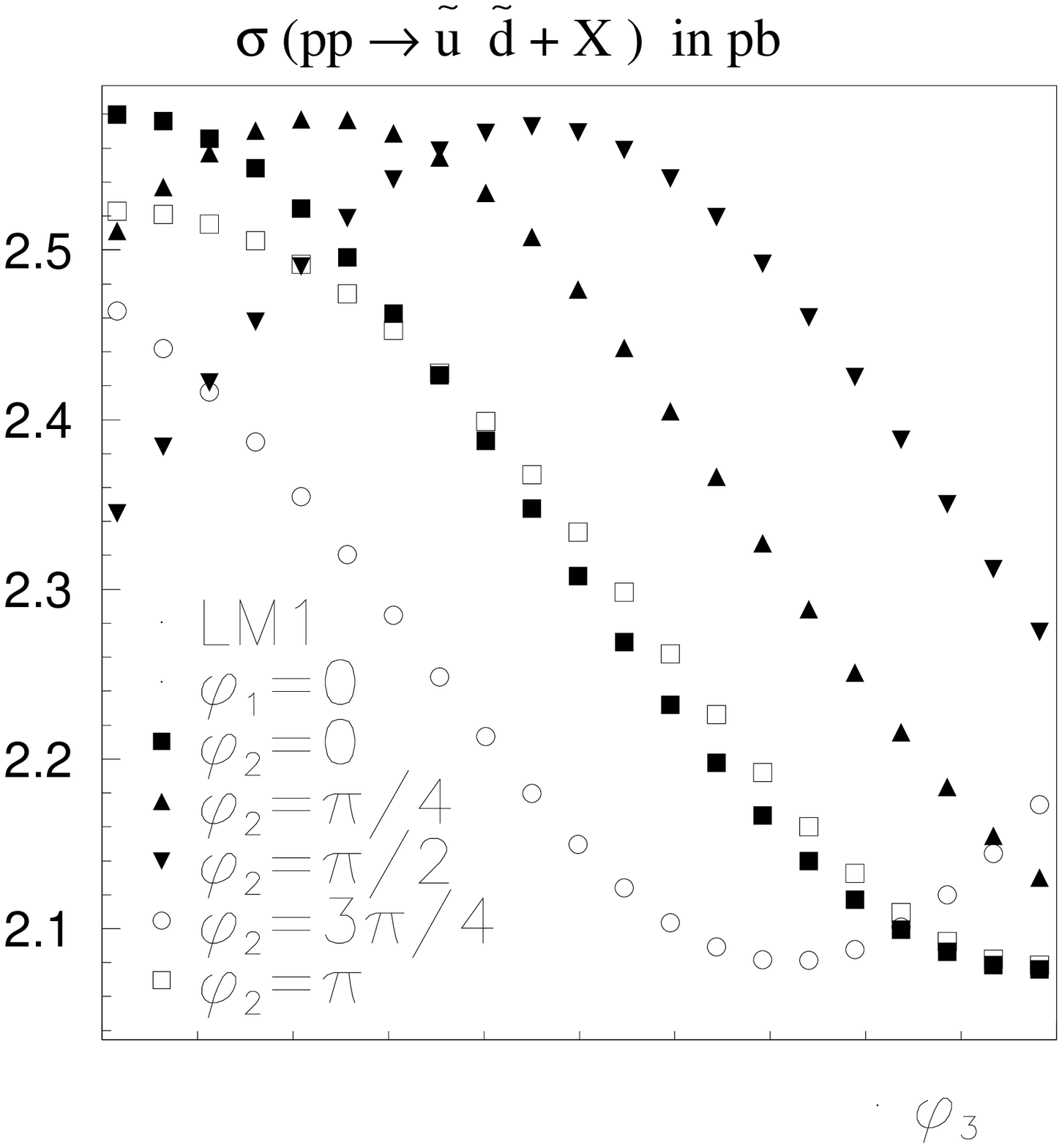}
\includegraphics[width=5cm]{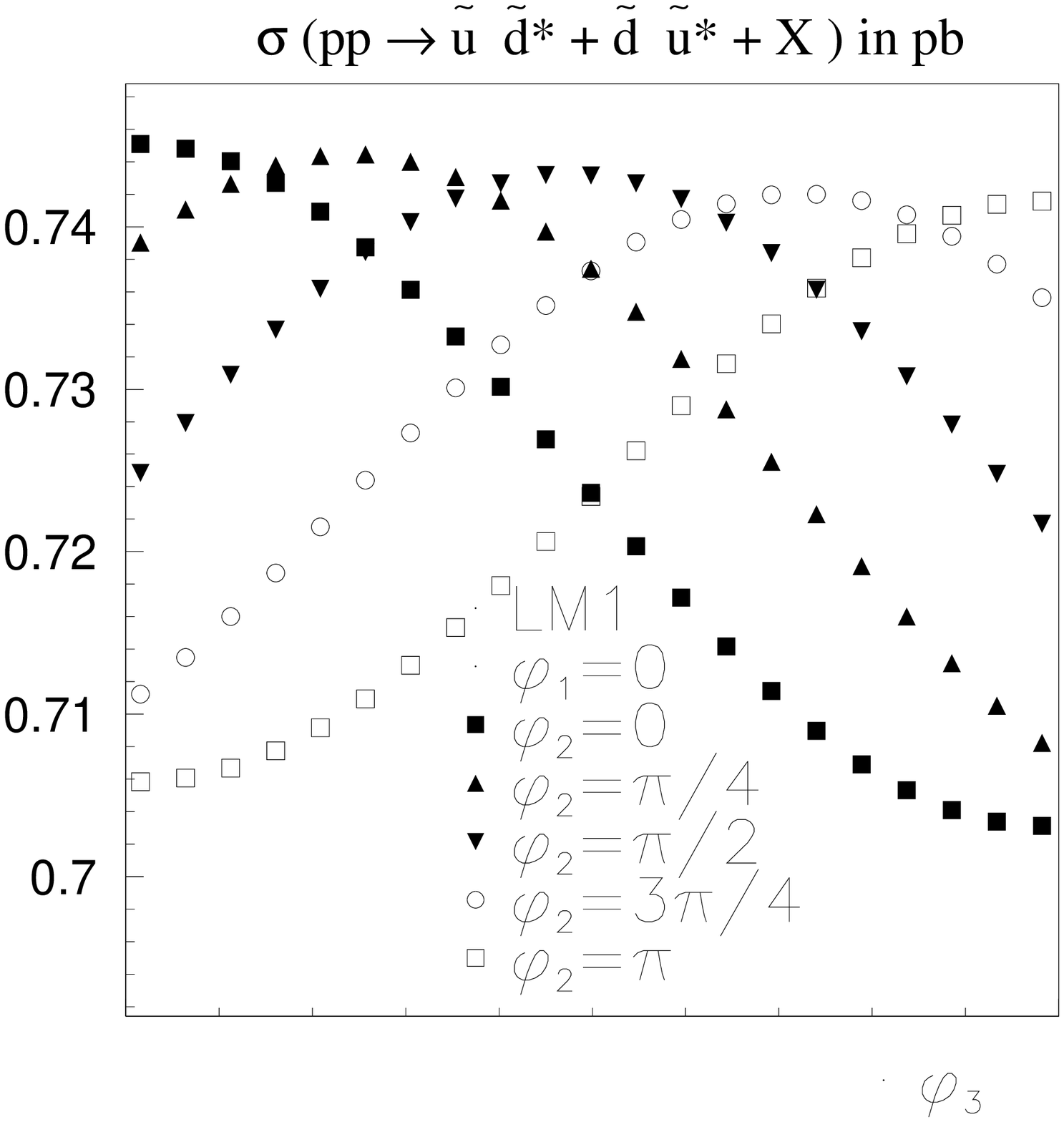}
\includegraphics[width=5cm]{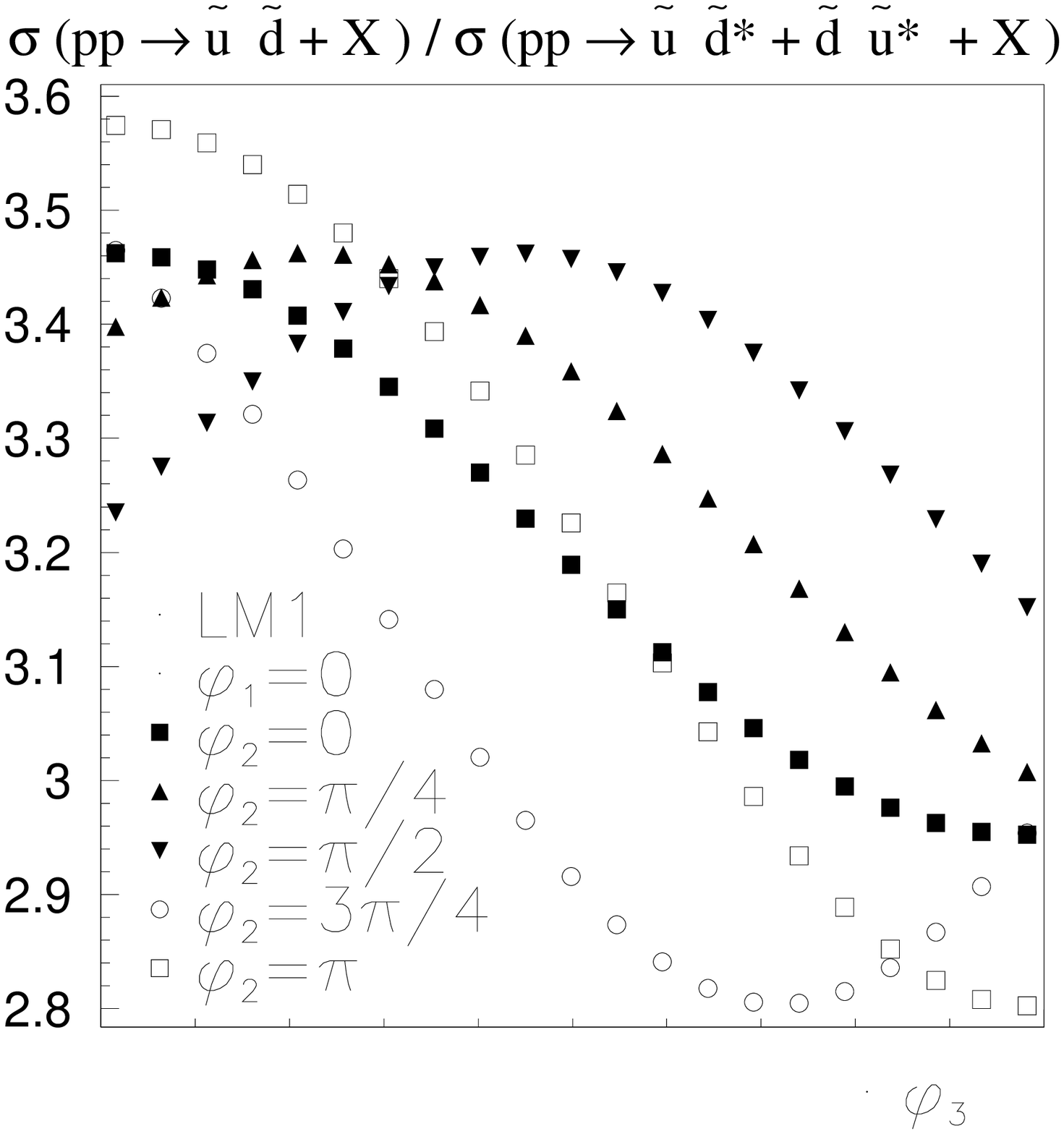}
\includegraphics[width=5cm]{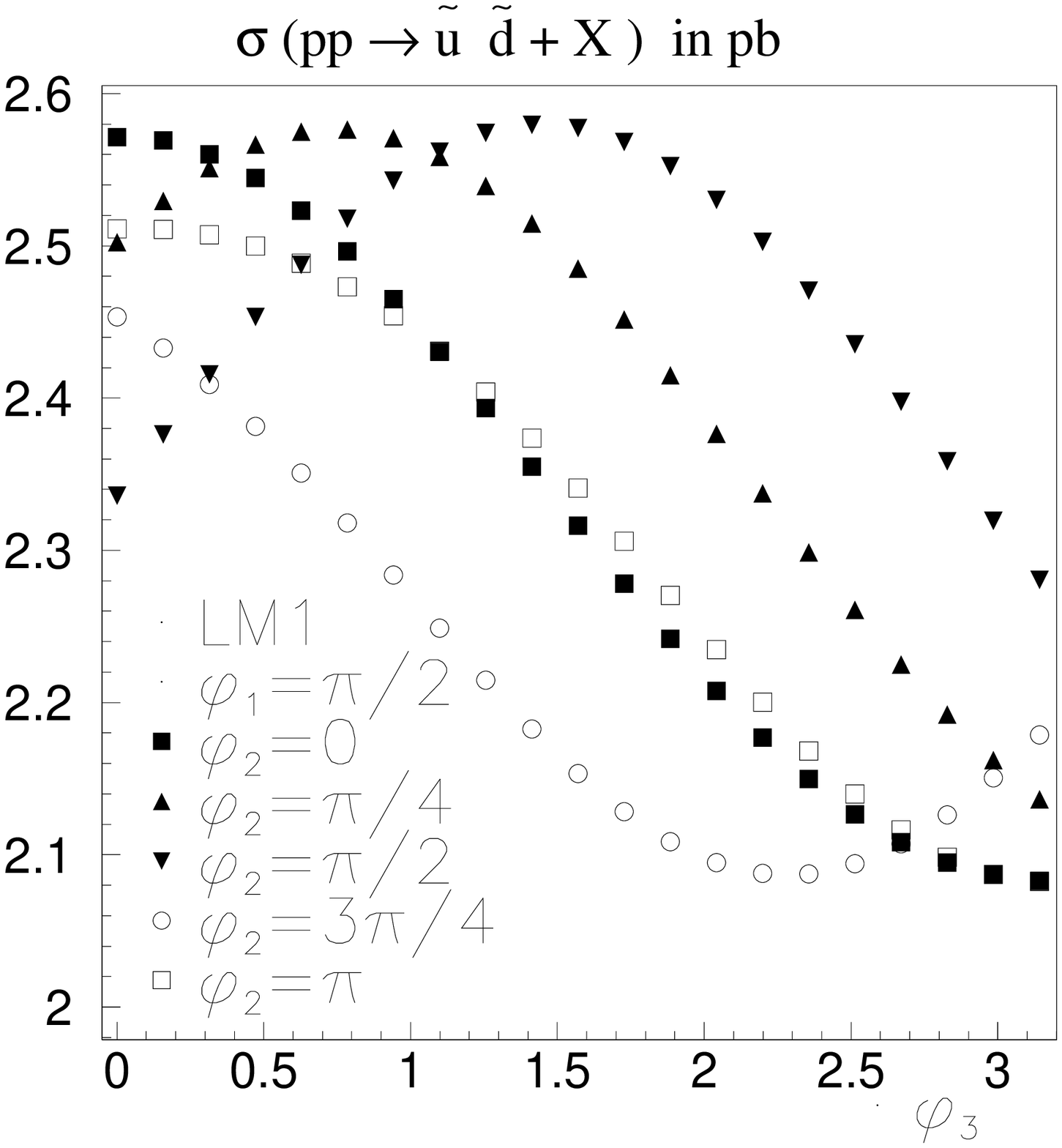}
\includegraphics[width=5cm]{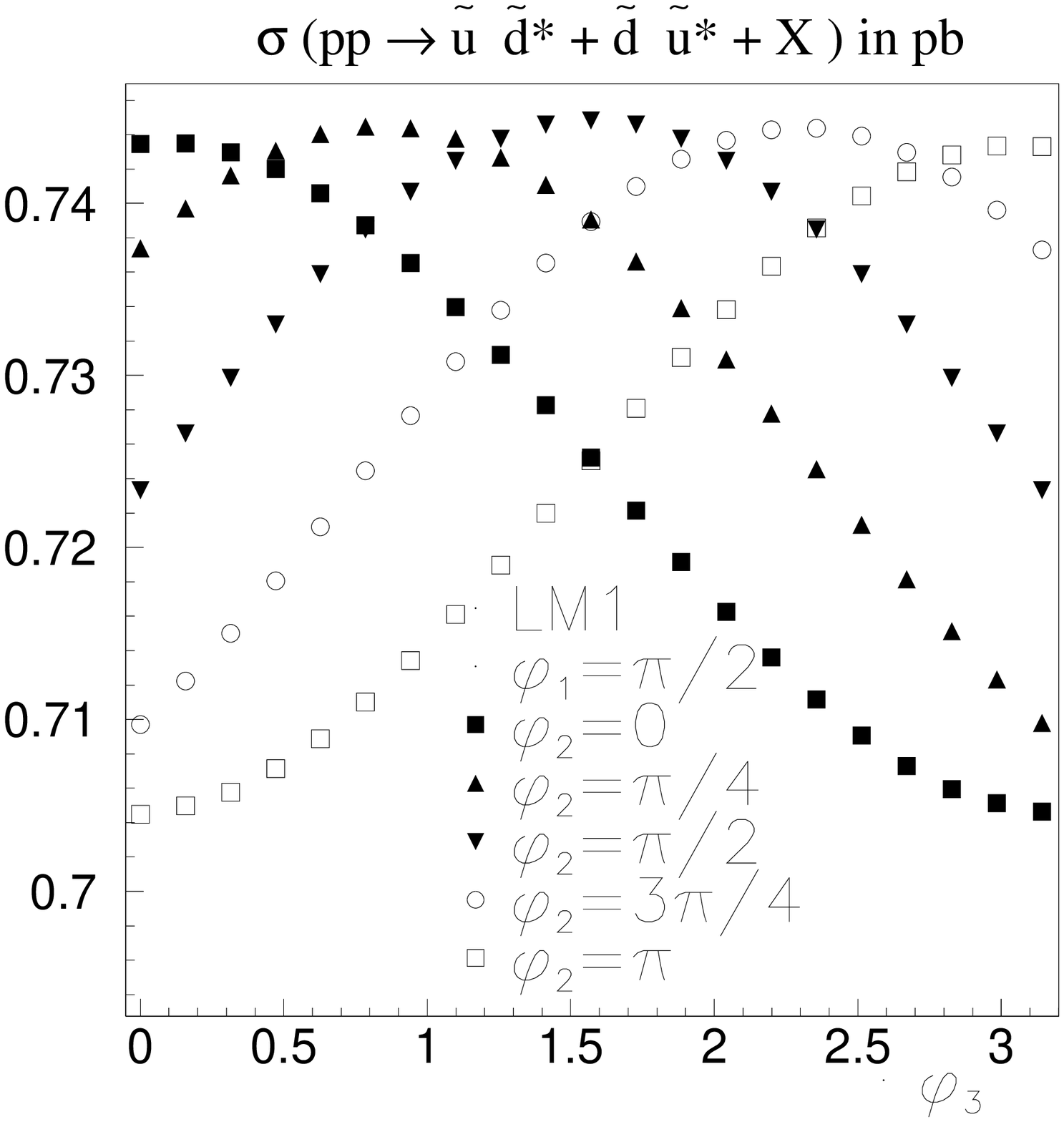}
\includegraphics[width=5cm]{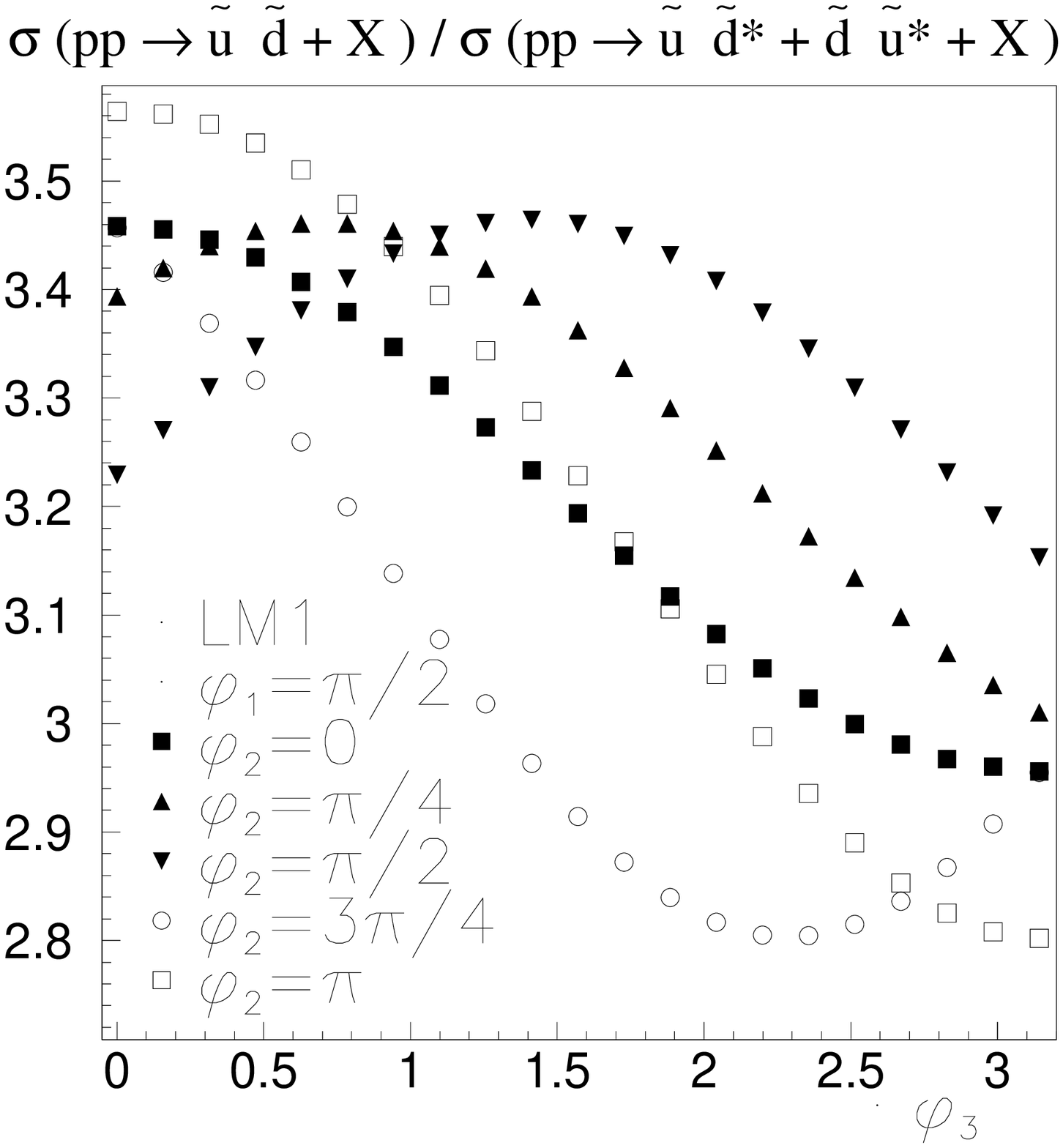}

\caption{The chirality-blind up-and-down squark production cross
sections with the conventions employed in Fig. \ref{fig-uptotal}.}
\label{fig-updowntotal}
\end{figure}

In this subsection we perform a chirality-blind analysis of the
squark pair-production by summing over all chirality combinations
allowed. Depicted in Fig. \ref{fig-uptotal} are $\sigma(p\, p
\rightarrow \widetilde{u} \widetilde{u}) = \sum_{X=L,R; Y=L,R}
\sigma(p\, p \rightarrow \widetilde{u}_X \widetilde{u}_{Y})$ (left
panel) and $\sigma(p\, p \rightarrow \widetilde{u}
\widetilde{u}^{\star}) = \sum_{X=L,R; Y=L,R} \sigma(p\, p
\rightarrow \widetilde{u}_X \widetilde{u}_{Y}^{\star})$ (middle
panel) and their ratios being given in the right panel. Similar
structures with explanations in figure captions are given in Fig.
\ref{fig-downtotal} (for down squark pair production) and in Fig.
\ref{fig-updowntotal} (for associated production of up and down
squarks). In all three figures the upper panels stand for
$\varphi_1=0$ and lower panels for $\varphi_1=\pi/2$. From these
figures we infer that:
\begin{itemize}
\item The chirality-blind cross sections are at the $pb$ level, and
therefore, given the planned luminosity at the LHC, one expects a
large number of events with high-enough statistics for examining
the CP violation effects.

\item The chirality-blind cross sections as well as their ratios do exhibit
significant variations with CP-odd soft phases. If possible
experimentally, detectors could infer if there are CP violation
sources in the underlying model by comparing $\widetilde{q}\,
\widetilde{\hat{q}}$ and $\widetilde{q}\,
\widetilde{\hat{q}}^{\star}$ productions within a specific model,
say, the MSSM.

\item The figures suggest that typically
$\sigma(p\, p \rightarrow \widetilde{u} \widetilde{u})$ /
$\sigma(p\, p \rightarrow \widetilde{u} \widetilde{u}^{\star})
\preceq 3$, $\sigma(p\, p \rightarrow \widetilde{d}
\widetilde{d})$ / $\sigma(p\, p \rightarrow \widetilde{d}
\widetilde{d}^{\star}) \preceq 1.2$, and $\sigma(p\, p \rightarrow
\widetilde{u} \widetilde{d})$ / $\sigma(p\, p \rightarrow
\widetilde{u} \widetilde{d}^{\star}) \preceq 3.5$. These ratios
vary slightly with $\varphi_1$ and significantly with $\varphi_2$
and $\varphi_3$. They give an idea of what contamination of
$\widetilde{q}\, \widetilde{\hat{q}}$ pairs are expected to be in
$\widetilde{q}\, \widetilde{\hat{q}}^{\star}$ signal and vice
versa.
\end{itemize}

The material in this subsection could be useful in situations
where experimentalist does care only on two promt transverse jets.
However, for the plots Fig. \ref{fig-uptotal}, \ref{fig-downtotal}
and \ref{fig-updowntotal} to be useful one needs a precise
knowledge of what net charge the debris in two regions of the
barrel carries. It is with this information that one can compare
cross sections plotted above with a specific model.

\subsection{Squark Pair-Production:  Indefinite Flavor and
Indefinite Chirality}

\begin{figure}
\includegraphics[width=8cm]{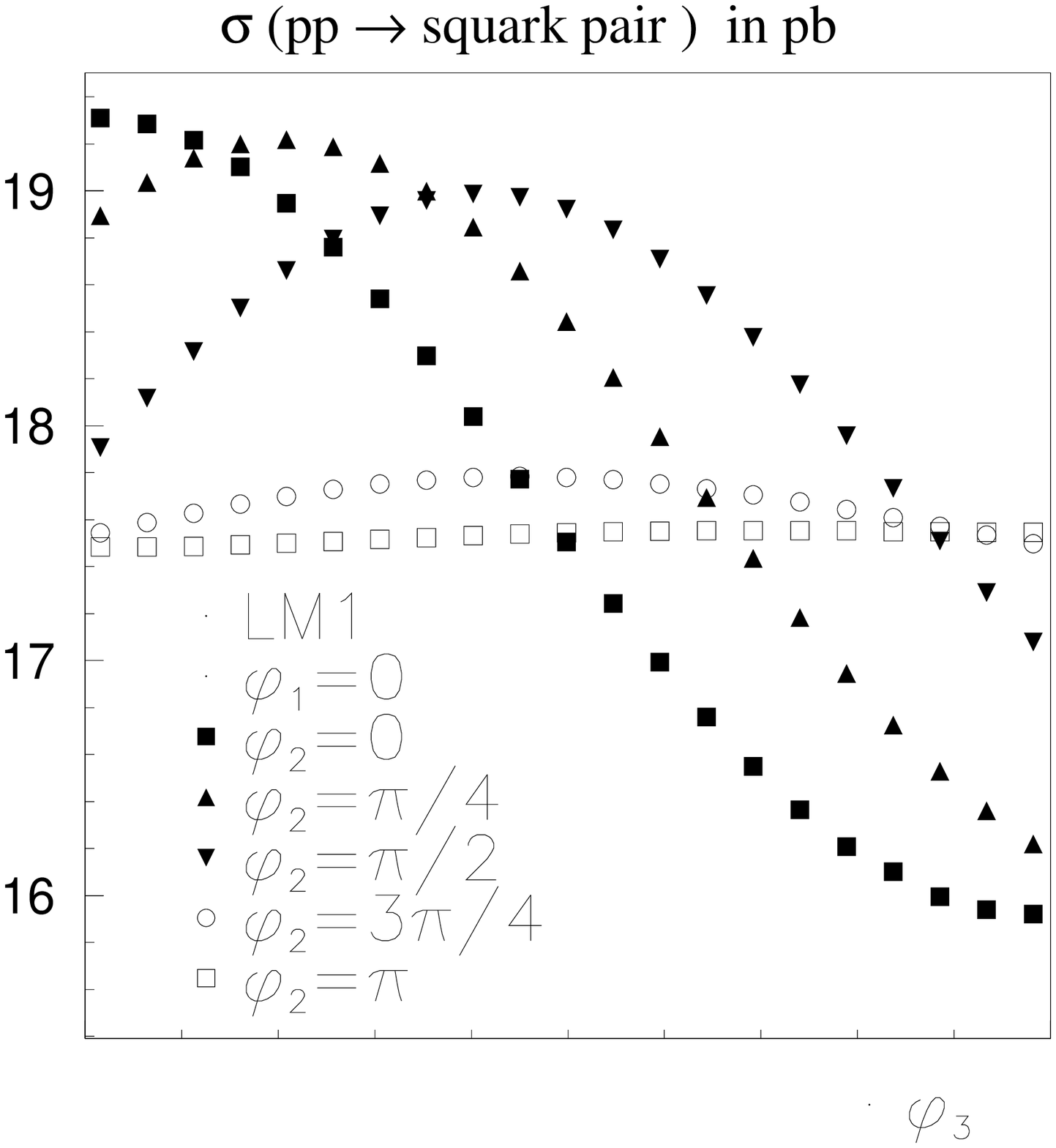}
\includegraphics[width=8cm]{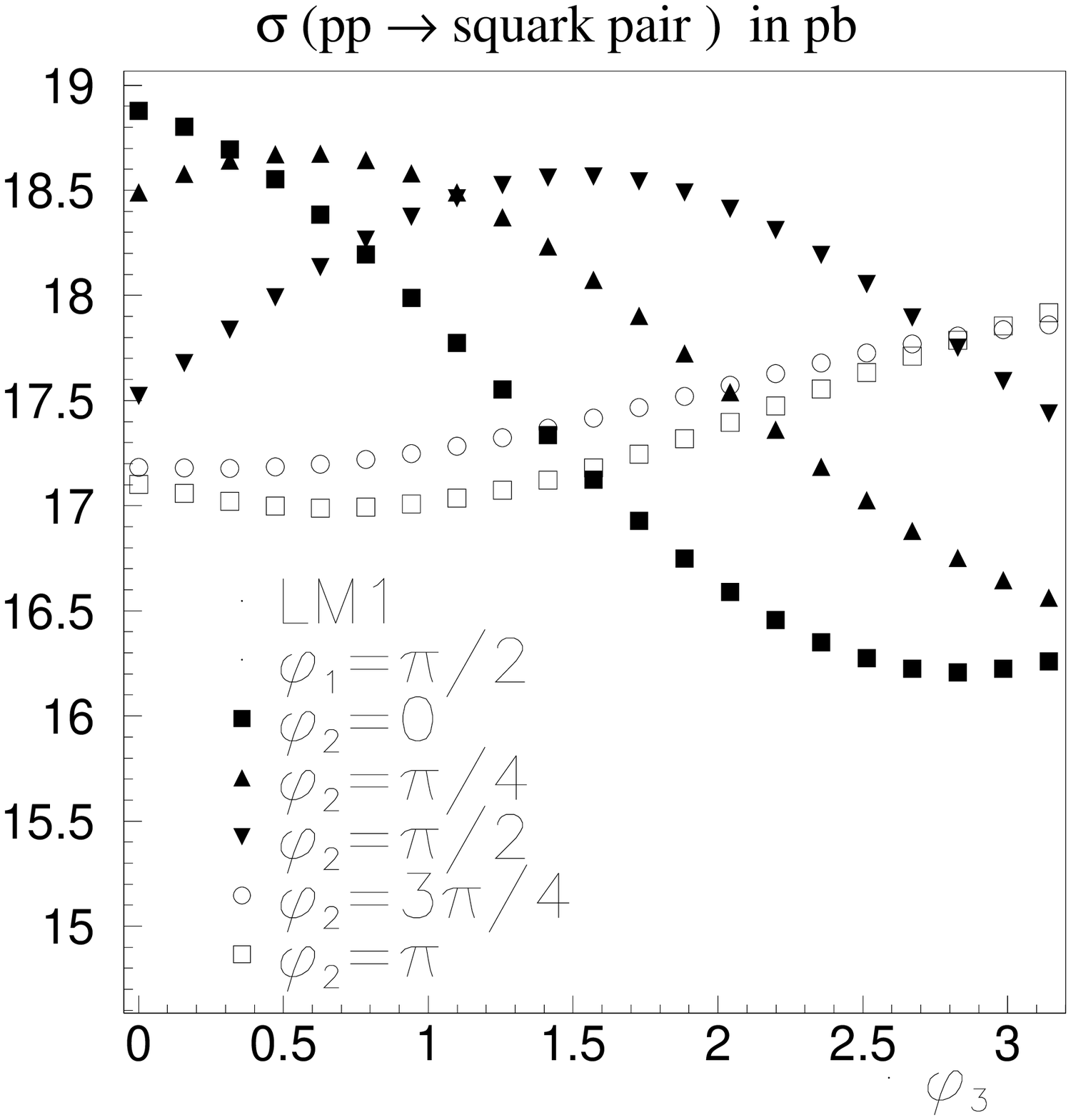}
\caption{Dependence on the soft phases of the total squark
pair-production cross section defined in (\ref{total-sq})
in the text. The left-panel stands for $\varphi_1=0$ and right-panel does for
$\varphi_1=\pi/2$.}\label{fig-total}
\end{figure}

In this subsection we perform a flavor--and--chirality--blind
analysis in that we examine situations in which experimentalist
measures only the rate of producing two high-$\slashchar{P}_T$
jets (disregarding all the leptons and other stuff accompanying
the jet). In this case, direct counting of number of such events
can give an idea about CP-violation sources in the underlying
model. This is exemplified in Fig. \ref{fig-total} by plotting the
total squark pair-production cross section
\begin{eqnarray}
\label{total-sq}
\sigma(p\, p \rightarrow \mbox{squark pair}) = \sum_{
\left\{q,\hat{q}\right\} \in \left\{u,c,d,s\right\}; X=L,R; Y=L,R}
\left[ \sigma(p\, p \rightarrow \widetilde{q}_X
\widetilde{\hat{q}}_{Y}) + \sigma(p\, p \rightarrow
\widetilde{q}_X \widetilde{\hat{q}}_{Y}^{\star})\right]
\end{eqnarray}
which is completely blind to what flavors with what charges and
chiralities are being produced.

This dependence on the soft phases implies that LHC events started
by two high-$\slashchar{P}_T$ jets (disassociating into secondary,
tertiary jets plus leptons plus missing $\slashchar{E}_T$) are
already sensitive to variations in CP-odd phases in gluino and
neutralino sectors of the theory. The advantageous aspect of this
kind of search is that experimentalist does not need to identify
jet charges, leptons, chiralities, missing energy {\it etc}. At
this point a crucial question arises: How does one know that two
high-$\slashchar{P}_T$ events are originating from squarks but not
from two gluinos or from a gluino and a squark? This is indeed a
non-trivial question to answer, and its answer lies in
identification of the final-state particles at the level of Subsection A,
above. Nevertheless, pair-production of squarks differs from those of the gluino
and associated production of gluino and squark in one crucial aspect: Gluino-gluino
and gluino-squark productions are independent of the CP-odd soft
phases. This is a highly advantageous property of squark-pair production
events over the other two since while fitting experimental data to a
specific model, say MSSM, it can be inferred from event rates
(illustrated in Fig. \ref{fig-total}) whether the model
accommodates CP violation sources or not.

\section{Squark Pair-Production and EDM Bounds}
In the previous section we have determined sensitivities of
the cross sections on the soft CP-odd phases by letting
phases to vary in their full ranges and by taking soft
mass parameters at the LM1 point. However, given that EDMs
of electron, neutron and atoms can put stringent limits
on the sizes of CP-odd phases, it is
necessary, for completeness of the analysis, to give
a discussion of the EDM bounds. The material in this section
parallels that of \cite{cancel2} where
authors provide a dedicated study of the EDM bounds,
and compute certain CP-violating observables at a
linear collider within the EDM-favored parameter
regions.

The EDM bounds on CP-odd phases depend crucially
on what values are taken for soft mass parameters
themselves \cite{cancel,cancel1,cancel2}. Gaugino phases,
which are the prime CP-odd parameters for squark (as well as slepton)
pair-production, become relevant only when gaugino masses are
non-universal (at least in phase, as assumed in Sec. IV, above).
The pair-production cross sections of the squarks in first and second generations
are independent of what values are assigned to triliear
couplings: They can be universal or non-universal. These
two cases have been analyzed as 15-parameter MSSM and 23-parameter
MSSM scenarios in \cite{cancel2}. For non-universal gaugino
masses one finds a rather wide parameter space in which
EDMs cancel out with ${\cal{O}}(1)$ values for gaugino phases.
Indeed, setting $\varphi_2=0$ by using the U(1)$_R$
freedom of the MSSM, one finds that $\varphi_{\mu}$ is imprisoned
to lie close to $0$ or $\pi$ whereas all the rest of the phases wander
in $\left[0,2\pi\right]$ interval when gaugino and sfermion masses as well
as trilinear couplings are allowed to vary within $\left[0, 1\ {\rm TeV}\right]$
band (see Figs. 4-9 of \cite{cancel2}). Consider, for instance, Fig.4 or Fig.7 of
\cite{cancel2}. These figures suggest that $\varphi_{\widetilde{B}}$ and
$\varphi_{\widetilde{g}}$ are not constrained at all; they vary in their
full range. In such regions, as expected from the analysis in previous
section, the squark pair-production cross sections will exhibit
strong variation with the phases.

The parameter regions depicted in Figs. 4-9 of \cite{cancel2} and
subsequent discussion of CP-violating observables at a linear collider
indicate that a similar analysis can be carried out for hadron
colliders, in particular, the LHC. In analyzing the correlation
between EDMs and pair-production processes (of sleptons or
squarks) one keeps in mind that latter is sensitive to
relative phase among the gauginos, only. On the other hand,
EDMs depend generically on relative phase between gaugino masses and
trilinear couplings (or $\mu$ parameter) \cite{thomas}.

\section{Conclusion and Future Prospects}
In this work we have analyzed systematically effects of finite
CP-odd phases of soft-breaking parameters on squark
pair-production in $p\,p$ collisions at the LHC energies. Our
observations and results can be summarized as follows:
\begin{itemize}
\item Out of all pairs (squark-squark,
gluino-gluino, squark-gluino) of colored particles, only the
squark production exhibits an explicit dependence on the SUSY
CP-odd phases. Out of all pairs of squarks, only those belonging
to first and second generations exhibit a significant dependence
on the phases.

\item Pair-productions of squarks in first and second generations
are sensitive to CP-odd phases in neutralinos and gluinos, only.
They thus enable one to examine CP violation sources in the ino
sector besides the processes that directly probe inos (neutralino
or chargino pair productions).

\item Depending on the chirality, flavor and electric charge of a given pair of
squarks, pair-production rates change (see Table \ref{table1}),
and this change needs not be small (see
Figs.\ref{fig-sig-up_0}--\ref{fig-updownratio}). In particular,
squark--squark and squark--anti-squark production rates differ
significantly for up-type squark pairs and associated production
of up-and-down squarks.

\item The cross sections exhibit significant variations with the
phases even if flavor, chirality and electric charge of the squark
pairs are left unmeasured (see Figs. \ref{fig-uptotal},
\ref{fig-downtotal}, \ref{fig-updowntotal} and especially Fig.
\ref{fig-total}). In Fig.\ref{fig-total},  the total swings of
the cross sections $i.e.$ the difference between their
extrema vary between $\approx 0.5\ {\rm pb}$ to $ 4\ {\rm pb}$,
which should be a measurable signal at the LHC.

\item The discussions in Sec.V show that there are rather
wide regions in SUSY parameter space where EDMs are sufficiently
suppressed (via cancellation of various contributions) with
${\cal{O}}(1)$ phases for gauginos. In such regions of the parameter
space, squark pair-production must feel phases significantly (similar
to ones shown in Sec.IV).

\end{itemize}
In light of these observations and results, we find squark
production processes as an important probe of CP violation sources
in the theory.

In spite of their clear and guiding aspects, the results above are
far from being sufficient for a definitive conclusion since:
\begin{itemize}
\item The analysis in Sec. IV is restricted to a specific benchmark point
LM1. It is necessary to cover different portions of the SUSY
parameter space, as wide as possible, so as to determine golden
regions for putting discovery limits.

\item The results above far from telling what will
actually happen in a given LHC detector. Indeed, detector
responses, background, jet identification, cuts, \dots all are to
be implemented before reaching a definite answer for signal
significance. The work in this direction is in progress
\cite{nasuf}.

\item The discussions of the EDM bounds in Sec.V, though
sufficient for having a 'proof of existence' of parameter
regions with ${\cal{O}}(1)$ CP-odd phases, must be rectified
with a full scan of the parameter space so as to determine
correlation among EDMs and cross sections. In any case,
the EDM bounds are to be incorporated by
assuming from the scratch that squarks of first two generations
are light enough to be pair-produced at the LHC.

\item It is necessary to rectify the LO cross sections discussed
here by incorporating  NLO QCD effects. They are expected to
stabilize results against variations in renormalization/decoupling
scale, and their contributions are expected to be ${\cal{O}}(20
\%)$ level.

\item In general, identification of sparticles at hadron colliders is a
nontrivial task as it involves the reconstruction of the masses,
couplings and chiralities from incomplete (due to missing energy
signals) final states comprising leptons and jets. Although
several studies of the supersymmetric parameter space have already
resulted in a set of benchmark points (see \cite{post-lep} and
references therein), a full and precise determination of the
spectrum calls for more general techniques for sparticle
identification \cite{CMS}, and might eventually require a
complementary lepton collider \cite{denegri}. Nevertheless, as
confirmed by the phase-dependencies of the total cross sections in
Figs. \ref{fig-uptotal}, \ref{fig-downtotal} and
\ref{fig-updowntotal}, it is possible to extract important
information about CP violation characteristics of the ino sector.

\end{itemize}
The results of this work, with reservations just listed, show that
squark pair-production is an important process to probe CP
violation sources in the theory in addition to testing various
aspects pertaining to flavor structures and scale of the new
physics.

\section{Acknowledgements}
D. A. D. thanks CERN Theory Division where part of this work was
done. The work of D. A. D. and K. C. were partially supported by
the Scientific and Technological Research Council of Turkey
through project 104T503. The work of D. A. D. was partially
supported by Turkish Academy of Sciences through GEBIP grant.


\begin{thebibliography}{99}

\bibitem{Chung:2003fi}
  D.~J.~H.~Chung, L.~L.~Everett, G.~L.~Kane, S.~F.~King, J.~D.~Lykken and L.~T.~Wang,
  Phys.\ Rept.\  {\bf 407}, 1 (2005)
  [arXiv:hep-ph/0312378].

\bibitem{nima}
N.~Arkani-Hamed, G.~L.~Kane, J.~Thaler and L.~T.~Wang,
  arXiv:hep-ph/0512190.

\bibitem{pape}
L.~Pape and D.~Treille,
  Rept.\ Prog.\ Phys.\  {\bf 69}, 2843 (2006).


\bibitem{flavor-cp}
 J.~S.~Hagelin, S.~Kelley and T.~Tanaka,
  Nucl.\ Phys.\ B {\bf 415}, 293 (1994);
F.~Gabbiani, E.~Gabrielli, A.~Masiero and L.~Silvestrini,
  Nucl.\ Phys.\ B {\bf 477}, 321 (1996)
  [arXiv:hep-ph/9604387];
S.~Pokorski, J.~Rosiek and C.~A.~Savoy,
  Nucl.\ Phys.\ B {\bf 570}, 81 (2000)
  [arXiv:hep-ph/9906206].


\bibitem{cross}
S.~Dawson, E.~Eichten and C.~Quigg,
Phys.\ Rev.\ D {\bf 31}, 1581 (1985).

\bibitem{cross1}
V.~D.~Barger, K.~Hagiwara, W.~Y.~Keung, R.~J.~N.~Phillips and
J.~Woodside,
  Phys.\ Rev.\ D {\bf 32}, 806 (1985).


\bibitem{Beenakker:1996ch}
W.~Beenakker, R.~Hopker, M.~Spira and P.~M.~Zerwas,
Nucl.\ Phys.\ B {\bf 492}, 51 (1997) [arXiv:hep-ph/9610490].

\bibitem{Beenakker3}
W.~Beenakker, R.~Hopker, M.~Spira and P.~M.~Zerwas,
  Phys.\ Rev.\ Lett.\  {\bf 74}, 2905 (1995)
  [arXiv:hep-ph/9412272].

\bibitem{tilman}
T.~Plehn, D.~Rainwater and P.~Skands,
  arXiv:hep-ph/0510144.

\bibitem{wyler}
T.~Gehrmann, D.~Maitre and D.~Wyler,
  Nucl.\ Phys.\  B {\bf 703}, 147 (2004)
  [arXiv:hep-ph/0406222].

\bibitem{beenakker2}
W.~Beenakker, M.~Kramer, T.~Plehn, M.~Spira and P.~M.~Zerwas,
  Nucl.\ Phys.\ B {\bf 515}, 3 (1998)
  [arXiv:hep-ph/9710451].

\bibitem{stoppair}
G.~Bozzi, B.~Fuks and M.~Klasen,
  Phys.\ Rev.\ D {\bf 72}, 035016 (2005)
  [arXiv:hep-ph/0507073].

\bibitem{flavor-cp2}
D.~A.~Demir,
  Phys.\ Lett.\ B {\bf 571}, 193 (2003)
  [arXiv:hep-ph/0303249];
J.~Foster, K.~i.~Okumura and L.~Roszkowski,
  Phys.\ Lett.\ B {\bf 609}, 102 (2005)
  [arXiv:hep-ph/0410323];
  JHEP {\bf 0508}, 094 (2005)
  [arXiv:hep-ph/0506146].

\bibitem{cancel}
A.~Bartl, T.~Gajdosik, W.~Porod, P.~Stockinger and H.~Stremnitzer,
Phys.\ Rev.\ D {\bf 60}, 073003 (1999)
[arXiv:hep-ph/9903402];

T.~Ibrahim and P.~Nath,
Phys.\ Rev.\ D {\bf 58} (1998) 111301 [Erratum-ibid.\ D {\bf 60}
(1999) 099902] [arXiv:hep-ph/9807501];
M.~Brhlik, G.~J.~Good and G.~L.~Kane,
Phys.\ Rev.\ D {\bf 59}, 115004 (1999) [arXiv:hep-ph/9810457].

\bibitem{cancel1}
S.~Abel, S.~Khalil and O.~Lebedev,
  Nucl.\ Phys.\  B {\bf 606}, 151 (2001)
  [arXiv:hep-ph/0103320].

\bibitem{cancel2}
V.~D.~Barger, T.~Falk, T.~Han, J.~Jiang, T.~Li and T.~Plehn,
  Phys.\ Rev.\  D {\bf 64}, 056007 (2001)
  [arXiv:hep-ph/0101106].

\bibitem{thomas}
M.~E.~Peskin,
  Int.\ J.\ Mod.\ Phys.\ A {\bf 13}, 2299 (1998)
  [arXiv:hep-ph/9803279];
S.~D.~Thomas,
  Int.\ J.\ Mod.\ Phys.\ A {\bf 13}, 2307 (1998)
  [arXiv:hep-ph/9803420].

\bibitem{biz}
D.~A.~Demir, O.~Lebedev, K.~A.~Olive, M.~Pospelov and A.~Ritz,
  Nucl.\ Phys.\ B {\bf 680}, 339 (2004)
  [arXiv:hep-ph/0311314].

\bibitem{beyondmssm}
A.~T.~Alan,
  Phys.\ Rev.\  D {\bf 72}, 115006 (2005)
  [arXiv:hep-ph/0508252];
M.~Pospelov, A.~Ritz and Y.~Santoso,
  Phys.\ Rev.\ Lett.\  {\bf 96}, 091801 (2006)
  [arXiv:hep-ph/0510254].

\bibitem{2loop}
K.~A.~Olive, M.~Pospelov, A.~Ritz and Y.~Santoso,
  Phys.\ Rev.\ D {\bf 72}, 075001 (2005)
  [arXiv:hep-ph/0506106].


\bibitem{CTQ5}
H.~L.~Lai {\it et al.}  [CTEQ Collaboration],
Eur.\ Phys.\ J.\ C {\bf 12}, 375 (2000) [arXiv:hep-ph/9903282].


\bibitem{post-lep}
M.~Battaglia {\it et al.},
  Eur.\ Phys.\ J.\ C {\bf 22}, 535 (2001)
  [arXiv:hep-ph/0106204].

\bibitem{post-wmap}
M.~Battaglia, A.~De Roeck, J.~R.~Ellis, F.~Gianotti, K.~A.~Olive
and L.~Pape,
  Eur.\ Phys.\ J.\ C {\bf 33}, 273 (2004)
  [arXiv:hep-ph/0306219].


\bibitem{CMS}
S.~Abdullin {\it et al.}  [CMS Collaboration],
  J.\ Phys.\ G {\bf 28}, 469 (2002)
  [arXiv:hep-ph/9806366].

\bibitem{tarek}
T.~Ibrahim and P.~Nath,
  Phys.\ Rev.\ D {\bf 67}, 095003 (2003)
  [Erratum-ibid.\ D {\bf 68}, 019901 (2003)]
  [arXiv:hep-ph/0301110].

\bibitem{bsgam}
D.~A.~Demir and K.~A.~Olive,
  Phys.\ Rev.\ D {\bf 65}, 034007 (2002)
  [arXiv:hep-ph/0107329].

\bibitem{pythia}
T.~Sjostrand, S.~Mrenna and P.~Skands,
  JHEP {\bf 0605}, 026 (2006)
  [arXiv:hep-ph/0603175].

\bibitem{nasuf}
K.~Cankocak, D.~A.~Demir and N.~Sonmez, CMS Physics Simulation
Study (in progress).

\bibitem{denegri}
D.~Denegri, priviate communication.

\end{thebibliography}
\end{document}